\documentclass[12pt,letterpaper]{lsuetd}
\usepackage{setspace,graphics,dsfont,verbatim,paralist,indentfirst}
\usepackage{graphicx}
\usepackage{mathtools}
\usepackage{mathrsfs}
\usepackage[hidelinks=false]{hyperref}
\usepackage{braket}
\usepackage{subfigure}
\usepackage{amsmath}
\usepackage{amssymb}
\usepackage{caption}
\usepackage{xspace}
\usepackage{ifpdf}
\usepackage[usenames,dvipsnames]{xcolor}
\usepackage{titlesec}
\usepackage{etoolbox}

\titleformat{\section}
  {\normalfont\fontsize{12}{12}\bfseries}{\thesection}{1em}{}
\titleformat{\subsection}
  {\normalfont\fontsize{12}{12}\bfseries}{\thesubsection}{1em}{}

\newcommand{\ip}[1]{\langle #1 \rangle}
\newcommand{\cc}[1]{\textsc{#1}}
\newcommand{\BS}[0]{\textsc{BosonSampling}\xspace}
\newcommand{\PP}[0]{\textbf{P}\xspace}
\newcommand{\coNP}[0]{\textbf{co-NP}\xspace}
\newcommand{\NP}[0]{\textbf{NP}\xspace}
\newcommand{\BPP}[0]{\textbf{BPP}\xspace}
\newcommand{\BQP}[0]{\textbf{BQP}\xspace}
\newcommand{\PostBQP}[0]{\textbf{PostBQP}\xspace}
\newcommand{\PH}[0]{\textbf{PH}\xspace}
\newcommand{\sharpP}[0]{\textbf{\#P}\xspace}
\newcommand{\Or}[0]{$\mathcal{O}$\xspace}
\newcommand{\PASSV}[0]{\textsc{PASSVSampling}\xspace}
\newcommand{\perm}[0]{\mathrm{perm}}
\makeatletter

\begin{document}
\renewcommand\@pnumwidth{1.55em}
\renewcommand\@tocrmarg{9.55em}
\renewcommand*\l@chapter{\@dottedtocline{0}{1.5em}{2.3em}}
\renewcommand*\l@figure{\@dottedtocline{1}{0em}{3.1em}}
\let\l@table\l@figure

\pagenumbering{roman}
\thispagestyle{empty}
\begin{center}
COMPLEXITY THEORY AND ITS APPLICATIONS IN LINEAR QUANTUM OPTICS

\vfill
\doublespacing
A Dissertation \\
\singlespacing
Submitted to the Graduate Faculty of the \\
Louisiana State University and \\
Agricultural and Mechanical College \\
in partial fulfillment of the \\
requirements for the degree of \\
Doctor of Philosophy\\
\doublespacing
in \\
                                       
The Department of Physics and Astronomy \\
\singlespacing
\vfill

by \\
Jonathan P. Olson \\
M.S., University of Idaho, 2012 \\
August 2016
\end{center}
\pagebreak



\chapter*{Acknowledgments}
\doublespacing
\vspace{0.55ex}
My advisor, Jonathan Dowling, is apt to say, ``those who take my take my advice do well, and those who don't do less well.''  I always took his advice (sometimes even against my own judgement) and I find myself doing well.  He talked me out of a high-paying, boring career, and for that I owe him a debt I will never be able to adequately repay.

My mentor, Mark Wilde, inspired me to work hard without saying a word about what I ``should'' be doing, and instead leading by example.  His impressive work ethic and productivity continues to set a bar for me that I continue to strive toward.

My best friend, Masaki Ikeda, was and still is a constant distraction.  But without his friendship, I surely never would have made it to this point without intense mental hospitalization.

My girlfriend Kat, whom I love, spent many evenings with me as I worked at nearly every coffee shop in the city of Baton Rouge.  I'm not sure I could have graduated on time without her support and presence.

My many-time coauthors, Peter Rohde and Keith Motes, made almost all of the work in this thesis possible.  I hope for the continued success and collaboration with the Thunder from Down Under. 

Dr. Giulia Ferrini, who has made many helpful comments and contributions over the course of my Ph.D. career, thank you.

Finally, my family, friends, and thesis commitee affected me in so many positive ways that I cannot begin to enumerate them.  Thank you all.

\addcontentsline{toc}{chapter}{\hspace{-1.5em} {ACKNOWLEDGMENTS} \vspace{12pt}}
\pagebreak

\chapter*{Preface}
\doublespacing
\vspace{0.55ex}
The topics presented in this thesis span a number of fields in physics and computer science.  With each field comes a certain set of common misconceptions and ``well-known'' results that are actually almost impossible to find or decipher from the literature.  Because of my many struggles trying to put together and comprehend all of these little pieces, I am writing this thesis with the intention of providing a resource for those who are new to one or more of these topics.  I try my best to make this work as hierarchical as possible, so that if the reader is in any way confused, he is always pointed to the right place to answer his questions.

If you are unfamiliar with the mathematical formalisms that pervade the literature, of course the preliminary sections of Chapter 1 and Chapter 4 are a good place to start.  However, if you have some experience, you should feel comfortable skipping these sections with the understanding that any substantive discussions in these sections will be referenced when they become relevant.

\addcontentsline{toc}{chapter}{\hspace{-1.5em} {PREFACE} \vspace{12pt}}
\pagebreak

\singlespacing
\tableofcontents
\pagebreak



\renewenvironment{abstract}{{\hspace{-2.2em} \huge \textbf{\abstractname}} \par}{\pagebreak}
\addcontentsline{toc}{chapter}{\hspace{-1.5em} ABSTRACT}
\begin{abstract}
\vspace{0.55ex}
\doublespacing
This thesis is intended in part to summarize and also to contribute to the newest developments in passive linear optics that have resulted, directly or indirectly, from the somewhat shocking discovery in 2010 that the \BS problem is likely hard for a classical computer to simulate.  In doing so, I hope to provide a historic context for the original result, as well as an outlook on the future of technology derived from these newer developments.  An emphasis is made in each section to provide a broader conceptual framework for understanding the consequences of each result in light of the others.  This framework is intended to be comprehensible even without a deep understanding of the topics themselves.

The first three chapters focus more closely on the \BS result itself, seeking to understand the computational complexity aspects of passive linear optical networks, and what consequences this may have.  Some effort is spent discussing a number of issues inherent in the \BS problem that limit the scope of its applicability, and that are still active topics of research.  Finally, we describe two other linear optical settings that inherit the same complexity as \BS. 

The final chapters focus on how an intuitive understanding of \BS has led to developments in optical metrology and other closely related fields.  These developments suggest the exciting possibility that quantum sensors may be viable in the next few years with only marginal improvements in technology.  Lastly, some open problems are presented which are intended to lay out a course for future research that would allow for a more complete picture of the scalability of the architecture developed in these chapters.

\end{abstract}

\pagenumbering{arabic}
\addtocontents{toc}{\vspace{12pt} \hspace{-1.8em} CHAPTER \vspace{-1em}}

\singlespacing
\setlength{\textfloatsep}{12pt plus 2pt minus 2pt}
\setlength{\intextsep}{6pt plus 2pt minus 2pt}
\chapter{Linear Quantum Optics}
\doublespacing
\section{Historical Introduction} \label{sec:history}
The origins of quantum optics parallel the birth of quantum theory itself, and may be said to have begun with Einstein's discovery of the photoelectric effect in 1905.  Since then, the understanding that nature has a kind of dual nature, where particles and waves can exist in tandem, has increasingly pervaded popular culture.  Numerous technological applications of quantum optics -- most notably,  lasers -- are now so integral to our society that it would be difficult to imagine operating without them.

Yet, shockingly many aspects of the quantum nature of light still remain poorly understood.  One such area, the connection of optics to quantum computing and complexity theory, is indeed the entire motivation of this thesis.  But before we delve into the intricacies of this topic, it is helpful to understand what we \textit{do} know, as this context helps inform why there has been a great deal of recent research interest.

Although optical networks have been used for interferometry for many years, the latest push in research is due at least in part to the advent of quantum computing.  Although previous quantum algorithms had shown impressive speedups over their classical counterparts, in 1994 Peter Shor demonstrated a \textit{useful} quantum algorithm (integer factorization) that gives an exponential speedup over the best classical algorithm available \cite{bib:ShorAlg}.  Because common encryption algorithms such as RSA rely on the hardness of factoring large numbers \cite{bib:RSA}, the field of quantum computing was suddenly taken very seriously (though the older Simon's algorithm has recently found new application in cryptography).   

It had been known prior to 1994 that optical networks employing nonlinearities were capable of universal quantum computation, but the possibility that efficient linear optical quantum networks could perform the same seemed far fetched.  It had been shown throughout the 1990s that linear interferometers could perform integer factorization, solve \NP-complete problems and perform universal quantum computation, but not without an exponential overhead in either the energy or spatial dimension of the system \cite{bib:cerny, bib:clauser, bib:cerf}.

This changed in 2000, when Knill, LaFlamme, and Milburn (KLM) devised a scheme which allows for universal quantum computing with only a polynomial overhead \cite{bib:KLM01}.  Subsequent improvements on the KLM protocol have since been discovered \cite{bib:kok2007linear}, but still the linear optical quantum computing (LOQC) architecture seems to be an unlikely candidate for a truly scalable implementation of universal quantum computing.  The difficulty lies mostly in the challenge of providing ancillary resources, teleportation, error correction, and maintaining a coherent optical quantum memory \cite{bib:lapaire2003conditional}.  Some of these additional components are referred to as ``active" or ``adaptive'' components of an optical network, since they require a certain interaction within the network based on measurements made during the computation.

Passive linear optical networks are instead characterized by modes which use only beam splitters and single-mode phase shifters.  Although certainly still nontrivial to implement, passive networks greatly reduce the complexity inherent when scaling the size of the network.  However, a number of these passive networks had been known to be efficiently classically simulable, and thus incapable of showing any kind of interesting quantum advantage.  For instance, it is known that a network of Gaussian-state inputs together with Gaussian measurements are classically simulable \cite{bib:BartSand03}.  It is also known that the probability for measuring a particular basis state in an $n$-photon linear optical experiment can be approximated efficiently \cite{bib:Gurvits}.

It was a great surprise, then, when Arkhipov and Aaronson showed that a particular sampling problem (\BS ) based on an $n$-photon passive linear optical network could likely not be efficiently simulated by a classical computer \cite{bib:AA10}.  The essence of the complexity of this problem is in tying the output probabilities to the computation of a matrix permanent, which is known to be exceptionally hard to calculate in the exact case.  Excepting the verification problem (discussed in Sec.~\ref{sec:limit}), the primary criticism of \BS is that simulating the output of such a system \textit{does not actually solve any problem of interest}.  In a sense, \BS mirrors the kind of quantum algorithms that were discovered before the advent of Shor's algorithm.

This thesis, which focuses on the developments of post-\BS linear optics, is generally split into two parts.  The first is to describe other passive linear optical networks which share the same complexity as \BS, in hopes of better understanding what aspects of the network make a hard sampling problem.  The second is to describe attempts to adapt the ideas of \BS into a protocol that either directly solves a problem of interest, or exploits some of the ``resources'' inherent in \BS to create useful quantum technologies.

\section{Preliminaries} \label{sec:prelim}
The following sections are designed to introduce the reader to the mathematical framework that is used throughout the rest of this thesis.  I will also explicitly define a consistent notation that is used throughout, though most readers already experienced with the framework will likely follow the notation already, as it is quite standard in the field.  

Also, while I would love to discuss the formal treatment of ``classical randomness" in this thesis, it is not a necessary component for understanding the rest of this thesis.  The theory of density operators is pivotal to understanding most of the facets of quantum theory, and I am somewhat shocked that for the work presented here, it happens to be unnecessary.  While the following section attempts to give some idea of how quantum theory results in a more powerful model of computing than a classical one, a true comparison needs to consider mixed states, purification, and the partial trace.  The reader can find these concepts in \cite{bib:NielsenChuang00, bib:Wilde}, if interested.  The resulting section here may be considered to be a major simplification of both classical and quantum theory, but it is intended to be as such.

\subsection{Quantum vs.~Classical State Spaces} \label{subsec:quantum}
In this section, we give a short introduction to some general concepts that are meant to give the reader a good understanding of the difference between a quantum theory, and a classical one.  Although there is much to be said on the topic, we will restrict ourselves to what is necessary to understand the remainder of this thesis.    A firm grasp of linear algebra is the only mathematical prerequisite, though some concepts from probability and information theory will certainly be permeated within.  The ideas presented in this section are mostly summaries of chosen topics from \cite{bib:NielsenChuang00, bib:Wilde}, which are an excellent source for the reader if they are interested in a more in-depth discussion.

A natural place to begin is what might be considered the axioms, or postulates, of quantum theory.  We take these postulates directly from \cite{bib:NielsenChuang00}, and expand on their relevance to quantum theory and to the mathematical framework within this thesis. 
\begin{itemize}
    \item[{}] \textbf{Postulate 1:} Associated to any isolated physical system is a complex vector space with inner product (that is, a Hilbert space) known as the \textit{state space} of the system. The system is completely described by its \textit{state vector}, which is a unit vector in the system's state space.
\end{itemize}

The first postulate provides the setting in which any quantum theory resides -- a Hilbert space.  One of the consequences of quantum theory residing in a vector space is that, for any space with dimension greater than one, there exists unit vectors that are linear combinations, or superpositions, of its basis states.  These superpositions of states are themselves entirely valid states.  Mathematics and physical intuition both seem to suggest that the system which these states describe is simultaneously in more than one basis state at the same time.  We can see this effect directly once we define the other two postulates.

It is important to note that throughout this discussion, we take the existence of the systems which underly these state spaces somewhat for granted.  That is, what we refer to as the classical and quantum descriptions of these state spaces are taken to compare classical probability theory with what is consistent with quantum theory.  We do not, for example, compare the state spaces that are derived from a Hamiltonian in classical mechanics directly with those of quantum mechanics.  The claims we make in this section should be understood to be in this context.  For an alternate formulation of quantum theory involving more (but as the author argues, simpler) axioms, see Ref. \cite{bib:Hardy01}.

Note that throughout this thesis, we will use the standard ``bra-ket notation" or ``Dirac notation", where $\ket{\cdot}$ indicates a state vector.  We use calligraphic capital letters, such as $\mathcal{H}$, to indicate a state space.  If you are unfamiliar with this notation, please refer to Section 2.1 of \cite{bib:NielsenChuang00}.  

\begin{itemize}
    \item[{}] \textbf{Postulate 2:} The evolution of a \textit{closed} quantum system is described by a \textit{unitary transformation}. That is, the state $\ket{\psi}$ of the system at time $t_1$ is related to the state $\ket{\psi'}$ of the system at time $t_2$ by a unitary operator $U$ which depends only on the times $t_1$ and $t_2$,
\end{itemize}
\begin{equation}
\ket{\psi'}=U\ket{\psi}.
\end{equation}

A unitary operator is a bounded linear operator that satisfies $UU^\dagger=U^\dagger U=I$, where $U^\dagger$ is the Hermitian adjoint of $U$.  Because $U^\dagger$ is also unitary, the condition that $UU^\dagger=U^\dagger U=I$ can be thought of as the time reversibility of quantum evolution.  Unitary operators perserve the \textit{inner product}---denoted $\ip{\cdot,\cdot}$---on $\mathcal{H}$, so that for two vectors $x,y\in\mathcal{H}$, $\ip{x,y}=\ip{Ux,Uy}$.  The latter condition can be thought of as a unitary transformation being a kind of ``rotation" of the vectors in Hilbert space.  

It is important to stress that the evolution is necessarily unitary \textit{only if the system is closed}.  Of course, we obviously care about how a system may evolve if it is not closed, and we partially answer this with the third postulate (related to measurement).  One answer (though perhaps a bit unsatisfying) is to say that the open system is only a part of a larger system that \textit{is} closed -- even if we have to consider the state space of the entire universe.  A more complete answer is that evolution of an open system can be described by a CPTP map (completely positive and trace preserving), often referred to as a \textit{quantum channel}.  We do not need the tools related to CPTP maps for the purposes of this thesis, but a curious reader can refer to \cite{bib:Wilde} for more information.

\begin{itemize}
    \item[{}] \textbf{Postulate 3:} Quantum measurements are described by a collection $\{M_m\}$ of \textit{measurement operators}. These are operators acting on the state space of the system being measured. The index $m$ refers to the measurement outcomes that may occur in the experiment. If the state of the quantum system is $\ket{\psi}$ immediately before the measurement then the probability that result $m$ occurs is,
\end{itemize}
\begin{equation}
\bra{\psi}M_m^\dagger M_m \ket{\psi},
\end{equation}
\begin{itemize} \item[{}] and the state of the system after the measurement is, \end{itemize}
\begin{equation}
\frac{M_m\ket{\psi}}{\sqrt{\bra{\psi}M_m^\dagger M_m\ket{\psi}}}.
\end{equation}
\begin{itemize} \item[{}] The measurement operators satisfy the completeness equation, \end{itemize}
\begin{equation}
\sum_{m} M_m^\dagger M_m=I.
\end{equation}

Postulate 3 is crucial because it explains the way in which we interact with a quantum system and the consequence of that interaction.  It is at this point where the difference between a quantum theory and classical one takes shape.  First, note that measurement is non-unitary, which is particularly evident in the fact that it is \textit{not} time-reversible.  While a classical theory purports that the universe behaves independently of an observer, a quantum theory is inextricably tied to the observer.  It might seem at first that this is a more restrictive condition than classical theory, but together with Postulates 1 and 2, we will see that it allows for a much richer (albeit counterintuitive) context for computing.

Let us consider the example of a qubit -- short for ``quantum bit'' -- which is the simplest non-trivial example of a state space.  The Hilbert space $\mathcal{H}=\mathbb{C}^2$ is spanned by a two-dimensional basis $B=\{\ket{0},\ket{1}\}$ where,
\begin{equation}
\ket{0}= \left[
\begin{array}{l}
      1 \\ 0 
\end{array} 
\right] \qquad
\ket{1}= \left[
\begin{array}{l}
      0 \\ 1 
\end{array} 
\right].
\end{equation}
There are many physical systems that can be described by a qubit, but for the purposes of this example, we will consider a two-level atom whose basis states  represent the ground state $\ket{0}$ and the excited state $\ket{1}$.  Hence, any \textit{quantum} state $\ket{\psi}$ in this system can be written as $\ket{\psi}=\alpha\ket{0}+\beta\ket{1}$ for some $\alpha,\beta\in\mathbb{C}$ satisfying $|\alpha|^2+|\beta|^2=1$ (since Postulate~1 requires that states correspond to unit vectors).

If we use the same two-level atom to describe a classical bit $b$, what possible states can it correspond to?  Classical theory requires that states, in principle, should always be distinguishable from each other.  We define distinguishability in the following sense: two states $\ket{\sigma},\ket{\tau}$ are distinguishable from one another if there exists a measurement $M_{dis}\in\{M_m\}$ such that $\bra{\sigma}M_{dis}^\dagger M_{dis} \ket{\sigma}=1$ and $\bra{\tau}M_{dis}^\dagger M_{dis} \ket{\tau}=0$ (or vice versa).  In other words, we must have a measurement that, given an input restricted to $\{\ket{\sigma},\ket{\tau}\}$, always fails in one case and succeeds in the other.  It is not hard to see that this reduces to the case that $\ip{\sigma|\tau}=0$; we say, in this case, that $\sigma$ and $\tau$ are \textit{orthogonal} states.

If we choose the classical bit $b=0$ to correspond to the state $\ket{0}$, then the only state orthogonal to $b$ is necessarily $\ket{1}$, which thus must be our choice for the classical bit $b'=1$.  Of course, the choice that $b$ corresponds to $\ket{0}$ was arbitrary; we could choose any $\ket{b}=\alpha\ket{0}+\beta\ket{1}$.  However, for any such choice, there exists some $U_b$ such that we can represent $\ket{b}=U_b\ket{0}$, and the corresponding orthogonal state $\ket{b'}=U_b\ket{1}$, since $\ip{b'|b}=\bra{1}U_b^\dagger U_b \ket{0}=\ip{1|0}=0$.  In other words, the unitary $U_b$ is only equivalent to a change of basis on the Hilbert space, and a ``clever selection" of the basis state does not enable us to do anything more.

We can now see the tip of the iceberg suggesting that quantum computation might be fundamentally more powerful than classical computation.  The only single bit operations we can possibly perform on a classical bit $b$ is to apply the identity operator (do nothing), or to flip the bit (in circuit terminology, a NOT gate).  For a qubit, our valid set of transformations is the entire set of $2\times 2$ unitary operators, $U(2)$ (if the reader is curious about unitaries as a mathematical group, see \cite{bib:DummitFoote}).  A qubit, then, is clearly a generalization of a classical bit.  This is itself not so convincing that a quantum computer might be more powerful, since one could make an argument that a classical computer can be equipped with randomization to do just as well.  Of course there is not much one can do with a single bit/qubit, and so we should consider the difference in composite systems.  We will see that \textit{entanglement} is a resource in quantum systems that no classical system, even with randomization, can possess.

If $\mathcal{H}$ is a Hilbert space of a single qubit, then we can consider an $n$-qubit system $\mathcal{H}'$ defined by the tensor product of $n$ copies of $\mathcal{H}$, i.e., \begin{equation}
\mathcal{H}'=\underbrace{\mathcal{H}\otimes\mathcal{H}\otimes ... \otimes\mathcal{H}}_{n}.
\end{equation}
The tensor product is multiplicative in dimension so that $\dim(\mathcal{H}')=2^n$, hence a unitary acting on a state vector in $\mathcal{H}'$ resides in $U(2^n)$.  Naturally, the corresponding classical system has $2^n$ states, one for each vector comprising an orthonormal basis of $\mathcal{H'}$. Since we wish to discuss properties of a composite quantum system that are unique, let us for simplicity consider only a two-qubit system that we denote by $\mathcal{H}^{(2)}=\mathcal{H}_1\otimes\mathcal{H}_2=\mathcal{H}\otimes\mathcal{H}$.  The basis vectors we will choose according to the standard basis, also known as the \textit{computational basis}, which are given by,
\begin{equation}
\ket{00}= \left[
\begin{array}{l}
      1 \\ 0 \\ 0 \\ 0 
\end{array} 
\right] \qquad
\ket{01}= \left[
\begin{array}{l}
      0 \\ 1 \\ 0 \\ 0
\end{array} 
\right] \qquad
\ket{10}= \left[
\begin{array}{l}
      0 \\ 0 \\ 1 \\ 0
\end{array} 
\right] \qquad
\ket{11}= \left[
\begin{array}{l}
      0 \\ 0 \\ 0 \\ 1
\end{array} 
\right].
\end{equation}
In general, if a state such as $\ket{\psi}\in\mathcal{H}_1\otimes\mathcal{H}_2$ can be written in the form $\ket{\psi}=\ket{\sigma}\otimes\ket{\tau}$ for some $\ket{\sigma}\in\mathcal{H}_1, \ket{\tau}\in\mathcal{H}_2$, it is said to be a \textit{product} state.  For example, consider the joint state of two qubits, each in a local state of $\ket{+}=\frac{1}{\sqrt{2}}(\ket{0}+\ket{1})$.  In the computational  basis of the joint system, the state $\ket{\psi}=\ket{+}\otimes\ket{+}$ has the form,
\begin{equation}
\ket{+}\otimes\ket{+}=\frac{1}{\sqrt{2}}(\ket{0}+\ket{1})\otimes\frac{1}{\sqrt{2}}(\ket{0}+\ket{1})=\frac{1}{2}(\ket{00}+\ket{01}+\ket{10}+\ket{11}).
\end{equation}

It is trivial to see that the basis vectors of $\mathcal{H}^{(2)}$ are all product states, and hence any representation of a classical bit is also product. The notion of this independence really underlies the ideas of classical systems in the first place; while individual systems can be correlated, the system never becomes more than ``a sum of its parts."  Quantum systems, on the other hand, exhibit something quite different.  Consider the state, 
\begin{equation}
\ket{\Phi^+}=\frac{1}{\sqrt{2}}(\ket{00}+\ket{11}),
\end{equation}
often referred to as the \textit{Bell state}.  This state (among many others) has the property that \textit{it cannot be decomposed into a tensor product of two states}. This statement has a short proof, so we provide it here.
\begin{itemize}
    \item[{}] \textbf{Proof:} Suppose to the contrary that $\ket{\Phi^+}$ is a product state.  Then there exist two states $\ket{\sigma}=\alpha\ket{0}+\beta\ket{1}$ and $\ket{\tau}=\gamma\ket{0}+\delta\ket{1}$ for some $\alpha,\beta,\gamma,\delta\in\mathbb{C}$ such that $\ket{\Phi^+}=\ket{\sigma}\otimes\ket{\tau}$.  Carrying out the tensor product, \end{itemize} \begin{eqnarray}
\ket{\Phi^+}&=&\ket{\sigma}\otimes\ket{\tau} \nonumber \\
&=&(\alpha\ket{0}+\beta\ket{1})\otimes(\gamma\ket{0}+\delta\ket{1}) \nonumber \\
&=&\alpha\gamma\ket{00}+\alpha\delta\ket{01}+\beta\gamma\ket{10}+\beta\delta\ket{11}. 
\end{eqnarray}
\begin{itemize}
    \item[{}] By definition of $\ket{\Phi^+}$, the coefficients on $\ket{00}$ and $\ket{11}$ are non-zero, and this implies that all of $\alpha,\beta,\gamma,\delta$ should be non-zero.  But since the coefficients on $\ket{01}$ and $\ket{10}$ are zero, we find that one of $\{\alpha,\delta\}$ and one of $\{\beta,\gamma\}$ must be zero.  However, this is a contradiction, and so $\ket{\Phi^+}$ cannot be a product state. $_{\square}$ 
\end{itemize}

Pure states that are not product are called \textit{entangled states}.  These states exhibit a special kind of correlation that do not fit into a classical picture.  Not even the addition of randomness allows classical theory to properly describe the behavior of these states.  In fact, Bell proposed in 1964 that one could devise an experiment using states of this form to prove that no local hidden variable theory (or ``local realism") can adequately explain the predictions of quantum mechanics \cite{bib:Bell1964} (only recently has this experiment been conducted to the level of accuracy that enables no ``loopholes" to explain away the discrepancy \cite{bib:LFBT}).  The exploitation of entangled states is a necessary condition for every quantum algorithm that beats its classical counterpart.  Entanglement itself has been the subject of a great deal of research; still, no universally accepted quantifying measure has been adopted by the community as an adequate description for every case \cite{bib:Horodecki09,bib:Christandl06} (however for pure bipartite states, it seems resolved).  Regardless, it is important that the reader understand that entanglement is a major theme in this thesis (as it generally is whenever one discusses quantum devices). 

We conclude this section with a brief overview of what we have discussed.  The three postulates of quantum mechanics gave us the means to construct a state space where we could compare the behavior of classical states to quantum states.  We saw that quantum theory generalizes classical theory, and produces a large set of states that have no efficient classical description.  This culminated in showing the existence of entangled states, which can possess correlations that no classical theory can describe.  Although the reader may feel a little shortchanged on seeing explicit examples of quantum supremacy so far, there will be no shortage of examples of devices and algorithms later that utilize entanglement to beat the best known classical strategy.  First, however, we must move on to the quantum mechanics of light to fully explain the phenomena presented in this thesis.
\\
\subsection{Quantum States of Light}\label{subsec:optics}
In this section, our goal is to understand the specific quantum setting we will be using throughout the thesis.  While the previous section gave us a general set of rules for any quantum system, we must first understand the behavior of the photon and the nature of light before we can explain the system's dynamics.  In this section, we will discuss various important quantum states of light and the mechanism by which they arise.  We then describe what we mean by an ``optical network'' and the kinds of operations that we can perform on the system.  The primary mathematical prerequisite for this section is, as before, a solid understanding of linear algebra (though there are a few statements that might require some knowledge of operator algebras).  We try to avoid the language of Hamiltonians whenever possible, because it is largely unnecessary for the work presented later.  The concepts in this section are largely inspired by and often relying on calculations in \cite{bib:GerryKnight05}, which the reader should refer to for more in-depth consideration.

Photons arise from the quantization of the electromagnetic field.  For a single-mode field, the quantization is similar to that of the one-dimensional quantum harmonic oscillator of frequency $\omega$.  The energy levels of the quantum harmonic oscillator are discrete, with equal separations between consecutive levels.  Furthermore, the ground state of the quantum harmonic oscillator has a non-vanishing energy (called the \textit{vacuum energy} or \textit{zero-point energy}).  This allows us to represent the eigenvectors of these states in a convenient form, called the \textit{Fock basis}.  Namely, $\ket{n}$ is the eigenvector corresponding to the $n$th excitation of the field.  We say a field has $n$ photons (of frequency $\omega$) if its energy corresponds to the $n$th such excitation; we call a quantum state in this form a \textit{Fock state}.  The energy level of the $n$th excitation is formally,
\begin{equation}
E_n=\hbar\omega(n+\frac{1}{2}).
\end{equation}

The ladder operator method is a useful tool for representing evolutions of the field.  First, we define two operators, $\hat{a}^\dagger$ and $\hat{a}$, called the \textit{creation operator} and \textit{annihilation operator}, respectively.  These operators are formally defined with respect to the position and momentum quadrature operators of the electromagnetic field, but here we define them in terms of their action on the eigenstate $\ket{n}$, \begin{eqnarray}
\hat{a}\ket{n}&=&\sqrt{n}\ket{n-1} \\
\hat{a}^\dagger\ket{n}&=&\sqrt{n+1}\ket{n+1}.
\end{eqnarray}
Together, these operators form the \textit{number operator} $\hat{n}=\hat{a}^\dagger\hat{a}$, which has the convenient property that $\hat{n}\ket{n}=n\ket{n}$.  Note that the operators $\hat{a}$ and $\hat{a}^\dagger$ are neither Hermitian nor unitary, while $\hat{n}$ is Hermitian but not unitary.  For any state $\ket{\psi}$ which is a superposition of Fock states, the mean number of photons can be computed by evaluating the quantity, 
\begin{equation}
\bar{n}_{\psi}=\bra{\psi}\hat{n}\ket{\psi}. \label{eq:mean}
\end{equation} 
The uncertainty $\Delta n$ of a state is thus,
\begin{equation}
\Delta n=\sqrt{\ip{\hat{n}^2}-\ip{\hat{n}}^2}. \label{eq:deviation}
\end{equation}

A surprising result that occurs when evaluating the expectation of the electric field operator for a Fock state is that, because Fock states have a uniform phase distribution, the expectation causes the magnitude of the field to vanish.  This is suggestive of the idea that classical macroscopic systems cannot be explained by a simple scaling of the Fock states to large photon numbers.  It is also suggestive of the idea that number and phase are complementary variables and so obey an uncertainty relation,
\begin{equation}
\Delta n \Delta \varphi \geq 1. \label{eq:uncertainty}
\end{equation}
We will leave out a full discussion of quantum phase since there is much to say before arriving at a satisfying level of understanding, but the reader can consult \cite{bib:GerryKnight05}.  Instead, we will simply take the above uncertainty relation somewhat for granted, and investigate the implications thereof.

Consider the set of eigenstates of the annihilation operator, i.e., the set of states satisfying,
\begin{equation}
\hat{a}\ket{\alpha}=\alpha\ket{\alpha}.
\end{equation}
with $\alpha\in\mathbb{C}$.  We can solve for the coefficients of $\ket{\alpha}$ by expanding in terms of the Fock basis and noting that the coefficients must satisfy a recurrence relation.  Formally,
\begin{eqnarray}
\ket{\alpha}&=&\sum_{n=0}^{\infty}C_n\ket{n}\quad \textrm{together with}\quad \hat{a}\ket{\alpha}=\alpha\ket{\alpha} \Rightarrow \nonumber \\
C_n\sqrt{n}&=&\alpha C_{n-1} \quad\Rightarrow \nonumber \\
C_n&=&\frac{\alpha^n}{\sqrt{n!}}C_0 \quad\Rightarrow \nonumber \\
\ket{\alpha}&=&C_0\sum_{n=0}^{\infty} \frac{\alpha^n}{\sqrt{n!}}\ket{n}.
\end{eqnarray}
where $C_0$ is a normalization constant that can be easily computed from the requirement that $\ip{\alpha|\alpha}=1$,
\begin{eqnarray}
1&=&\Big[ C_0^*\sum_{n=0}^{\infty} \frac{(\alpha^*)^n}{\sqrt{n!}}\bra{n}\Big]\Big[C_0\sum_{n=0}^{\infty} \frac{\alpha^n}{\sqrt{n!}}\ket{n}\Big] \nonumber \\
1&=&|C_0|^2\sum_{n=0}^\infty \frac{|\alpha|^{2n}}{n!}\quad\Rightarrow \label{eq:expsum} \\
|C_0|&=&\mathrm{exp}(-|\alpha|^2/2) \label{eq:csnorm}\quad\Rightarrow \\
\ket{\alpha}&=& e^{-|\alpha|^2/2} \sum_{n=0}^{\infty} \frac{\alpha^n}{\sqrt{n!}}\ket{n}, \label{eq:cs}
\end{eqnarray}
where the sum in Eq.~(\ref{eq:expsum}) easily results from the expansion of the exponential function.  We call the state $\ket{\alpha}$ a \textit{coherent state}.  Note that $\alpha=0$ corresponds to the vacuum state, which is an eigenstate of the annihilation operator with eigenvalue zero.  The average photon number $\bar{n}=\ip{\alpha|\hat{n}|\alpha}$ (see Eq.~(\ref{eq:deviation})) of the coherent state is $\bar{n}=|\alpha|^2$, with an uncertainty $\Delta\bar{n}=\sqrt{\bar{n}}$ (the distribution is Poissonian).  This gives the coherent state a very nice representation in terms of the complex number $\alpha=|\alpha|e^{i\varphi}$ which defines it---the magnitude $|\alpha|$ determines the photon number, while $\varphi=\arg(\alpha)$ determines the phase.

Coherent states have a number of properties that make them the most ``classical'' quantum states of light in the sense that they behave more like a classical electromagnetic field (less so in the sense of the previous section where we discuss the notion of a classical state space).  Namely, \cite{bib:GerryKnight05} succinctly summarizes,

\begin{itemize}
    \item[{}] \textit{The coherent states $\ket{\alpha}$ are quantum states very close to classical states because (i) the expectation value of the electric field has the form of the classical expression, (ii) the fluctuations in the electric field variables are the same as for a vacuum, (iii) the fluctuations in the fractional uncertainty for the photon number decrease with the increasing average photon number, and (iv) the states become well localized in phase with increasing average photon number.}
\end{itemize}

While we will not concern ourselves with the first two items, the latter two have important consequences for some of the results in this thesis.  First,  
item (iii) refers to the fractional uncertainty in the average number of photons,
\begin{equation}
\frac{\Delta \bar{n}}{\bar{n}}=\frac{\sqrt{\bar{n}}}{\bar{n}}=\frac{1}{\sqrt{\bar{n}}}.
\end{equation}
Item (iv) relates to the expression of the uncertainty relation in Eq.~(\ref{eq:uncertainty}), where since $\Delta\bar{n}=\sqrt{\bar{n}}$,
\begin{equation}
\Delta \varphi\geq \frac{1}{\sqrt{\bar{n}}}. \label{eq:csuncertainty}
\end{equation}
The inequality in Eq.~(\ref{eq:csuncertainty}) is actually \textit{saturated} for all coherent states $\ket{\alpha}$.  In fact, an alternate way of defining the coherent state is in terms of minimizing an uncertainty product (\textit{not} the one in Eq.~(\ref{eq:uncertainty}); see \cite{bib:GerryKnight05}) such that both are equal in amplitude.  There is another family of minimum uncertainty states, called \textit{squeezed} states, where the products are not equal between quadratures (plotting the uncertainty in phase space generates an ellipse, hence the term ``squeezed'').  We will discuss these states shortly.

Our final comment on coherent states is that they can be generated by an operator acting on the vacuum.  The \textit{displacement operator} $\hat{D}(\alpha)$ is defined by,
\begin{equation}
\hat{D}(\alpha)=e^{\alpha\hat{a}^\dagger-\alpha^*\hat{a}},
\end{equation}
so that,
\begin{equation}
\ket{\alpha}=\hat{D}(\alpha)\ket{0}.
\end{equation}
For a derivation, see \cite{bib:GerryKnight05}.  It is easy to see from the definition of $\hat{D}(\alpha)$ that,
\begin{equation}
\hat{D}^\dagger(\alpha)=D(-\alpha),
\end{equation}
which implies that $\hat{D}(\alpha)$ is a \textit{unitary} operator.  The commutation relations of $\hat{D}(\alpha)$ with the annihilation operator (which can be Hermitian-conjugated to produce that of the creation operator) can be shown to be,
\begin{eqnarray}
\left[a^\dagger, \hat{D}(\alpha)\right]=\alpha^*\hat{D}(\alpha).
\end{eqnarray}

We now consider another kind of state, the squeezed state $\ket{\xi}$, which exemplifies some very non-classical behavior.  First, we define the \textit{squeezing operator} in a way that looks very similar to the displacement operator,
\begin{equation}
\hat{S}(\xi)=\exp\Big[\frac{1}{2}(\xi \hat{a}^2-\xi \hat{a}^{\dagger 2})\Big].
\end{equation}
It has the same property that,
\begin{equation}
\hat{S}^\dagger(\xi)=\hat{S}(-\xi),
\end{equation}
and is also unitary.  When acting on the vacuum, it generates a \textit{squeezed vacuum state} \cite{bib:GerryKnight05},
\begin{equation}
\ket{\xi}=\hat{S}(\xi)\ket{0}=\frac{1}{\sqrt{\cosh r}}\sum_{m=0}^\infty (-1)^m \frac{\sqrt{(2m)!}}{2^m m!}e^{i m \theta}(\tanh r)^m\ket{2m}. \label{eq:squeezedvac}
\end{equation}
for $\xi=re^{i\theta}\in\mathbb{C}$.  It is particularly notable that the amplitude of every odd-numbered Fock state is zero, so that $\ket{\xi}$ always consists of an even number of photons.  Analogous to the coherent state, it is apparent from the form of Eq.~(\ref{eq:squeezedvac}) that the ``intensity" of the squeezing is given by the magnitude of the squeezing parameter $r=|\xi|$, and possesses a phase determined by $\theta$.  Indeed, the average photon number for $\ket{\xi}$ is,
\begin{equation}
\ip{\hat{n}}_{\xi}=\sinh^2 r.
\end{equation}

We conclude this section with a few notes about the states we have discussed.  A well-known quasi-probability distribution, know as the Wigner distribution function \cite{bib:Wig1932}, is often used to characterize quantum states of light.  Coherent states and squeezed vacuum states (among a few others) have the property that their Wigner distributions have a Gaussian form, granting them the title of \textit{Gaussian states}.  Due to some of the ``nice" properties of these states, they have been studied in some depth. Ref.~\cite{bib:AA10} gives a synopsis of what is known about simulating these and other states in optical networks.

There are, of course, many other quantum states of light to be discussed in the broader context of quantum optics.  The ones presented in this section are, as usual, restricted to those which will be relevant later in this thesis.  More on other states, such as thermal states, two-mode squeezed states, and Schroedinger cat states, can be found in \cite{bib:GerryKnight05}.

\subsection{Linear Optical Networks}\label{subsec:networks}
The previous section dealt with different states of light in a single mode.  In this section, we wish to look at how the evolution of a composite state on multiple modes takes place.  Although it is possible to treat composite photonic systems in the same vein as the first section of this chapter (that is, by considering the tensor product $\mathcal{H}_1\otimes\mathcal{H}_2$), it is very difficult to \textit{physically} implement these kinds of general transformations.  One approach is to use a beam splitter to combine two spatial modes, and though the resulting space of transformations by comparison is much smaller, this approach is much easier to realize.  We thus refer to a ``linear optical network" as a collection of modes and beamsplitter operations between them.  Before we proceed, note that the creation and annihilation operators on different modes commute.  That is, 
\begin{eqnarray}
\lbrack \hat{a}_i,\hat{a}_j^\dagger \rbrack &=&\delta_{ij}\;, \\
\lbrack \hat{a}_i^\dagger,\hat{a}_j^\dagger \rbrack&=&0, \end{eqnarray} 
where the index refers to a mode labeling.

\begin{figure}[h]
\centering
\includegraphics[scale=0.7]{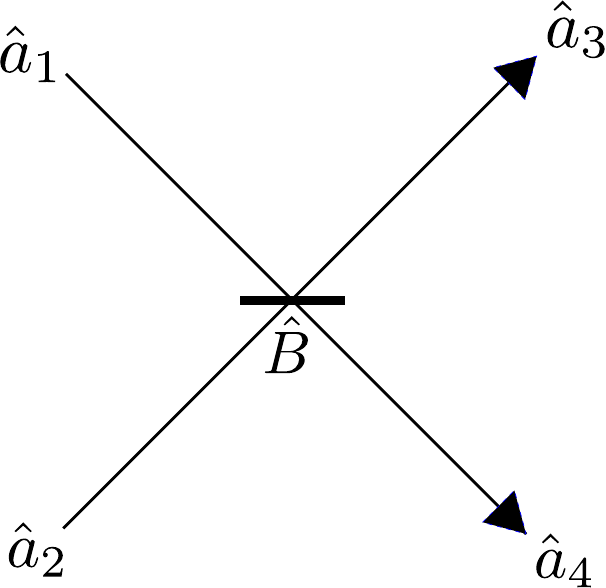}
\caption{Two modes $\hat{a}_1$ and $\hat{a}_2$ are incident on a beam splitter $\hat{B}$.  After the action of $\hat{B}$, the two modes propagate away in modes $\hat{a}_3$ and $\hat{a}_4$.} \label{fig:bs}
\end{figure}

The action of the beam splitter can be viewed in terms of a transformation on the annihilation operators $\hat{a}_i$ in mode $i$.  Consider, for example, the labellings in Figure \ref{fig:bs}, where modes 1 and 2 are incident on a beam splitter $\hat{B}$.  The operators $\hat{a}_3$ and $\hat{a}_4$ for the output modes are given by,
\begin{equation}
\left[
\begin{array}{l}
      \hat{a}_3 \\ \hat{a}_4
\end{array} 
\right]
=\hat{B}
\left[
\begin{array}{l}
      \hat{a}_1 \\ \hat{a}_2
\end{array} 
\right],
\end{equation}
where $\hat{B}$ is a $2\times2$ unitary matrix.  This matrix is often written in the general form,
\begin{equation}
\hat{B}=
\frac{1}{\sqrt{2}}\left[
\begin{array}{ll}
      t' & r \\
      r' & t 
\end{array} 
\right],
\end{equation}
where the relations,
\begin{equation}
|r|=|r'|, |t|=|t'|, |r|^2+|t|^2=1, r^*t'+r't^*=0, \textrm{ and } r^*t+r't'^*=0,
\end{equation}
are required due to energy conservation  \cite{bib:GerryKnight05}.
For example, a 50:50 beam splitter has the form,
\begin{equation}
\hat{B}_{50:50}=
\frac{1}{\sqrt{2}}\left[
\begin{array}{ll}
      1 & i \\
      i & 1 
\end{array} 
\right].
\end{equation}

The set of unique beam splitter transformations on two modes is the group SU(2).  But how do we generalize the notion of a beam splitter to more than two modes?  One very straightforward approach for $n$ modes is to consider the group generated by a concatenation of beam splitters between any two arbitrary modes.  It turns out that (conveniently) any network consisting of only two-mode beam splitter operations can be represented by a single unitary matrix in SU($n$).  The converse is also true, so that SU($n$) completely characterizes an $n$-mode network of beam splitters \cite{bib:Reck94}.  We can write the action of a general $n$-mode unitary on the creation operators as,

\begin{equation} \label{eq:unitary_trans}
\hat{U}\hat{a}_i^\dag\hat{U}^\dag \to \sum_j U_{i,j} \hat{a}_j^\dag.
\end{equation}
We will see in the next chapter how matrix permanents can be used to more easily describe the evolution of Fock states under these general unitaries.

A property of these transformations is that they preserve the photon number (or average photon number, in the case of an indeterministic number of photons).  For this reason, beam splitter operations are often called \textit{passive} linear optics in the respect that they do not inject or remove any photons from the network.  Throughout this thesis, we will almost exclusively be concerned with these kinds of systems because of the relative ease by which they can be physically implemented when the number of photons is small (the same often cannot be said about \textit{active} optical systems).


\subsection{Complexity Theory}\label{subsec:complexity}
We now turn to what may seem at first a disconnected topic, since the roots of complexity theory lie in the realm of computer science rather than physics. On the other hand, the notion of complexity can be restated in a very physical way, where we wish to determine what the ``logical" similarities between two physical systems may be.  We saw earlier in Sec.~\ref{subsec:quantum} that a quantum state space and a classical state space are certainly not equivalent.  Still, one may wonder if they are at least ``close" in the sense that a slightly larger classical system could be prepared and evolved such that the classical system simulated the quantum one.  The goal of this section is to introduce some of the formal ideas of complexity theory so that we have some understanding of what can be said, and how this relates to quantum optics.  The only prerequisites to understanding this section are elementary algebra and set theory.  The ideas presented in this section have more rigorous meanings in the domain of formal language theory, but these are largely unnecessary for a reasonable understanding of the topic.  

The first concept we would like to discuss is that of a \textit{decision problem}.  Informally, a decision problem is a question which, depending upon a particular input, always has a ``yes or no'' answer.  A simple example of a decision problem is \cc{Primes}, which asks, ``Given an integer $n>1$, decide if $n$ is a prime number or not."  The input to a decision problem---in this example, the integer $n$---is called an \textit{instance} of the problem.  A \textit{YES}-instance to the problem is an instance of the problem in which the answer is yes (e.g., the integer $n=5$), and a \textit{NO}-instance where the answer is no (e.g., $n=6$).

There are other kinds of problems than decision problems; for instance (no pun intended), a \textit{functional} problem is one in which the answer is allowed to be a more complicated than a simple yes or no (e.g., ``for an integer $n$, compute the largest prime factor of $n$").  For our purposes, \textit{sampling problems} will be of particular interest.  In these kinds of problems, the output of the problem is a sample from an instance of a probability distribution (often within some specified error tolerance of the statistical distance to the true distribution).  Also relevant is a \textit{counting} problem, which searches through a particular space of possible solutions to a relation and returns the number of correct solutions.  

In order to characterize the difficulty of solving different kinds of problems, we place them into sets which we call \textit{complexity classes}.  These classes must be defined relative to some system or device that is capable of solving the system, so that a well-defined notion of ``difficulty" can be assessed.  Generally, these classes group problems according to the number of steps, the physical space, or the time required on such a device to solve the problem.  The devices normally referred to in these classes are called \textit{Turing machines}, to which there is a rich history including (but not limited to) code-breaking in World War 2.  In the interest of brevity, we forgo rigorous definitions and interesting anecdotes; instead, we give some informal definitions that should  provide the reader with a sufficient understanding.

Let $x$ be the \textit{input size} to a problem $Z$, i.e., the number of bits needed to represent a problem instance.  We will say that a \textit{classical computer} is a device that is capable of solving the problem $Z$ with some finite number of bits in a finite time, where we assume each step in the computation (i.e., a change in the computer's internal state) takes a constant time.  For simplicity, we will refer to the number of bits as the ``size" of the computer.  We say that a classical computer is \textit{deterministic} if each step of the computer's calculation occurs with definite probability.  We say a computer is \textit{probabilistic} if the computer may rely on some randomness to arrive at a solution.  We will use the same definition for a \textit{quantum computer}, only replacing the role of the probabilistic bit with a qubit.  Finally, we will say that a computer can solve a problem \textit{efficiently} if it can do so when the computation time is bounded by a polynomial function of $x$.

We now define several classical complexity classes of decision problems:
\begin{itemize}
    \item[{}] \textbf{P}:  A decision problem $Z$ is in the complexity class \PP if it can be solved efficiently by a deterministic classical computer.
\end{itemize}
The class \PP is meant to capture ``easy" problems for a computer.  That is, even for a large input size $x$, the length of the calculation does not increase more than polynomially in $x$.
\begin{itemize}
    \item[{}] \textbf{BPP}:  A decision problem $Z$ is in the complexity class \BPP if it can be solved efficiently by a probabilistic classical computer, with a success probability of at least 2/3.
\end{itemize}
The class \BPP is meant to capture easy problems for a computer that relies on some kind of randomness to solve a problem.  It is trivial to see that \PP $\subseteq$ \BPP.  It is believed by the majority of the complexity community that $\PP=\BPP$, but no formal proof yet exists to show this.  Also, although a success probability of 2/3 might seem somewhat arbitrary, there are known methods to boost the probability of success arbitrarily close to 1 without violating the efficiency condition; changing the success probability to any constant above 1/2 is equivalent.  Importantly, \BPP is the class of problems that is understood to be realistically scalable for a classical computer to solve (of course, one can always consider solving problems from harder complexity classes when the input size is small).  
\begin{itemize}
    \item[{}] \textbf{NP}:  A decision problem $Z$ is in the complexity class \NP if, for YES-instances of the problem, there is a polynomial-size witness string $w$ which a deterministic classical computer can use to efficiently verify the solution.
\end{itemize}
The class \NP in some sense classifies ``provable problems."  Like a mathematical theorem, though it may be initially hard to prove, once one has access to the proof (in analogy with the witness string $w$), it can be verified easily.  It is easy to see that \PP $\subseteq$ \NP since, if a problem is in \PP, a verifier can simply ignore the witness string and prove it efficiently by himself. 
\begin{itemize}
    \item[{}] \textbf{co-NP}:  A decision problem $Z$ is in the complexity class \coNP if, for NO-instances of the problem, there is a polynomial-size witness string $w$ which a deterministic classical computer can use to efficiently verify the solution.
\end{itemize}
The class \coNP is much like \NP, classifying what might be thought of as efficiently ``falsifiable problems."  

Complexity theory can be described as the study of complexity classes, their relations to one another, and the problems which reside inside them.  In order to say something constructive about complexity classes, we need some way of identifying and comparing the kinds of problems they have.  The first comparison tool we will discuss is that of \textit{polynomial reducibility}.  There are multiple (often non-equivalent) ways to define reducibility. We will say that a problem $Z$ is polynomial reducible to a problem $Y$ (denoted $Z\leq Y$) if an instance of $Z$ can be efficiently transformed into an instance of $Y$ such that the solution to both instances is the same.  The following example gives a simple illustration of this idea.

\begin{itemize}
    \item[{}] \textbf{Example (polynomial reducibility):} Let \cc{Even} be the decision problem defined as, ``Given an integer $n$, decide if $n$ is even."  Let \cc{Zero} be the decision problem defined as, ``Given an integer $m$, decide if the last digit of $m$ is zero."  Suppose we had a machine that could solve \cc{Zero}.  Could we use that machine in some way to efficiently evaluate the solution to \cc{Even} as well?  Indeed, we can -- simply check \cc{Zero}($5n$).  We must be sure that, given the input $n$, we can efficiently calculate $5n$ to input to the machine; the standard multiplication algorithm works for this.  Thus \cc{Even} $\leq$ \cc{Zero}.
\end{itemize}

Polynomial reductions are important for complexity theory because they tell us that some problems have structural similarities.  We can exploit this relation to help classify many problems and classes.  Let \textbf{C} be a complexity class.  A problem $Z$ is said to be \textit{hard for} \textbf{C} (denoted \textbf{C}\textit{-hard}) if, for every problem $Y\in\textbf{C}$, $Y\leq Z$.  This is a useful idea since it suggests that $Z$ is at least as hard as any problem in \textbf{C}.  This also allows us to establish that certain problems characterize the hardness of a class very well.  We say a problem $Z$ is \textbf{C}\textit{-complete} if $Z\in$ \textbf{C} and $Z$ is \textbf{C}-hard.

We would like to note a bit of a caveat when dealing with problems in complexity classes.  Namely, a complexity class is a kind of \textit{worse-case} classification for a problem.  This is because classes are generally defined in terms of arbitrary instances of a problem.  Hence, if there is even a tiny subset of instances to a problem that that are hard to solve, the problem will be classified according to these instances.  There is another notion of \textit{average-case} complexity, but we will not discuss this further in this thesis (mostly because it seems that the exceptional cases for matrix permanents are ones which are easy to compute, rather than hard).

We now have a simple recipe for proving inequalities between many complexity classes, based on complete problems.  This follows because most of the complexity classes we define are in terms of computers being able to solve (or verify) some problem efficiently.  For instance, if we want to prove the relation \NP$\subset$ \PP, we need only show that a single \NP-complete problem resides in \PP.

Incidentally, the conjecture \PP$\stackrel{?}{=}$ \NP is arguably the most important problem in computer science.  It is one of the famous Millennial Prize problems, for which a solution grants the discoverer a US \$1,000,000 prize.  There is a heap of evidence suggesting that  \PP$\neq$ \NP, and for this reason it is almost universally thought by the complexity community that this is the case.  It is so motivated that some results rely on a dichotomy -- either a particular result is true, or else \PP$=$ \NP{} -- to suggest that the result is likely true.  We will later see that this is the same kind of reasoning behind believing that \BS is a classical intractable problem.  In order to do so, we must introduce a new tool, the \textit{oracle machine}.

A complexity class \textbf{C} \textit{relative to an oracle} \Or (denoted $\textbf{C}^{\mathcal{O}}$) is defined as the class of problems that are solvable in \textbf{C} with access to a ``black box" that can provide a solution to a problem in \Or with only a single step of the computer.  The oracle \Or can be either a problem, or an entire complexity class.  They are important in understanding the definition of the polynomial hierarchy and why we expect that \BS$\not\subset$ \BPP.

\begin{figure}[h]
\centering
\includegraphics[scale=0.7]{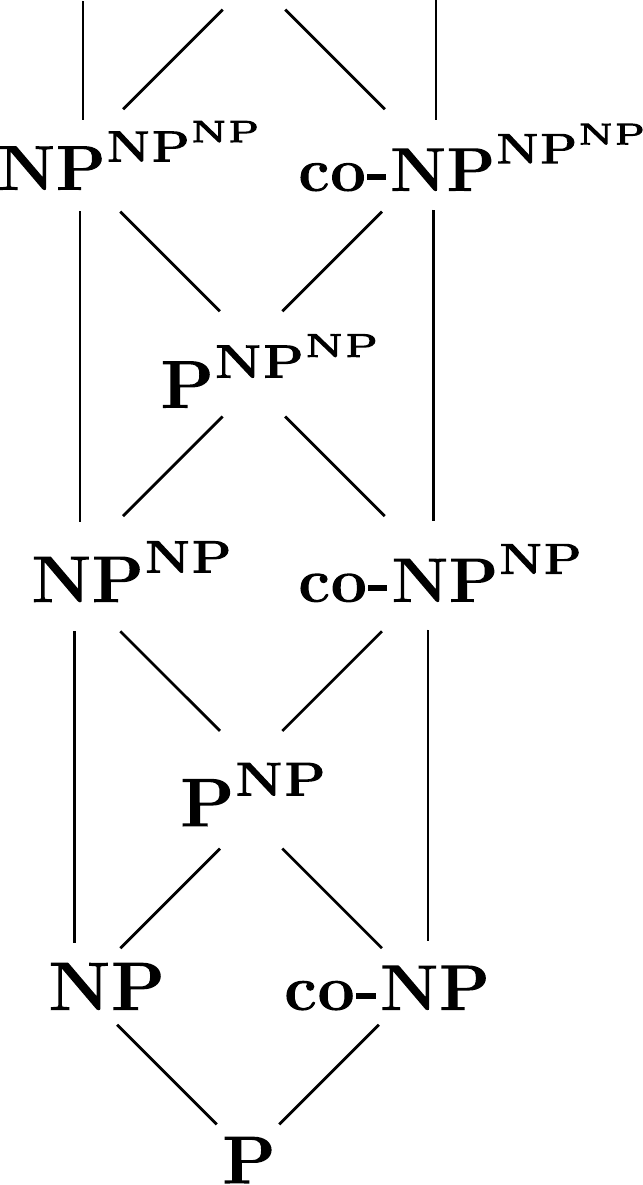}
\caption{The polynomial hierarchy \PH.  Complexity classes higher in the figure denote more difficult classes, and lines indicate containment.  \PH is the union of all such classes.} \label{fig:ph}
\end{figure}

The polynomial hierarchy (denoted \PH) is defined as the union of a recursive chain of complexity classes defined in the following way \cite{bib:Stockmeyer76}:
\begin{itemize}
    \item[{}] Initialize $\PP=\Pi^P_0=\Sigma^P_0=\Delta^P_0$.  Define:
\begin{eqnarray}
\Pi^P_{i+1}&=&\coNP^{\Sigma^P_i} \\
\Sigma^P_{i+1}&=&\NP^{\Sigma^P_i} \\
\Delta^P_{i+1}&=&\PP^{\Sigma^P_i} \\
\PH&=&\bigcup_{i=0}^{\infty}\Sigma^P_{i} \cup \Pi^P_{i} \cup \Delta^P_{i}.
\end{eqnarray}
\end{itemize}
A more intuitive graphical representation of \PH can be found in Figure \ref{fig:ph}.  The polynomial hierarchy is defined with the idea in mind that each ``level" of the hierarchy is expected to be a strict containment.  Formally, if for some $k$, the equality $\Sigma^P_{k+1}=\Sigma^P_{k}$ holds, then the equality must hold for all $i\geq k$.  This ``collapses" the polynomial hierarchy to the $k$th level, so that \PH is a union of only finitely many classes.  Note that if \PP= \NP, the polynomial hierarchy completely collapses and \PH= \PP.  Hence, the expectation that \PP$\neq$ \NP in some sense generalizes the notion that a collapse of \PH should not occur.  This is the dichotomy that suggests \BS should be a hard problem; if there is an efficient classical algorithm for estimating \BS, then \PH will collapse (to the third level).  We discuss this in more detail in Sec.~\ref{sec:exact} and \ref{sec:approx}.

We now introduce one final complexity class, which will be helpful for discussing the capabilities of quantum computers in general.
\begin{itemize}
    \item[{}] \textbf{BQP}:  A decision problem $Z$ is in the complexity class \BQP if it can be solved efficiently by a quantum computer, with a success probability of at least 2/3.
\end{itemize}
\BQP is the quantum generalization of \BPP.  A quantum computing architecture is said to be \textit{universal} if it is capable of computing problems in \BQP.  Although there are a number of potential architectures for implementing universal quantum computing, we will often refer to linear optical quantum computing (LOQC) as a specific example.  The LOQC model exemplifies the gap between the realities of current technology and the requirements for building a fully universal, fault-tolerant quantum computer \cite{bib:KLM01}.  Even with improvements to the original protocol, a large scale demonstration of LOQC is likely decades away \cite{bib:kok2007linear}.  This thesis presents the case that, alternatively, it may be possible to utilize passive linear optics to perform some task in a complexity class that is outside of \BPP, without the need for full universal quantum computing.  



\pagebreak

\singlespacing
\chapter{Boson Sampling}
\doublespacing
In this chapter, we will review the seminal result due to Arkhipov and Aaronson (referred to henceforth as AA) \cite{bib:AA10}, which defines \BS and shows a dichotomy---either the polynomial hierarchy collapses, or \BS is a hard problem to simulate classically.  To begin, we will informally define the \BS problem so that the discussions throughout are more motivated.  In the first section, we will review the underlying mathematical details that are necessary for understanding the root of the complexity in \BS.  In the following sections, we formally define \BS and summarize the main results of Ref.~\cite{bib:AA10}, and finally discuss some important obstacles in utilizing \BS experiments to implement a truly post-classical computation. 

\begin{figure}[t]
\centering
\includegraphics[scale=1.8]{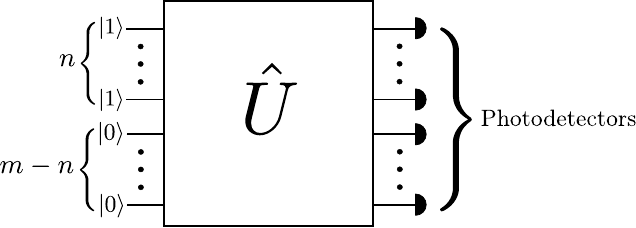}
\caption{An optical network implementation of \BS.} \label{fig:bsarch}
\end{figure}

\begin{itemize}
    \item[{}] \textbf{Definition} (\BS, informal): Let $m$ be the number of modes in a linear optical network, whose input consists of $n$ single photon Fock states (without loss of generality, in the first $n$ modes) and $m-n$ vacuum states.  Let $\hat{U}\in SU(m)$ be a random unitary matrix acting on all $m$ modes.  Let $P(\hat{U})$ be the probability distribution corresponding to the joint measurements of all $m$ modes in the Fock basis.  Sample from $P(\hat{U})$ to within some error $\epsilon$ of the total variation distance. 
\end{itemize}
Figure \ref{fig:bsarch} shows the architecture of the \BS model.  For the case that an instance of \BS has the input photons in other modes, one can consider a relabelling of the indices and a permutation of the rows of $\hat{U}$ such that all input photons are in the first $n$ modes,
\begin{eqnarray} \label{eq:bsin}
\ket{\psi_{\rm{in}}} &=& \hat{a}_1^\dag\dots \hat{a}_n^\dag \ket{0_1,\dots,0_m} \nonumber \\
&=& \ket{1_1,\dots,1_n,0_{n+1},\dots,0_m}.
\end{eqnarray}
The permutation preserves the randomness of $\hat{U}$ as well as the probability distribution $P(\hat{U})$, modulo that permutation.  We can then succintly write the output of the device in Fig. \ref{fig:bsarch} as,
\begin{equation} \label{eq:bsout}
\ket{\psi_{\rm{out}}}= \hat{U}\left[\hat{a}_1^\dag\dots\hat{a}_n^\dag\right]\ket{0_1,\dots,0_m}.
\end{equation}
Note, however, that the desired output of the \BS problem stated above is a \textit{sample} from a distribution, \textit{not} a computation of the output distribution itself.  

Since we would like to say something about the power of quantum computing, we would like to check that the circuit described in Figure \ref{fig:bsarch} can be implemented efficiently, so that we verify \BS $\in$ \BQP.  The input to the \BS problem is an $m\times n$ matrix $A$, which has $m\cdot n$ many elements.  Clearly, then, the number of modes $m$, photons $n$, and hence the number of photodetectors required to implement an instance of \BS is efficient in the input size.  The only question that remains is whether the matrix $A$ can be implemented with a polynomial number of optical elements.  Indeed, Reck \textit{et al.}~showed that any $n\times n$ unitary can be decomposed into a series of at most $O(n^2)$ beam splitters \cite{bib:Reck94}, making such a construction efficient.  Thus, \BS $\in$ \BQP.

\section{Complexity of Matrix Permanents} \label{sec:permanent}
In this section, our primary discussion relates to the complexity of matrix permanents and their connection to \BS.  First, we show how the output amplitudes of evolved Fock states in a linear optical network can be computed via matrix permanents (equivalent to propagating the field operators in Section \ref{subsec:optics}), shown by Scheel in 2004 \cite{bib:Scheel04perm}.  We then discuss permanent complexity, with regard to both exact computation and approximation.

Let us begin by first defining the permanent of a matrix $M$.

\begin{itemize}
    \item[{}] \textbf{Definition} (Permanent): Let $M$ be an $n\times n$ matrix with complex entries $m_{i,j}\in\mathbb{C}$.  The \textit{permanent} of $M$ is defined by,
\begin{equation}
\perm(M)=\sum_{\sigma\in S_{n}} \prod_{i=1}^n m_{i,\sigma(i)}.
\end{equation}
where $S_n$ is the symmetric group on $n$ elements.
\end{itemize}
Note that the permanent of a matrix is very similar to the determinant, with the exception that sgn($\sigma$) is missing from formula.  Put simply, the permanent is equal to the determinant ``with all + signs''.  One can easily see that computing a permanent from the definition alone will not be efficient, since the group $S_n$ contains $n!$ elements.  We leave a discussion of complexity for later, once we have discussed its connection to linear optics.

Let us consider the evolution of an $n$-mode Fock state through an $n$-mode unitary $\hat{U}$ (where the $ij$th entry is denoted by $u_{ij}$), with a total of $K$ photons in the input modes.  Let $k$ denote the $n$-tuple corresponding to the input configuration, i.e. $k=(k_1,\dots,k_n)$.  Namely,
\begin{eqnarray}
\ket{\psi_{\rm{in}}}&=&\ket{k}=\ket{k_1,\dots, k_n}=(\hat{a}_1^\dagger)^{k_1}\dots(\hat{a}_n^\dagger)^{k_n}\ket{0,\dots, 0} \\
\ket{\psi_{\rm{out}}}&=&\hat{U}\ket{\psi_{\rm{in}}}=\sum_{s\in S}\gamma_{s}\ket{s}, \label{eq:permout}
\end{eqnarray}
where $S$ denotes the set of all $n$-tuple configurations of $K=k_1+\dots+k_n$ photons, and $s=(s_1,\dots,s_n)$ denotes a particular configuration.  The cardinality of $S$ is given by,
\begin{equation}
|S|=\binom{n+K-1}{K},
\end{equation}
which is the number of ways to configure $K$ indistinguishable objects into $n$ distinguishable bins, also called the number of ``stars and bars" (i.e. the number of ways to configure $n$ $|$'s and $K$ $\star$'s in a lineup).

We now wish to determine $\gamma_s$.  Let the $i$th row vector of $\hat{U}$ be denoted $\textbf{u}_i$.  Define the row vector $\textbf{u}_{i,s}$ to be the row vector consisting of $s_j$ copies of the $j$th element of $\textbf{u}_i$.  For example, if $\textbf{u}_1=(u_{11}\; u_{12}\; u_{13})$ and $s=(2,1,1)$ then $\textbf{u}_{1,s}=(u_{11}\; u_{11}\; u_{12}\; u_{13})$.  We then define the matrix $\hat{U}_{k,s}$ to be the matrix consisting of $k_i$ copies of the row vector $\textbf{u}_{i,s}$.  For example, if $k=(1,0,3)$ and $s$ is as before, then,
\begin{equation}
\hat{U}_{k,s}=\left[
\begin{array}{l}
      \textbf{u}_{1,s} \\ \textbf{u}_{3,s} \\ \textbf{u}_{3,s} \\ \textbf{u}_{3,s}
\end{array} 
\right]=
\left[
\begin{array}{llll}
      u_{11} & u_{11} & u_{12} & u_{13} \\
      u_{31} & u_{31} & u_{32} & u_{33} \\
        u_{31} & u_{31} & u_{32} & u_{33} \\
      u_{31} & u_{31} & u_{32} & u_{33} \\
\end{array} 
\right].
\end{equation}

We claim that, for the input configuration $k$ and output configuration $s$, the amplitude of the state $\ket{s}$ is equal to the permanent of $\hat{U}_{k,s}$,
\begin{equation}
\gamma_s=\perm(\hat{U}_{k,s}).
\end{equation}
A proof of this fact can be found in \cite{bib:Scheel04perm}. 
Note that in the case that the input and output states consist of only single photon Fock states, $\hat{U}_{k,s}$ is simply a submatrix of $\hat{U}$.

Now that we have made the connection of matrix permanents to state amplitudes, we wish to turn our attention to the computational complexity of computing permanents in general.  This is relevant to our attempt at simulating \BS because, from the state amplitudes, we can immediately infer the probability distribution $P(\hat{U})$.  If we cannot easily compute these amplitudes, then we must seek another way of trying to produce $P(\hat{U})$.

A common way to compute the determinant that works analogously for the permanent is the method of Laplace decomposition;  one can write the entire permanent of $M$ as a sum of permanents of sub-matrices multiplied by elements of one row or column of $M$.  Specifically, if $M'_{i,j}$ is the submatrix of $M$ generated by deleting the $i$th row and $j$th column of $M$, then for any $i\in\{1,\dots,n\}$,
\begin{equation}
\perm(M)=\sum_{j=1}^n m_{i,j}\cdot\perm(M'_{i,j}).
\end{equation}
Using the Laplace decomposition, we trade computing the permanent of one $n\times n$ matrix to computing $n$ permanents of $(n-1)\times(n-1)$ matrices.  It is easy to see that such a decomposition is in general inefficient to compute, because each step reducing the matrix size has a cost of the size of the matrix.  We would need at least $n!$ steps to compute in this way.

A reader familiar with linear algebra may recall another way of computing determinants---the method of Gaussian elimination.  This technique involves using elementary row operations to reduce $M$ to row-echelon form, at which point the diagonal is the only non-zero term.  This method is indeed much more efficient---it can be solved with only $\mathcal{O}(n^3)$ number of steps.  However, Gaussian elimination relies on the multiplicative property of determinants that is not shared by permanents, and hence one cannot use this technique here.  Interestingly, if one replaced the bosonic Fock states in the statement of \BS with fermionic Fock states (unsurprisingly, this problem is called \cc{FermionSampling}), the state amplitudes $\gamma_s$ correspond to a \textit{determinant} of the submatrix $\hat{U}_{k,s}$.  It is not so surprising, then, that it can be shown \cc{FermionSampling} $\in$ \PP \cite{bib:AA13response}.

In hopes of more easily classifying the permanent problem, one might consider a simplification; suppose the entries of the matrix $M$ take on only binary values, $m_{ij}\in\{0,1\}$.  Does $\perm(M)$ now admit an efficient classical algorithm?  The answer is somewhat nuanced.  To get to the bottom of this question, we first define a new complexity class.

\begin{itemize}
    \item[{}] \textbf{\#P}:  Let $Z\in\NP$. The functional problem $Z'$ is in the complexity class \sharpP if, for an instance of $Z$, $Z'$ outputs the number of satisfying assignments of that instance.
\end{itemize}
Let us clarify some of the language used in this definition.  Many \NP problems take the form of a satisfiability clause---for example, 3SAT (an \NP-complete problem) asks if a particular kind of boolean string has an assignment of variables such that the string evaluates to TRUE (called a \textit{satisfying assignment}).  The corresponding \sharpP problem would then be, \textit{how many satisfying assignments does an instance of $\mathrm{3SAT}$ have?}  It is trivial to see that \NP $\subset$ \sharpP, since if you can count the number of assignments of an \NP problem, then you know whether it has no satisfying assignments (a NO-instance) or at least one satisfying assignment (a YES-instance).

It was proven by Valiant in 1979 (in the same paper in which \sharpP was defined) that the permanent of a matrix with only binary entries is in fact \sharpP-complete \cite{bib:Valiant79}.  This is a somewhat shocking result, especially when combined with a later development of Toda in 1991 \cite{bib:Toda91}.  His theorem (later earning him the Godel prize in 1998) shows that $\PH \subset \PP^{\sharpP}$, which has the shocking implication that a classical computer with access to an oracle for finding binary permanents would contain the entire polynomial hierarchy.  This should give the reader a sense of how difficult \sharpP problems are expected to be, and thus how surprisingly hard even a simple case of computing permanents may be (it should be noted that there is an efficient way of transforming an integer matrix of a corresponding matrix with only \{0,1\} entries that has the same permanent).   

The result changes quite dramatically when one considers an \textit{approximation} of $\perm(M)$.  It was shown in Ref.~\cite{bib:Jerrum04} that an efficient approximation algorithm exists for matrices with non-negative real entries.  Since the \BS problem involves the permanent of complex-valued entries, a natural question is whether this result could be extended to such matrices.  In fact, this question is addressed in the same paper, where it is stated that if such an approximation algorithm did exist for matrices with even a single negative entry, it could be used to compute the \textit{exact} permanent of binary matrices, thereby implying \BPP= \sharpP, an even stronger statement than \PP= \NP.  

The existence of an efficient approximation algorithm for non-negative entries has an interesting implication for ``classical" \BS.  If the matrix $\hat{U}$ represents a classical probability distribution (which must have strictly non-negative entries, corresponding to probability amplitudes rather than state amplitudes), then the output can be efficiently approximated.  This is again evidence that quantum systems are fundamentally different from classical systems, and gives a glimpse at what kind of advantage post-classical computers might provide.

Are we able to conclude then, from the hardness of calculating permanents, that the \BS problem is classically intractable?  There are two subtle points to consider.  The first is that our goal is to approximate the probability distribution $P(\hat{U})$, and that the \textit{probability} of a measurement finding the state $\ket{s}$ is equal to $|\gamma_s|^2$, not $\gamma_s$ directly.  It may be that the former is fundamentally easier to compute than the latter, and thus there may be a way to produce this approximation without computing permanents at all.  Indeed, the ability to compute state amplitudes or probabilities is a \textit{sufficient} condition to efficiently approximate $P(\hat{U})$, it is not a \textit{necessary} condition.

Another consideration is a kind of converse of the previous one; would finding $P(\hat{U})$ allow one to compute $\gamma_s$?  From the definition of \BS, the reader might have guessed that our intention was always to implement the network itself (which can be done efficiently using the network of Figure \ref{fig:bsarch}).  Does this mean that we could reverse-engineer this as a method for approximating $|\gamma_s|^2$ or even $\gamma_s$ directly?  Is \BS a permanent-finding machine?  These questions can be restated in some form as complexity theoretic questions.  We can see that \BS $\in$ \sharpP, but is \BS \sharpP-complete? Or does there exist a smaller class \textbf{C} such that \BS $\in$ \textbf{C} but \textbf{C} $\subsetneq$ \sharpP?  We will answer most of these questions throughout the following sections; we first need some additional tools at our disposal, which are introduced where appropriate.

\section{Exact Case} \label{sec:exact}
We now give a formal statement of the definition of \BS, which is taken directly from Ref.~\cite{bib:AA10}:

\begin{itemize}
    \item[{}] \textbf{Definition} (\BS, formal): 

The input to the problem will be an $m \times n$ column-orthonormal matrix $A \in U_{m,n}$.  Given $A$, together with a basis state $S\in \Phi_{m,n}$---that is, a list $S = (s_1, . . . , s_m)$ of nonnegative integers, satisfying $s_1 +\dots+s_m = n$---let $A_S$ be the $n \times n$ matrix obtained by taking $s_i$ copies of the $i$th row of $A$, for all $i \in [m]$.  Let $\mathcal{D}_A$ be the probability distribution over $\Phi_{m,n}$ defined as follows:
\begin{equation}
\textrm{Pr[S]}_{\mathcal{D}_A}=\frac{|\textrm{perm}(A_s)|^2}{s_1!\dots s_m!}
\end{equation}
The goal of \BS is to sample either exactly or approximately from $\mathcal{D}_A$, given $A$ as input.
\end{itemize}
Throughout this thesis, we generally refer to the matrix $A$ as $\hat{U}$, but for this section we keep the notation consistent with the above definition from Ref.~\cite{bib:AA10} for clarity. We also require the definition of a \BS oracle, again from Ref.~\cite{bib:AA10}:

\begin{itemize}
    \item[{}] \textbf{Definition} (\BS oracle): 

Let $\mathcal{O}$ be an oracle that takes as input a string $r\in\{0,1\}^{\mathrm{poly}(n)}$, an $m\times n$ matrix $A\in U_{m,n}$, and an error bound $\epsilon > 0$ encoded as $0^{1/\epsilon}$.  Also, let $\mathcal{D}_{\mathcal{O}}(A,\epsilon)$ be the distribution over inputs $\mathcal{O}$ if $A$ and $\epsilon$ are fixed but $r$ is uniformly random.  We call $\mathcal{O}$ an exact \BS oracle if $\mathcal{D}_{\mathcal{O}}(A,\epsilon)=\mathcal{D}_A$ for all $A\in U_{m,n}$.  Also, we call $\mathcal{O}$ an approximate \BS oracle if $\| \mathcal{D}_\mathcal{O}(A,\epsilon)-\mathcal{D}_A\|\leq\epsilon$ for all $A\in U_{m,n}$ and $\epsilon >0$. 
\end{itemize}

Note that the norm in the above definition is the total variation distance between two probability distributions $\mathcal{P}$ and $\mathcal{Q}$ over a finite set $X$ defined by $\|\mathcal{P}-\mathcal{Q}\|=\frac{1}{2}\sum_{x\in X}|P(x)-Q(x)|$. In this section, we will concern ourselves with the problem of being able to exactly sample from the distribution $\mathcal{D}_A$, referred to as exact \BS.  Our goal is not to prove the results of AA, but to give general insight into the problem.  We separate this section from the approximate case because the two problems seem to admit very different complexities; the exact proof is straightforward, while the approximate case requires a deeper analysis.  Because \BS has so many intricate properties, much can be learned from both.

We will first state the result from Ref.~\cite{bib:AA10} (summarized):
\begin{itemize}
    \item[{}] \textbf{Theorem} (Exact \BS): 
The exact \BS problem is not efficiently solvable by a classical computer, unless $\PP^\sharpP=\BPP^\NP$ and the polynomial hierarchy collapses to the third level.  More generally, let $\mathcal{O}$ be an exact \BS oracle.  Then $\PP^\sharpP\subseteq\BPP^{\NP^\mathcal{O}}$.
\end{itemize}
Before we talk about proving this theorem, let us review why $\PP^\sharpP=\BPP^\NP$ collapses the polynomial hierarchy.  In the previous section, we saw Toda's theorem which states $\PH\subseteq \PP^\sharpP$ \cite{bib:Toda91}.  Looking back at Figure \ref{fig:ph}, we can see that $\NP^\NP$ is the third level of the polynomial hierarchy.  Since $\BPP^\NP \subseteq \NP^\NP$ as a result of $\BPP\subseteq\NP$, this would mean that together with Toda's theorem,
\begin{equation}
\PH\subseteq \PP^\sharpP\subseteq \NP^\NP\subseteq \PH.
\end{equation}
Thus $\PH=\NP^\NP$, which by definition is a collapse of the polynomial hierarchy to the third level.

The theorem above is proven in two ways by AA.  The first is by showing that approximating $|\gamma_s|^2$ to within a multiplicative constant is \sharpP-hard, and furthermore that an efficient classical \BS simulator would allow one to compute precisely that in the class $\BPP^\NP$.  Thus, $\PP^\sharpP\subseteq\BPP^{\NP^\mathcal{O}}=\BPP^{\NP}$ since $\NP^\BPP=\NP$.  The details of this proof can of course be found in Ref.~\cite{bib:AA10}, which are mostly mathematical in form.  Instead of discussing them in detail, we would rather like to give our attention to the second proof method, which comprises most of this remaining section.  This second method is not only much simpler, but utilizes the powerful complexity tool of \textit{postselection}, and deals more closely with linear optics and quantum computing as a whole.

First, we would like to discuss the role of postselection.  We will do so informally here, as a full description is lengthy but does not add much to the reader's intuition.  A  complexity class \textbf{C} with postselection (generally denoted by \textbf{PostC}) allows one to draw on a particular subset of data, which (though only polynomial in size) could have taken an exponential amount of time to generate.  A very straightforward example of the power of postselection comes from the class \textbf{PostBPP}, which is easily seen to contain \NP.  A \textbf{PostBPP} machine can simply guess the answer to an \NP problem, and then check to see if it is true. The machine then postselects only on accurate guesses.  Clearly, this is not an efficient approach for a classical computer, since it may take an exponential number of guesses before a correct one is chosen.

Earlier, we briefly mentioned the LOQC model and the fact that linear optics with adaptive measurements was universal for quantum computation, \BQP.  Along the way, KLM also showed that the capabilities of a postselected linear optical computer, \textbf{PostBosonP}, was also equivalent to postselected universal quantum computing \PostBQP.  Together with some other previously known results, AA shows the following chain (with their particular contribution indicated), assuming that an exact \BS oracle $\mathcal{O}$ is classically efficient:
\begin{equation}
\textbf{PP}=\PostBQP=\textbf{PostBosonP}\stackrel{AA}{\subseteq} \textbf{PostBPP}^{\mathcal{O}}\subseteq \BPP^{\NP^\mathcal{O}}.
\end{equation}
The reader can consult \cite{bib:AA10} for the definitions of these other classes, where the containment shown above follows almost immediately.  Importantly, the containment $\textbf{PP}\subseteq\BPP^{\NP^\mathcal{O}}$ is also known to collapse the polynomial hierarchy via Toda's theorem.

\section{Approximate Case} \label{sec:approx}

Having discussed the proofs of the previous section regarding exact \BS, one may wonder why we bother discussing the approximate result.  The reason is two-fold.  First, as we saw in Section \ref{sec:permanent}, there exist efficient algorithms for approximating certain kinds of permanents, whereas the exact permanent problem remains \sharpP-complete.  It would be poor form to base our belief that \BS is classically intractable on the results of the exact case alone, since this may be a kind of mathematical artifact or singularity resulting from demanding an exact algorithm.  This is especially true because, two, any physical implementation of a \BS device would only produce an approximation of the sampling distribution $\mathcal{D}_A$ since one could never hope to implement the matrix $A$ with infinite, error-free precision.   

As a disclaimer for this section, note that the result of AA for the approximate case is \textit{not a proof}.  Of course, the earlier result was in some sense not a proof that $\BS\notin\BPP$, but rather a dichotomy theorem suggesting that it is far more likely that $\BS\notin\BPP$ than the alternative.  Here, however, the dichotomy theorem relies on two (strong) conjectures about permanents---the permanent anti-concentration conjecture (PACC) and permanent of Gaussians conjecture (PGC).  Provided these hold, then we have a proof of a similar form as the exact case.

We now state the relevant definition and result from Ref.~\cite{bib:AA10}:
\begin{itemize}
    \item[{}] \textbf{Problem} ($|\textbf{GPE}|^2_{\pm}$): 
Given as input a matrix $X \in \mathcal{N}(0,1)_{\mathbb{C}}^{n\times n}$ of i.i.d. Gaussians, together with error bounds $\epsilon,\delta>0$, estimate $|\perm(X)|^2$ to within additive error $\pm\epsilon\cdot n!$, with probability at least $1-\delta$ over $X$, in poly($n,1/\epsilon,1/\delta$) time.
\end{itemize}

\begin{itemize}
    \item[{}] \textbf{Theorem} (Approximate \BS): 
Let $\mathcal{D}_A$ be the probability distribution sampled by a boson computer $A$. Suppose there exists a classical algorithm $C$ that takes as input a description of $A$ as well as an error bound $\epsilon$, and that samples from a probability distribution $\mathcal{D}'_{A}$ such that $\|\mathcal{D}'_{A}-\mathcal{D}_A\|\leq \epsilon$ in poly($|A|,1/\epsilon$) time.  Then the $|\textbf{GPE}|^2_{\pm}$ problem is solvable in $\BPP^{\NP}$.  Indeed, if we treat $C$ as a black box, then $|\textbf{GPE}|^2_{\pm}\in\BPP^{\NP^{C}}$.
\end{itemize}

Again, we will not explicitly prove the statement, but discuss a general proof strategy.  The method here is quite clever.  Essentially, one can hide a Gaussian permanent that they want to compute randomly inside of $A$ as a submatrix without dramatically changing the sampling probabilities.  Of course, one might guess that the size of the hidden submatrix must be relatively small compared to $A$.  Hence, there is a price one must pay in terms of the size of the matrix.  That is, to be sure that an instance of \BS is truly post-classical, then we need $n\leq m^{1/6}$, and the matrix $A$ should be chosen randomly (what is precisely meant by ``random" we will discuss momentarily).

The requirement $n\leq m^{1/6}$ is a restriction coming from the Haar-Unitary Hiding Theorem, but AA believe that a better analysis can show the restriction to be looser, likely up to $m=O(n^2)$  \cite{bib:AA10}.  This is distinct from another issue where $m$ must be bounded from below by $m=\Omega(n^2)$ to ensure that the probability of detecting more than a single photon in a single output port is negligible.  The scaling responsible for this second condition is a result of the \textit{bosonic birthday paradox}, which gets its name from the famously counterintuitive answer to the question: how many people need to be in a room such that there is a 50\% probability of at least two of them sharing the same birthday (assuming birthdays are uniformly distributed)? While generally one might guess this occurs around 100 people, or perhaps as low as 50, few expect the answer to be as low as 23.  The analogy here is simple--- there are $m$ possible ``birthday" output modes for each of the $n$ input photons, and we wish to avoid any two photons exiting through the same mode.

A vital assumption for these theorems is that the matrix $A$ is chosen randomly.  However, the question of how to choose a random unitary matrix is not immediately obvious.  One way to generate a random matrix is by considering a general factored form of a unitary $U(n)$ in terms of $n(n-1)/2$ rotations on a two dimensional subspace.  There is a natural mapping of these rotations onto the Reck decomposition of beam splitters \cite{bib:Reck94, bib:Dita01}, which individually can be generated by choosing two variables $\eta\in[0,1], \tau\in[0,2\pi)$ uniformly at random, corresponding to the transmissivity of the beam splitter and an additional phase.  One should be careful to check that this is truly a volume invariant way to randomly choose over $U(n)$ in the sense that each unitary matrix should have equal measure over the set.  As there is a unique such measure over $U(n)$---the Haar measure \cite{bib:RepLinearGroups}---one can establish that indeed this approach is Haar-random.

Importantly, the restrictions on the number of modes, total number of photons, and randomness do not necessarily mean that sampling is easy otherwise.  It only implies that the proofs from Ref.~\cite{bib:AA10} do not apply.  It may be that a more general case of \BS remains hard even for $m=O(n)$, for example, or specific sets of unitaries.  Still, until the result is strengthened, experimental implementations of \BS are likely to maintain these assumptions.

\section{Verification} \label{sec:limit}
In this section, we discuss a major obstacle toward \BS being implemented as a post-classical computational problem.  The motivation of the \BS problem is to show that a quantum computer is capable of performing a task that is intractable for a classical computer.  The trouble here is that the output of a quantum device that implements \BS is a probability distribution based on the unknown permanents of submatrices of $\hat{U}$.  Because there is no known classical way to simulate \BS, then how can we be sure that the device's output is correctly sampling from $\hat{U}$?  For example, suppose an optical interferometer does not properly synchronize the input photons from two different modes to arrive at a beam splitter simultaneously.  Because the photons are temporally mismatched, no interference would occur at the beam splitter, and this would change the output distribution of the device.  If we instead had a result showing that \BS could solve \NP problems, for example, this would be easy (to be clear, it is not expected that $\NP\subseteq\BQP$).  We could simply check whether the solution given by the machine is a satisfying assignment or not.  Is there, then, a way to classically \textit{verify} that the output of a \BS device is accurate?

Shortly after the \BS problem was introduced, it was suspected that the output distribution would be so diffuse relative to the entire state space that one could not distinguish the output (in a polynomial number of runs) from even the uniform distribution \cite{bib:Gogo13}.  It was shown in a followup by AA \cite{bib:AA13response} that these arguments were incomplete, by producing an efficient algorithm to distinguish between the two.  Still, this illustrates an important point; we can hope to compare the output distribution to some other distribution in hopes of disproving some hypothesis about what the machine may be doing.  This may be an entirely reasonable way to verify if one can narrow down the types of error to a specific type.  Of course, more must be known about the kind of distribution that an errant model might produce (e.g. for photons of differing spectral structure, see Ref.~\cite{bib:Rho14}).

Could there be an algorithm for verifying \BS under arbitrary assumptions?  It seems unlikely by the nature of the problem, and in fact is impossible for any \textit{fixed} polynomial sized circuit.  That is, for any $k$, one can efficiently create a distribution which is indistinguishable from \BS, when limited to $n^k$ classical operations \cite{bib:AA13response}.  Still, recent advancements have shown ways to verify \BS in some very general and practical settings which experimentalists (at least in the realm of quantum optics) find most problematic.  One such example is given in Ref.~\cite{bib:Shches16}, where a protocol is developed for distinguishing a \BS distribution from one where the photons are in some way distinguishable (and hence do not exhibit bosonic interference). So while verification does remain an open problem, it seems that the practical loopholes are rapidly shrinking to the point that only pathological errors might produce a distribution that is effectively unverifiable.

\pagebreak

\singlespacing
\chapter{Boson Sampling With Other States of Light}
\doublespacing

In this chapter, we will discuss other linear optical implementations of \BS.  First, we consider the states that differ from Fock states by a displacement operation---namely, displaced Fock states and photon-added coherent states. It is easy to show that the sampling problem associated with displaced single-photon Fock states and a displaced photon number detection scheme is in the same complexity class as boson sampling for all values of displacement. On the other hand, we show that the sampling problem associated with single-photon-added coherent states admits a transition from \BS-complexity in the small $\alpha$ regime to a trivial-to-simulate case for the large $\alpha$ regime.  This may indicate a complexity phase transition that has been seen in other problems thought to be outside of \PP \cite{bib:GentWalsh94}.

In the second model, we show that an analogous procedure implements the same problem, using photon-added or -subtracted squeezed vacuum states (with arbitrary squeezing), where sampling at the output is performed via parity measurements. The equivalence is exact and independent of the squeezing parameter, and hence provides an entire class of new quantum states of light in the same complexity class as boson sampling.  This model can even be viewed as a generalization of \BS, since in the limit as $\xi\rightarrow0$, the architecture reduces to that of \BS.

\section{Photon-Added Coherent States} \label{sec:pacs}

\footnote{This section previously appeared as: K. P. Seshadreesan, J. P. Olson, K. R. Motes,  P. P. Rohde, and J. P. Dowling. Boson sampling with displaced single-photon {F}ock states versus single-photon-added coherent states: The quantum-classical divide and computational-complexity transitions in linear optics.  \textit{Phys. Rev. A}, 91:022334, 2015. It is reprinted by permission of APS.}Here, we wish to investigate whether there are quantum states of light---other than Fock states---which when evolved through a linear-optical circuit and sampled using a suitable detection strategy, also implement likely classically hard problems similar to \BS.  This section summaries the results of Ref.~\cite{bib:Sesh15}.

Other recent results have shown that, in the case of Gaussian states (most generally displaced, squeezed, thermal states), sampling in the photon number basis can be just as hard as \BS~\cite{bib:Lund13}. To further elaborate, while the sampling of thermal states can be simulated efficiently by a classical algorithm~\cite{RLR_14}, it has been shown that the sampling of squeezed vacuum states is likely hard to efficiently simulate classically at least in some special cases~\cite{bib:PhysRevA.88.044301, bib:Lund13}. Among non-Gaussian inputs (other than Fock states),  generalized cat states---which are arbitrary superpositions of coherent states---with photon number detection have been shown to likely implement computationally hard sampling problems similar to \BS \cite{bib:RohdeCat}.

Here, we study the linear optics-based sampling problems associated with the quantum states of light that differ from Fock states by the displacement operater.  Namely, these are displaced Fock states and photon-added coherent states, together with a displaced photon number detection. Recall that the displacement operator (see Sec.~\ref{subsec:optics}) can be written as,
\begin{equation}
\label{disop}
\hat{D}(\alpha)=\exp\left(\alpha \hat{a}^{\dagger}-\alpha^*\hat{a}\right),
\end{equation}
where $\alpha$ is a complex amplitude that quantifies displacement in phase space, and $\hat{a}^\dagger$ is the photon creation operator for a single mode. The displaced single-photon Fock state (DSPFS) is the state $\hat{D}(\alpha)\hat{a}^\dagger|0\rangle$, while the single-photon-added coherent state (SPACS) has the reverse order of operators, $\hat{a}^\dagger\hat{D}(\alpha)|0\rangle$ (note that the latter state is not normalized). Although these input states are in practice more difficult to prepare than the single-photon Fock state, the associated sampling problems allow us to demonstrate a transition in the computational complexity of linear optics. It is easy to show that the DSPFS sampling problem (which we will refer to here as \cc{DisplacedSampling}) is in the same complexity class as \BS for any displacement $\alpha$. However, the SPACS, differing only in the ordering of the operators, presents an interesting case---we show that the sampling problem with SPACS is just as hard as \BS when the input coherent amplitudes are sufficiently small (subject to a bound that we derive explicitly), but transitions into a problem that is easy to simulate classically in the limit of large input coherent amplitudes.

\subsection{Sampling Displaced Fock states}
\label{dfssampling}

Consider the DSPFS in place of the single-photon Fock states in Eq.~(\ref{eq:bsin}) as inputs to a linear-optical interferometer. That is, consider an overall input state of the form,
\begin{equation} \label{eq:input_state_DFS}
\ket{\psi_\mathrm{in}}^{\mathrm{DSPFS}} =\left( \prod_{i=1}^{n}\hat{D}_i\left(\alpha^{(i)}\right)\hat{a}_i^\dagger\right) \ket{0_1,\dots,0_m},
\end{equation}
where $\hat{D}_i\left(\alpha^{(i)}\right)$ is the displacement operator of the $i$th mode, and $\alpha^{(i)}$ is the complex coherent amplitude for the displacement. A unitary operation $\hat{U}$ then transforms the state into $|\psi_{\rm out}\rangle^{\mathrm{DSPFS}}$,
\begin{align}
\label{UxformDFS}
&=\hat{U} \left(\prod_{i=1}^{n}\hat{D}_i\left(\alpha^{(i)}\right)\hat{a}_i^\dagger\right)\hat{U}^{\dagger}\hat{U}\ket{0_1,\dots,0_m},\nonumber\\
&=\hat{U} \left(\prod_{i=1}^{n}\hat{D}_i\left(\alpha^{(i)}\right)\right)\hat{U}^{\dagger}\hat{U}\left(\prod_{k=1}^{n}\hat{a}_k^\dagger\right)\hat{U}^{\dagger}\ket{0_1,\dots,0_m}\nonumber\\
&= \prod_{i=1}^{n}\left(\hat{U}\hat{D}_i\left(\alpha^{(i)}\right)\hat{U}^{\dagger}\right)\prod_{k=1}^{n}\left(\hat{U}\hat{a}_k^\dagger\hat{U}^{\dagger}\right)\ket{0_1,\dots,0_m}\nonumber\\
&=\left(\prod_{j=1}^{m}\hat{D}_j\left(\beta^{(j)}\right)\right)\left(\sum_{S}\gamma_S (\hat{b}_1^{\dagger})^{s_1}(\hat{b}_2^{\dagger})^{s_2}\dots(\hat{b}_m^{\dagger})^{s_m}\right)\ket{0_1,\dots,0_m},
\end{align}
where $\beta^{(j)} = \sum_i U_{i,j} \alpha^{(i)}$ is the new displacement amplitude in the $j$th mode, $\hat{b}_k^{\dagger}$ is the photon-creation operator of the  $k$th mode, and $s_k$ is the number of photons in the $k$th mode, associated with configuration $S$ at the output such that $\sum_{k=1}^m s_k=n$ for each $S$. In deriving Eq.~(\ref{UxformDFS}), we have used the following: $\hat{U}^{\dagger}\hat{U}=I$, $\hat{U}\ket{0_1,\dots,0_m}=\ket{0_1,\dots,0_m}$, Eq.~(\ref{eq:permout}), and the fact that the action of a unitary on a tensor product of coherent states results in another tensor product of coherent states as shown in Appendix A of~\cite{bib:RohdeCat}. The final expression is nothing but a displaced version of the usual \BS output state as given in Eq.~(\ref{eq:bsout}).

For any unitary operator $\hat{U}$, the new complex displacement amplitudes $\beta^{(j)}$ can be efficiently computed. Since $\hat{D}(-\alpha)\hat{D}(\alpha)=I$, a counter-displacement with amplitudes $-\beta^{(j)}$ could be applied to the $m$ output modes. The displacement operation could be performed using unbalanced homodyning~\cite{BW99,WLD07}. Upon such a displacement operation, the sampling problem associated with the output state reduces to the \BS output, which 
can subsequently be accessed using coincidence photon number detection (CPND). Thus, \cc{DisplacedSampling} with our modified measurement scheme at the output comprising of an inverse displacement followed by CPND has an identical output distribution to \BS, and hence clearly falls into the same complexity class. While this observation may appear trivial---since a product of displacement operators commutes through a linear-optical network to yield another product of displacement operators---it demonstrates that an entire class of quantum states of light yield a problem of equal complexity to \BS, with a suitable adaptation of the measurement scheme.

\subsection{Sampling Photon-Added Coherent States}
\label{pacssampling}

Now consider an input state comprising SPACS instead of the DSPFS. These states differ from the DSPFS only in the ordering of the operators. However, since the displacement operator of Eq.~(\ref{disop}) does not commute with the photon creation operator $\hat{a}^\dagger$, the SPACS and the DSPFS are distinctly different states.  We will refer to the sampling problem described below as \cc{PACSampling}.

A $k$-photon-added coherent state may be written as,
\begin{equation}
\label{pacsdefinition}
|\alpha,k\rangle=\mathcal{N}_k (\hat{a}^{\dagger})^k|\alpha\rangle,
\end{equation}
where $\alpha$ is the complex coherent amplitude and the normalization is,
\begin{equation}
\mathcal{N}_k=\frac{1}{\sqrt{k!L_k (-|\alpha|^2)}},
\end{equation}
$L_k$ being the Laguerre polynomial of order $k$. Such states were first discussed by Agarwal \& Tara \cite{Agarwal_91}. The SPACS we consider here thus corresponds to $\ket{\alpha,1}$ of Eq.~(\ref{pacsdefinition}). 

\begin{figure}[b]
\centering
\includegraphics[scale=0.7]{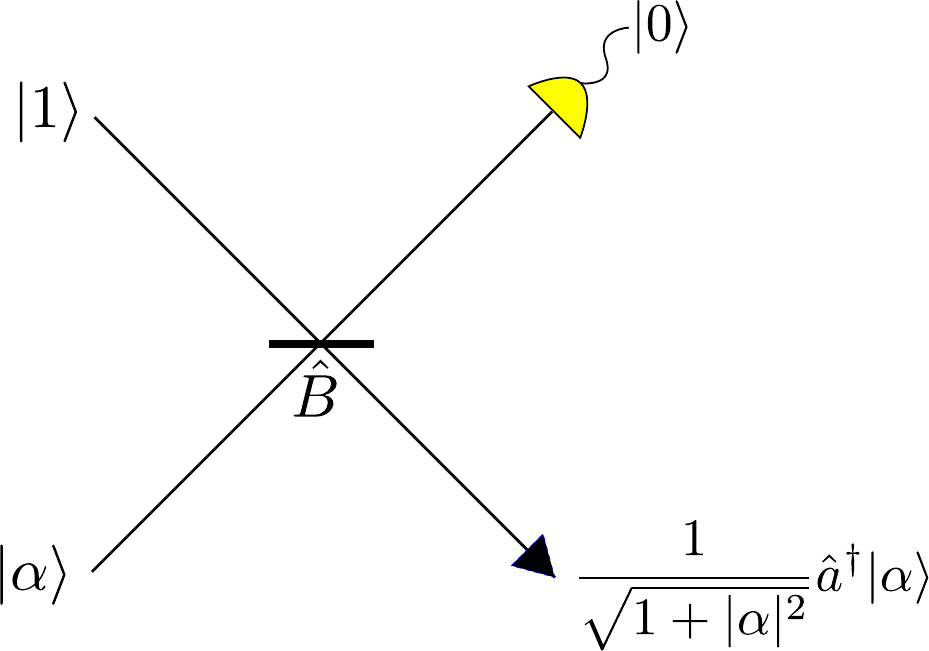}
\caption{ When a coherent state and a single photon state are mixed on a highly reflective beamsplitter, and no photon is detected in the transmitted mode, a SPACS is heralded in the transmitted mode.} \label{fig:PACS_prep}
\end{figure}

Consider a scheme where a single photon (e.g. prepared via heralded spontaneous parametric down-conversion) is mixed with a coherent state on a highly reflective beam splitter (Figure \ref{fig:PACS_prep}). When a single-photon detector placed in the transmitted mode detects vacuum, we know that the incident photon has been emitted into the other output port, and thus a SPACS has been heralded~\cite{Dakna_98, Dakna_98_2, Zavatta_04, Zavatta_05}.

\begin{figure}[!htb]
\centering
\includegraphics[scale=0.25]{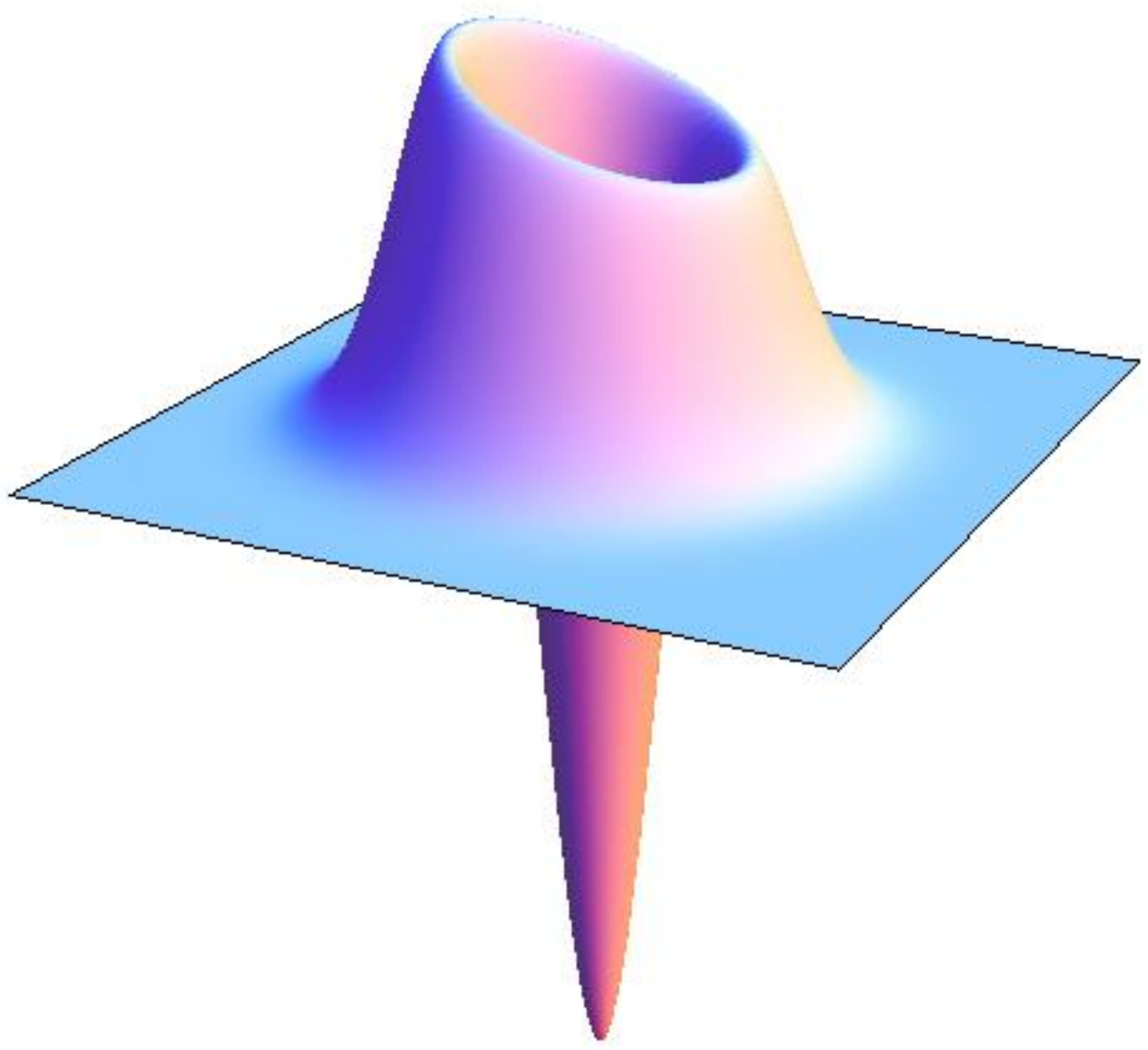} 
\includegraphics[scale=0.25]{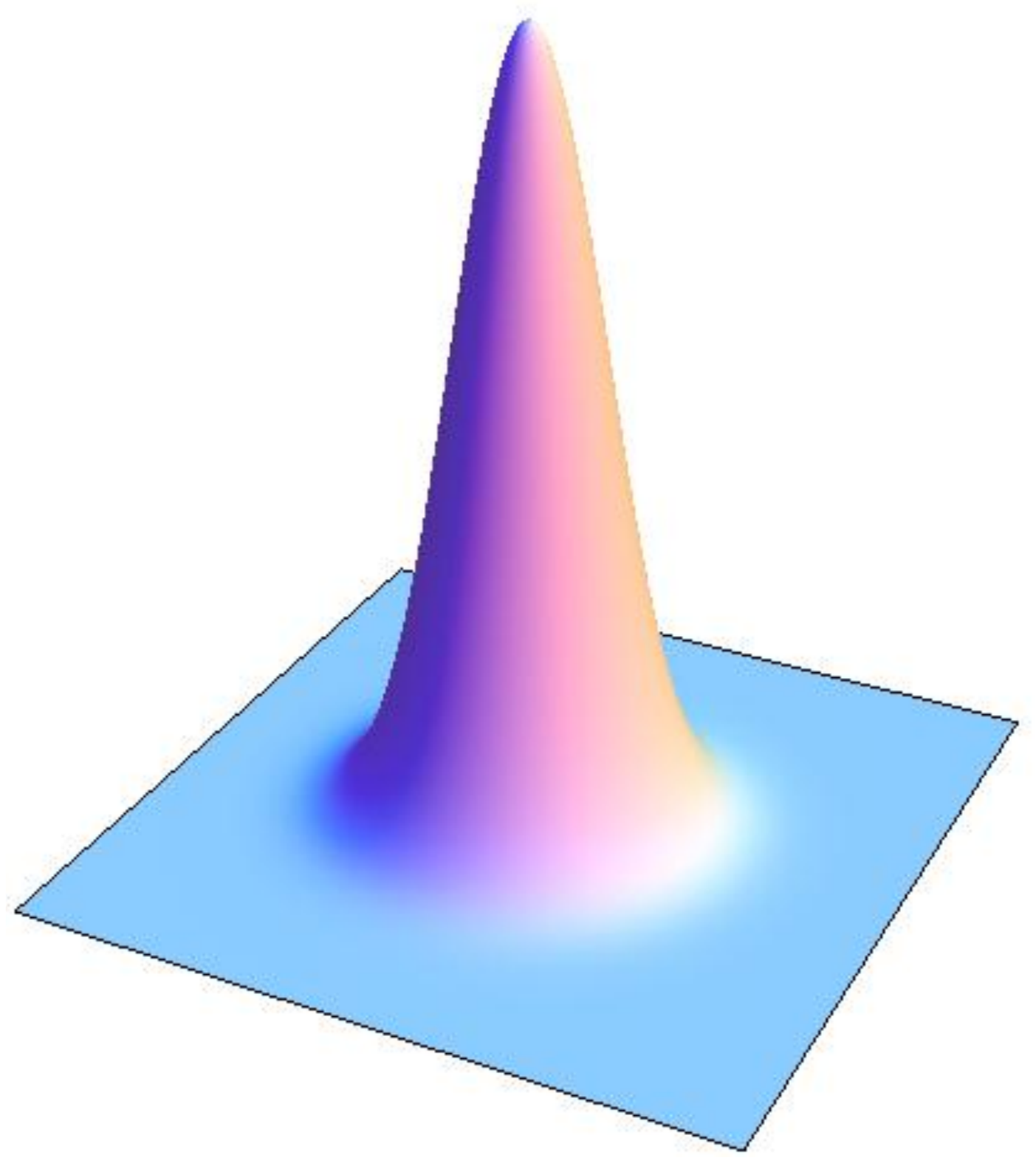}
\caption{Wigner function of (left) a SPACS, (right) a coherent state, with amplitude \mbox{$|\alpha|^2=0.01$}. The former is seen to take negative values close to the phase-space origin, while that of the latter is strictly positive everywhere. $W(0)$ is at the center of the plane. Sampling $W(0)$ would distinguish between a coherent state and a SPACS.} \label{wigner}
\end{figure}

\begin{figure}[!htb]
\centering
\includegraphics[scale=0.9]{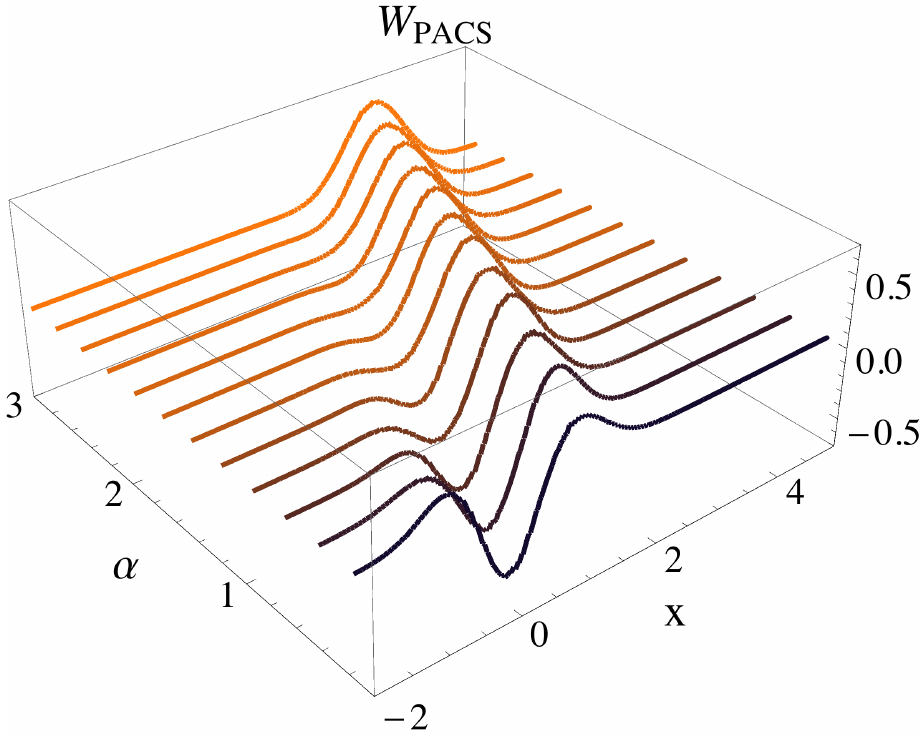} 
\caption{2-d slices of the Wigner function of SPACS across its major axis, as a function of the coherent amplitude $|\alpha|$. We see that the negativity vanishes, and the shape tends towards being a Gaussian for increasing values of $|\alpha|$.}\label{wignerslices}
\end{figure}

The SPACSs have been studied extensively in the context of demonstrating the quantum-classical transition, since they allow for a seamless interpolation between the highly nonclassical Fock state $|1\rangle$ ($\alpha\rightarrow 0$) and a highly classical coherent state $|\alpha\rangle$ ($|\alpha|\gg1$)~\cite{Zavatta_04}. The Wigner function of a SPACS can be expressed as \cite{Agarwal_91},
\begin{equation}
W(z)=\frac{2(|2z-\alpha|^2-1)}{\pi(1+|\alpha|^2)}e^{-2|z-\alpha|^2},
\end{equation}
where $z=x+iy$ is the phase-space complex variable, and $\alpha$ the coherent amplitude in the state. Figure \ref{wigner} shows the Wigner functions of a SPACS and a coherent state. The former attains negative values at points close to the origin in phase space, which is a demonstration of the nonclassical nature of the state. Figure \ref{wignerslices} shows a 2-d slice of the Wigner function of a SPACS across the major axis, as a function of the coherent amplitude $|\alpha|$. It can be seen that the Wigner function loses its negativity as $\alpha$ increases and tends towards being a Gaussian state. 

\newpage

The SPACS-based input that we consider to a linear-optical sampling device can be written as,
\begin{eqnarray}
\label{input}
|\psi_{\rm in}\rangle^{\mathrm{SPACS}}&=&\mathcal{N}\prod_{i=1}^{n}\hat{a}_i^\dagger\hat{D}_i\left(\alpha^{(i)}\right)\ket{0_1,\dots,0_m},\nonumber\\
\mathcal{N}&=&\prod_{j=1}^{n}\frac{1}{\sqrt{1+|\alpha^{(j)}|^2}}.
\end{eqnarray}
where $\alpha^{(i)}$ represents the complex coherent amplitude in the \mbox{$i$th} mode and $\mathcal{N}$ is the overall normalization factor. That is, the input to the first $n$ modes are SPACS, while the remaining $m-n$ modes are initiated in the vacuum state.  A unitary operation $\hat{U}$ then transforms the state into,
\begin{align}
\label{UxformPACS1}
|\psi_{\rm out}\rangle^{\mathrm{SPACS}}&=\hat{U}|\psi_{\rm in}\rangle^{\mathrm{SPACS}}\nonumber\\
&=\mathcal{N}\hat{U} \left(\prod_{i=1}^{n}\hat{a}_i^\dagger\hat{D}_i\left(\alpha^{(i)}\right)\right)\hat{U}^{\dagger}\hat{U}\ket{0_1,\dots,0_m}.
\end{align}
This state can be alternatively written as,
\begin{align}
\label{UxformPACS2}
&=\mathcal{N}\hat{U} \left\{\prod_{i=1}^{n}\left(\hat{D}_i\left(\alpha^{(i)}\right)\hat{a}_i^\dagger+{\alpha^{(i)}}^*\hat{D}_i\left(\alpha^{(i)}\right)\right)\right\}\hat{U}^{\dagger}\ket{0_1,\dots,0_m},
\end{align}
where we have used the commutation relation between the displacement operator and the photon-creation operator, namely,
\begin{align}
\left[a^\dagger, \hat{D}(\alpha)\right]=\alpha^*\hat{D}(\alpha).
\end{align} 
We can further simplify the state as,
\begin{align}
&=\mathcal{N}\hat{U} \prod_{i'=1}^{n}\hat{D}_{i'}\left(\alpha^{(i')}\right)\hat{U}^{\dagger}\hat{U}\prod_{i=1}^{n}\left(\hat{a}_i^\dagger+{\alpha^{(i)}}^*\right)\hat{U}^{\dagger}\ket{0_1,\dots,0_m},\nonumber\\
&=\mathcal{N}\prod_{i'=1}^{n}\left(\hat{U} \hat{D}_{i'}\left(\alpha^{(i')}\right)\hat{U}^\dagger\right)\prod_{i=1}^{n}\left(\hat{U}\hat{a}_i^\dagger\hat{U}^{\dagger}+{\alpha^{(i)}}^*\right)\ket{0_1,\dots,0_m}\nonumber\\
&=\mathcal{N}\prod_{j=1}^{m}\hat{D}_{j}\left(\beta^{(j)}\right) \prod_{i=1}^{n}\left(\hat{U}\hat{a}_i^\dagger\hat{U}^{\dagger}+{\alpha^{(i)}}^*\right)\ket{0_1,\dots,0_m},
\end{align}
where $\beta^{(j)} = \sum_{i'} U_{i',j} \alpha^{(i')}$ is the new displacement amplitude in the $j$th mode. Similar to the case of DSPFS sampling, we can now apply a counter-displacement operation of amplitude $\prod_{j=1}^{m}\hat{D}_j\left(-\beta^{(j)}\right)$ (again, this can be computed efficiently), so that the output state reduces to,
\begin{align}
\label{pacsundis_state}
\mathcal{N}\prod_{i=1}^{n}\left(\hat{U}\hat{a}_i^\dagger\hat{U}^{\dagger}+{\alpha^{(i)}}^*\right)\ket{0_1,\dots,0_m}.
\end{align}

Let us denote the state $\prod_{i=1}^{n}\left(\hat{U}\hat{a}_i^\dagger\hat{U}^{\dagger}\right)\ket{0_1,\dots,0_m}$, which corresponds to the usual \BS evolution as $|AA\rangle$ (in dedication to Arkhipov and Aaronson). Further, for simplicity, let us choose all the input coherent amplitudes to be equal to $\alpha$. Then, the output state in Eq.~(\ref{pacsundis_state}) can be written as,
\begin{align}
\mathcal{N'}\left(\sum_{i=0}^{n-1}{\alpha^*}^{n-i}\left(\hat{U}\hat{\mathcal{A}}^{(i)}\hat{U}^\dagger\right)\ket{0_1,\dots,0_m}+|AA\rangle\right),
\end{align}
where $\hat{\mathcal{A}}^{(i)}$ is defined for $i\in\{0,1,\cdots,n\}$ as,
\begin{equation}
\hat{\mathcal{A}}^{(i)}\equiv
\left\{
	\begin{array}{cl}
		\displaystyle\frac{1}{i!(n-i)!}\sum_{\sigma\in S_n}\prod_{k=1}^{i}\hat{a}^\dagger_{\sigma(k)},  & \mbox{if } i \geq 1 \\
		{\rm id}, & \mbox{if } i =0,
	\end{array}
\right.
\end{equation}
$S_n$ being the symmetric group of degree $n$, ${\rm id}$ being the identity operator, and $\mathcal{N'}=1/(\sqrt{1+|\alpha|^2})^n$. Now, if we perform photon number detection at the output, the set of all possible outcomes includes total photon numbers (from across all the $m$ output modes) ranging from zero to $n$. Detection events consisting of a total photon number of $n$ would correspond to sampling of the $|AA\rangle$ term from  the superposition. The probability of detecting a total of $i$ photons at the output can be written as,
\begin{align}
P_{i}=\mathcal{N'}^{2} {n\choose{i}}\left(|\alpha|^2\right)^{n-i}.
\end{align}
This is because there are ${n\choose{i}}$ terms in $\hat{\mathcal{A}}^{(i)}$, each with a weight of $\mathcal{N'}^{2}\left(|\alpha|^2\right)^{n-i}$.

We now ask the following question: how should $|\alpha|$ scale in terms of $n$---the total number of SPACS in the input (representative of the size of the sampling problem) so that the post-selection probability of detecting $n$ photons at the output of the interferometer scales inverse polynomially in $n$. This is a relevant question to ask, because such a scaling would guarantee the sufficiency of a polynomial number of measurements in order to sample the desired AA term in the output. For simplicity, let us consider ${\rm poly} (n)=n^k$, where $k\in \mathbb{Z}^+$ (the set of positive integers). Solving for $|\alpha|$ that satisfies the above scaling requirement in the limit of a large $n$, we have,
\begin{align}
\frac{1}{(1+|\alpha|^2)^n}&\geq \frac{1}{{\rm poly} (n)}\nonumber\\
\Rightarrow 1+|\alpha|^2&\leq ({\rm poly} (n))^{1/n}\nonumber\\
&\leq 1+\epsilon(n),
\label{psp1}
\end{align}
where the third inequality is due to the fact that for all $k\in \mathbb{Z}^+$,
\begin{align}
\lim_{n\rightarrow\infty}(n^k)^{1/n}&=\lim_{n\rightarrow\infty}e^{\frac{k}{n}\log n}\nonumber\\
&=\lim_{n\rightarrow\infty}e^{\frac{k}{n}}=e^{0^+}=1+\epsilon(n).
\label{psp2}
\end{align}
From Eq.~(\ref{psp1}), we have,
\begin{equation}
|\alpha|^2\leq \epsilon(n),
\label{psp3}
\end{equation}
and the large-$n$ expansion,
\begin{equation}
e^{\frac{k}{n}\log n}=1+\frac{k}{n}\log n+O\Big(\frac{1}{n^2}\Big),
\end{equation}
tells us that $\epsilon(n)\geq (k/n)\log n$. The chain of inequalities,
\begin{align}
\epsilon(n)\geq \frac{k\log n}{n}\geq\frac{1}{n}
\end{align}
thus implies $|\alpha|^2\leq 1/n$ is a sufficient condition on $|\alpha|$ to ensure that the post-selection probability of the AA term scales inverse polynomially in $n$. For $|\alpha|^2=1/n$, in the limit of large $n$, we find that the probability of the term $|AA\rangle$ being detected at the output is,
\begin{align}
P_n=\lim_{n\rightarrow\infty}\frac{1}{(1+\frac{1}{n})^n}=\frac{1}{e}\approx 36\%.
\end{align}
Further, the probability $P_n$ converges to one when $|\alpha|^2=1/n^2$; i.e., the considered sampling problem with SPACS inputs reduces to \BS without the need for post-selection. This result is consistent with the original result that \BS is robust against small amounts of noise. 

On the other hand, we could also ask the question: how should $|\alpha|$ scale, so that the photon number sampling almost always gives the $m$-mode vacuum. For $|\alpha|^2=n^2$, we find that the probability of the $m$-mode vacuum term being detected at the output is,
\begin{align}
P_0&=\lim_{n\rightarrow\infty}\frac{{(n^2)}^n}{(1+n^2)^n}\nonumber\\
&=\lim_{n\rightarrow\infty}\frac{1}{(1+\frac{1}{n^2})^n}=1.
\end{align}
That is, the considered sampling problem with SPACS inputs becomes classically simulable when $|\alpha|^2$ scales as $n^2$, or larger, in the sense that it always results in the detection of the $m$-mode vacuum at the output. 

Therefore, we see that the computational complexity of sampling the SPACS goes from being just as hard as \BS for coherent amplitudes $|\alpha|^2\leq 1/n$, to being classically simulable when $|\alpha|^2\geq n^2$, where $n$ is the total number of SPACS inputs.

As discussed in Sec.~\ref{pacssampling}, the SPACS is known to exhibit a quantum-classical transition in terms of the negativity of its Wigner function when the coherent amplitude is changed from small to large values. The results presented in this work indicate that \cc{PACSampling}, linear optics and a displaced CPND similarly demonstrates a transition in computational complexity. The complexity goes from being likely hard to simulate classically for small coherent amplitudes (similar to \BS), to being easy to simulate classically for large coherent amplitudes. This result is also consistent with a conjecture presented in Ref.~\cite{bib:Chapter} that computational complexity relates to the negativity of the Wigner function.

To summarize, a central open question is what class of quantum states of light yield linear-optical sampling problems that are likely hard to simulate efficiently on a classical computer. Here we have partially elucidated this question by considering two closely related classes of quantum states. We studied the linear-optical sampling of the DSPFS and the SPACS for a displaced CPND. We showed that while \cc{DisplacedSampling} remains likely hard to simulate efficiently for all values of the displacement, \cc{PACSampling} transitions from being likely hard to simulate efficiently for sufficiently small input coherent amplitudes to being efficiently simulable in the limit of large coherent amplitudes.


\section{Photon-Added or -Subtracted Squeezed Vacuum} \label{sec:pasv}

\footnote{This section previously appeared as: J. P. Olson, K. P. Seshadreesan,  K. R. Motes,  P. P. Rohde, and J. P. Dowling. Sampling arbitrary photon-added or photon-subtracted squeezed states is in the same complexity class as boson sampling.  \textit{Phys. Rev. A}, 91:022317, 2015. It is reprinted by permission of APS.}Here we will demonstrate that, in general, linear optical sampling using photon-added or -subtracted squeezed vacuum (PASSV) states and parity measurements yields a computational problem of equal complexity to \BS in \emph{all} parameter regimes (we will call this problem \PASSV). Importantly, because the mapping is exact, the robustness result for approximate sampling also holds.  Note that experimental implementation of \PASSV is not the focus of our result, as doing so is more difficult than \BS.  Our goal is to provide clarity on the theory of classifying the sampling complexity of quantum states.  In particular, we wish to demonstrate that Fock states are not unique---on the contrary, there are a plethora of other quantum states of light which yield sampling problems with similar complexity to \BS. Nevertheless, we believe it is still important to show that such a device is physically realizable.

\subsection{PASSV Sampling Model}
In order to show that the complexity of \BS also extends to \cc{PASSVSampling}, we prove that it implements the same logical problem, i.e. that the output of the device corresponds to the same matrix permanent sampling problem as in \BS.  The advantage of this method is that it allows us to avoid the very lengthy analysis comprising the original complexity proof, yet we can still apply all of the same results.  However, one must be careful to show equivalence throughout the entire problem.  

Both models employ a similar general setup; $m$ optical input modes are fed into a passive, linear interferometer and the resulting output is measured in each mode, with the joint distribution of the measurement constituting one sample.  However, the details differ in each step (which we will classify by \textbf{input}, \textbf{evolution}, \textbf{output}, and \textbf{measurement}).  To carefully guide the reader, we will first provide the details of each step of both models head-to-head, discussing the relevant differences.  We will then proceed to show that the two models implement the same sampling problem, and thus exhibit the same computational complexity.  For consistency and simplicity, we will consider the case of photon-added states throughout the comparison.

\textbf{Input:} 
The \BS model begins by preparing the first $n$ modes of a passive linear optical interferometer with single photons and the remaining \mbox{$m-n$} modes with vacuum states, where \mbox{$m=\Omega(n^2)$} (i.e. $m$ is asymptotically bounded below by some positive constant times $n^2$).  As conjectured by AA, this requirement ensures that the probability of more than one photon arriving at a given output mode is small (sometimes referred to as the `bosonic birthday paradox'). A stronger requirement of $m=\Omega(n^6)$ will suffice if one does not wish to adopt this additional conjecture. The input state is thus,
\begin{eqnarray} \label{eq:input_state}
\ket{\psi}_\mathrm{in}^\mathrm{AA} &=& \ket{1_1,\dots,1_n,0_{n+1},\dots,0_m} \nonumber \\
&=& \hat{a}_1^\dag\dots \hat{a}_n^\dag \ket{0_1,\dots,0_m},
\end{eqnarray}
where, as usual, subscripts denote mode number and $\hat{a}_i^\dag$ is the photonic creation operator on the $i$th mode.

In contrast, for PASSV boson sampling we prepare the first $n$ modes of a similar interferometer with PASSV states and the remaining \mbox{$m-n$} modes with squeezed vacuum (SV) states. We let the squeezing parameter $\xi$ be arbitrary, but ensure each mode has the same amount of squeezing. In the case of photon-added states, the input state is thus,
\begin{eqnarray} \label{eq:passv_input_state}
\ket{\psi}_\mathrm{in}^\mathrm{SV} &=& \hat{a}_1^\dag\hat{S}_1(\xi)\dots\hat{a}^\dag_n\hat{S}_n(\xi)\hat{S}_{n+1}(\xi)\dots \hat{S}_m(\xi)\ket{0_1,\dots,0_m} \nonumber \\
&=& \hat{a}^\dag_1\dots \hat{a}^\dag_n\ket{\xi_1,\dots,\xi_m},
\end{eqnarray}
where we have abbreviated $\hat{S}_i(\xi)\ket{0_i}=\ket{\xi_i}$ and again the subscript indicates mode number (not separate variables).  The state in Eq.~(\ref{eq:passv_input_state}) is not normalized, but this can be corrected by considering the state $\mathcal{N}\ket{\psi}_\mathrm{in}^\mathrm{SV}$ where,
\begin{equation} \label{eq:normalization}
\mathcal{N}= \Big[\sqrt{1+\sinh^2(\xi)}\;\Big]^{-n}.
\end{equation}
Since the normalization does not affect our result, we leave it out of subsequent equations for simplicity.  Recall from Sec.~\ref{subsec:optics},
\begin{equation}
\hat{S}(\xi)=\exp\left[\frac{1}{2}(\xi^*\hat{a}^2-\xi\hat{a}^\dag{}^2)\right],
\end{equation}
is the squeezing operator and $\hat{a}^\dag$ and $\hat{a}$ are the photon creation and annihilation operators respectively. In the Fock basis, if \mbox{$\xi=re^{i\theta}$}, then \mbox{$\hat{S}(\xi)\ket{0}=\ket{\xi}$} has the representation \cite{bib:GerryKnight05},
\begin{equation} \label{eq:sv}
\ket{\xi}=\frac{1}{\sqrt{\cosh(r)}}\sum_{m=0}^\infty (-1)^m \frac{\sqrt{(2m)!}}{{2^m m!}}e^{im\theta}\tanh^m(r)\ket{2m},
\end{equation}
and thus the SV state contains only even photon-number terms.  From the action of the creation or annihilation operator, a PASSV state then contains only odd photon-number terms. In the limit of vanishing squeezing, the SV state approaches the vacuum state, \mbox{$\lim_{\xi\to 0}\ket{\xi} = \ket{0}$}, and the photon-added SV state approaches the single-photon state, \mbox{$\lim_{\xi\to 0}\hat{a}^\dag\ket{\xi} = \ket{1}$}. Thus, we see that in the limit of vanishing squeezing, the input state for \PASSV reduces to \BS.

\begin{figure}[!htb]
\centering
\includegraphics[scale=0.8]{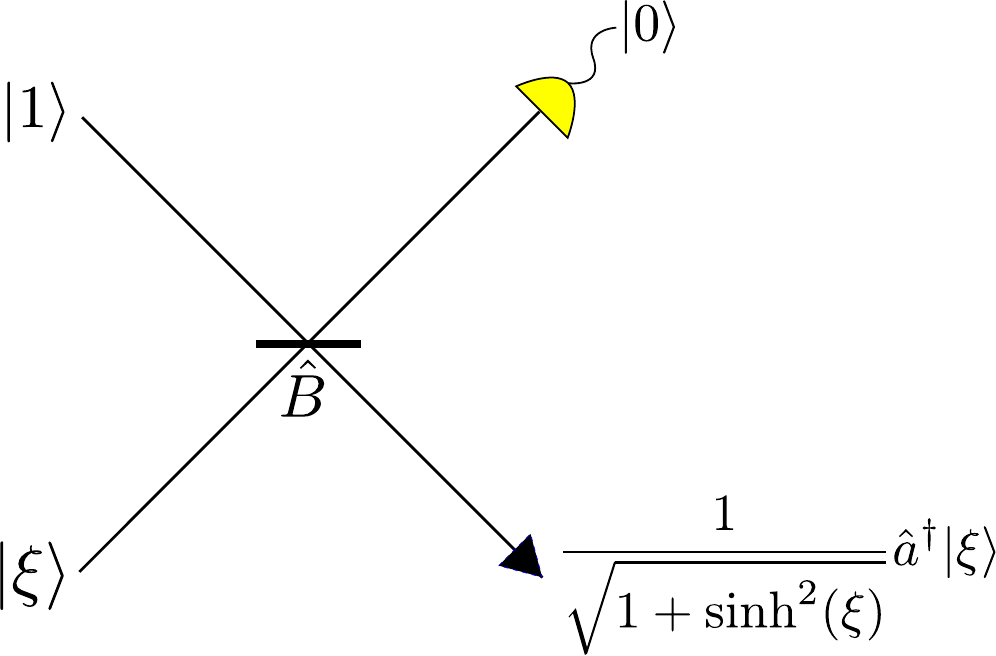}
\caption{Preparation of a photon-added SV state. A SV state is mixed with a single-photon state on a low reflectivity beamsplitter. The reflected mode is detected, and upon measuring the vacuum state we herald the preparation of the photon-added SV state in the other mode. The process is highly non-deterministic, but can be performed offline in advance.} \label{fig:prep}
\end{figure}

Photon-added SV states may be prepared (similar to the PACS state) by mixing a SV state (obtained from a degenerate parametric down-converter) with a single-photon state on a low reflectivity beamsplitter and post-selecting upon detecting the vacuum state in the reflected mode. Successful post-selection heralds the preparation of the photon-added SV state in the other mode. Thus, the preparation scheme is non-deterministic, but may be performed offline via trial-and-error in advance, enabling efficient state preparation. The preparation scheme is shown in Figure \ref{fig:prep}.  Photon-subtracted SV states may be prepared similarly by sending in a squeezed state and a vacuum state to the inputs and post-selecting on one photon in the reflected mode. 

\textbf{Evolution:}
In both models, the input state is fed into a passive linear optical interferometer consisting of beamsplitters and phaseshifters, which in general transforms the creation operators according to the linear map,
\begin{equation} \label{eq:unitary_map}
\hat{U}\hat{a}_i^\dag\hat{U}^\dag \to \sum_j U_{i,j} \hat{a}_j^\dag,
\end{equation}
where $\hat{U}$ is an {$m\times m$} matrix.  For \BS, $\hat{U}_{AA}$ is chosen to be a Haar-random, unitary matrix.

Unlike the Fock state model, for PASSV boson sampling we consider an interferometer consisting of \emph{real} beamsplitters which implements an orthogonal matrix (also chosen to be Haar-random). Thus, for Fock state boson sampling \mbox{$\hat{U}_\mathrm{AA}\in SU(m)$}, whereas for PASSV boson sampling \mbox{$\hat{U}_\mathrm{SV} \in SO(m)$}. Reck \emph{et al.} showed that for both cases, any \mbox{$m\times m$} unitary or orthogonal matrix can be implemented with at most $O(m^2)$ optical elements, and an efficient algorithm for finding the decomposition exists \cite{bib:Reck94}.

It is important to discuss the complexity of choosing an orthogonal matrix instead of a unitary because one should be concerned with the possibility of choosing a subset of matrices from $SU(m)$, whose permanent is efficiently simulable by a classical computer. If this were the case, the result would not be interesting, since the novelty of \BS is that it simulates a system which is classically intractable.  We will later prove (in Sec.~\ref{subsec:passvcomplex}) this is not the case and that, in fact, the associated complexities are equivalent. 

\textbf{Output:}
The output state for the Fock state model after passing through the interferometer is thus,
\begin{eqnarray} \label{eq:aaoutput1}
\ket{\psi}_\mathrm{out}^\mathrm{AA} &=& \hat{U}_\mathrm{AA}\ket{\psi}_\mathrm{in}^\mathrm{AA} \nonumber \\
&=& \hat{U}_\mathrm{AA}\left[\hat{a}_1^\dag\dots\hat{a}_n^\dag\ket{0_1,\dots,0_m}\right] \nonumber \\
&=& \left[\hat{U}_\mathrm{AA}(\hat{a}_1^\dag\dots\hat{a}_n^\dag)\hat{U}_\mathrm{AA}^\dag\right]\hat{U}_\mathrm{AA}\ket{0_1,\dots,0_m} \nonumber \\
&=& \left[\hat{U}_\mathrm{AA}(\hat{a}_1^\dag\dots\hat{a}_n^\dag)\hat{U}_\mathrm{AA}^\dag\right]\ket{0_1,\dots,0_m},
\end{eqnarray}
where the last equality holds because $\hat{U}_{AA}\ket{0}=\ket{0}$, i.e. $\hat{U}_{AA}$ represents passive optics elements and hence cannot generate new photons. Since the unitary transforms the creation operators according to Eq.~(\ref{eq:unitary_map}), the output of the interferometer can also be represented as,
\begin{equation} \label{eq:aaoutput2}
\ket{\psi}_\mathrm{out}^\mathrm{AA}=\sum_{S}\gamma_S \ket{S_1,\dots,S_m},
\end{equation}
where $S$ is an output configuration of the $n$ photons with $S_i$ photons in the $i$th mode, and $\gamma_S$ is the corresponding amplitude. Note that \mbox{$\gamma_S \propto \text{Per}(U_{S})$}, where $U_{S}$ is an \mbox{$n\times n$} sub-matrix of $\hat{U}_{AA}$ given as a function of the configuration $S$. The number of distinct configurations is
\begin{equation} \label{eq:config}
|S| = \binom{n+m-1}{n},
\end{equation}
which can be easily verified to be the number of ways to configure $n$ indistinguishable photons into $m$ distinct modes. This expression grows superexponentially with $n$ from the earlier requirement that $m=\Omega(n^2)$.

\begin{figure}[t]
\centering
\includegraphics[scale=.9]{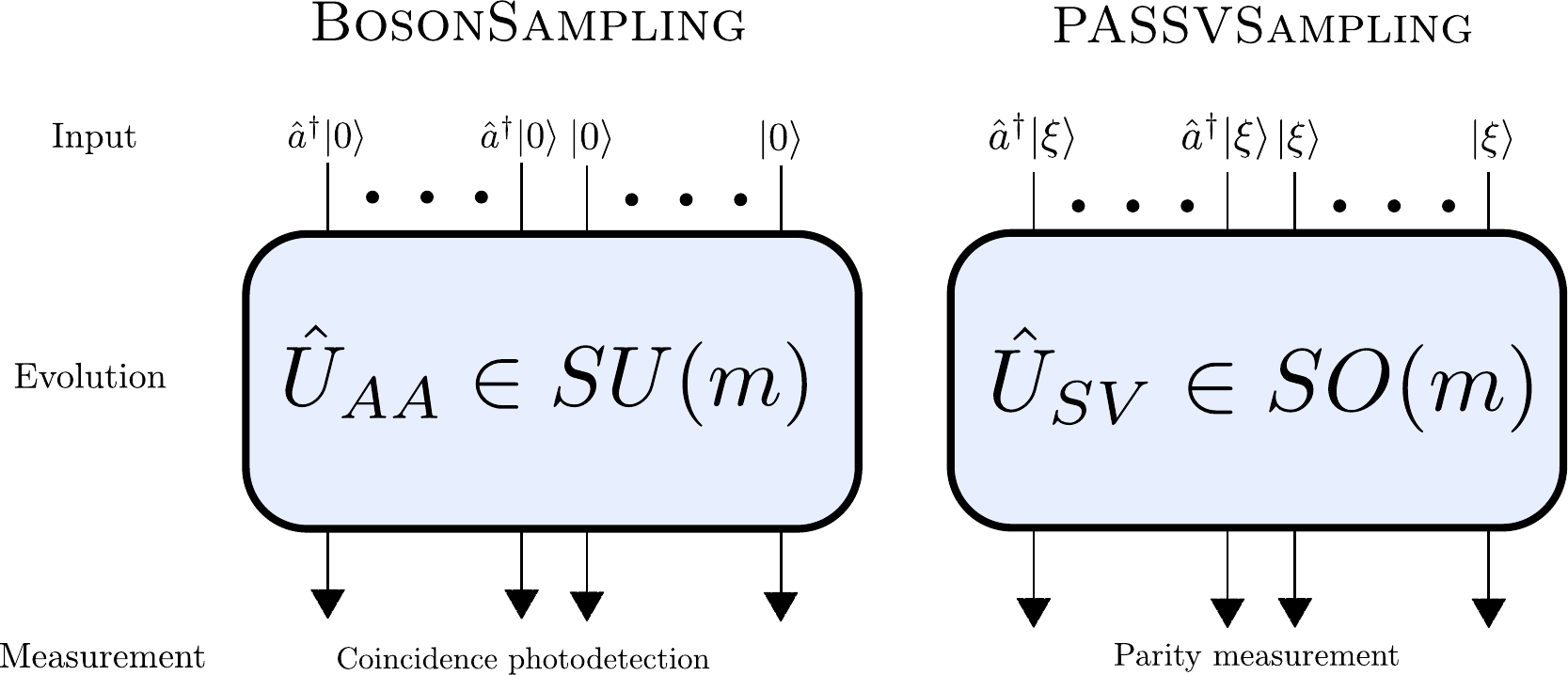}
\caption{(left) The \BS model. We feed an $m$-mode linear optics interferometer with $n$ single photons and \mbox{$m-n$} vacuum states. Following evolution, the state is sampled via coincidence number-resolved photodetection. (right) The \PASSV model. We prepare $n$ PASSV states and \mbox{$m-n$} SV states. Following evolution we perform coincidence parity measurement.}
\end{figure}

For \PASSV, we can use the same technique as in Eq.~(\ref{eq:aaoutput1}), such that the output state is,
\begin{eqnarray} \label{eq:svoutput1}
\ket{\psi}_\mathrm{out}^\mathrm{SV} &=& \hat{U}_\mathrm{SV}^\mathrm{}\ket{\psi}_\mathrm{in}^\mathrm{SV} \nonumber \\
&=& \left[\hat{U}_\mathrm{SV}(\hat{a}^\dag_1\dots\hat{a}^\dag_n)\hat{U}_\mathrm{SV}^\dag\right]\hat{U}_\mathrm{SV}\ket{\xi_1,\dots,\xi_m}. 
\end{eqnarray}
It was shown by Jiang \emph{et al.}~\cite{bib:PhysRevA.88.044301} that for a pure product state input to a linear optical network, the output is entangled unless the input is either a tensor product of coherent states or a tensor product of squeezed states (with the same squeezing), provided that the network does not mix the squeezed and anti-squeezed quadratures. The latter condition is equivalent to the network comprising real beamsplitters. This condition is satisfied since \mbox{$\hat{U}_{SV}\in SO(m)$} and thus,
\begin{equation}
\label{eq:svoutput1b}
\ket{\psi}_\mathrm{out}^\mathrm{SV}= \left[\hat{U}_\mathrm{SV}(\hat{a}^\dag_1\dots\hat{a}^\dag_n)\hat{U}_\mathrm{SV}^\dag\right]\ket{\xi_1',\dots,\xi_m'}.
\end{equation}
The leading operator corresponds to a configuration of $n$ creation operators as in Eq.~(\ref{eq:aaoutput1}). The output for a photon-added SV state input is therefore of the form,
\begin{equation} \label{eq:svoutput2}
\ket{\psi}_\mathrm{out}^\mathrm{SV} =\sum_{S}\gamma_{S}' \left[(\hat{a}_1^\dag)^{S_1} \dots (\hat{a}_m^\dag)^{S_m}\right]\ket{\xi_1',\dots,\xi_m'},
\end{equation}
where,
\begin{equation}
\gamma_S' = \frac{\gamma_S}{\sqrt{S_1!\dots S_m!}}=\frac{\textrm{Per}(U_S)}{\sqrt{S_1!\dots S_m!}},
\end{equation}
but in the binary regime \mbox{$\gamma_S'=\gamma_S$}. Recall from Eq.~(\ref{eq:sv}) that squeezed states represented in the Fock basis have only even photon-number terms. Thus, for a configuration $S$ where mode $i$ does not have a creation/annihilation operator acting on it, mode $i$ is a superposition of only even photon number states, whereas if $S$ applies a creation/annihilation operator to mode $i$ it contains only odd photon-number terms.

For photon-subtracted SV states the output is of the same form, replacing $\hat{a}^\dag_i$ with $\hat{a}_i$, but $\gamma_S$ will now relate to $\hat{U}^\dag_{SV}$ instead of $\hat{U}_{SV}$, which is also Haar-random, and thus has the same sampling complexity.  We exclude the case of the photon-subtracted states when $\xi=0$ since $\hat{a}\ket{0}=0$.

\textbf{Measurement:}
The last step is to measure the output distribution. For \BS, this may be implemented via number-resolved photodetection. However, since \mbox{$m=\Omega(n^2)$}, \mbox{$S_i=\{0,1\}\,\,\forall \,\,i$} in Eq.~(\ref{eq:aaoutput2}), on/off (or `bucket') detectors are sufficient to recover the configuration $S$. Repeating the sampling procedure multiple times yields partial information of the joint photon-number distribution \mbox{$P_S=|\gamma_S|^2$}, which was shown by AA to be a computationally difficult sampling problem.

For \PASSV, we perform a parity measurement capable of distinguishing only between odd and even photon-number. Such measurements are characterised by the measurement operators,
\begin{eqnarray}
\hat{\Pi}_+ &=& \ket{0}\bra{0} + \ket{2}\bra{2} + \ket{4}\bra{4} + \dots \\ \nonumber
\hat{\Pi}_- &=& \ket{1}\bra{1} + \ket{3}\bra{3} + \ket{5}\bra{5} + \dots
\end{eqnarray}
Most simply, one could implement this measurement using photon-number-resolving detectors. Measuring an even photon-number at output mode $i$ then implies that there was no creation/annihilation operator associated with that mode, whereas measuring an odd photon-number implies that there was.  This measurement thus perfectly recovers the configuration $S$, and hence continued sampling yields the desired distribution.  Since the squeezing parameter $\xi$ has no effect on the parity of the state, the sampling amplitudes are completely independent of the squeezing.  

More formally, in \BS we are sampling from a set of strings, 
\begin{equation} 
s_i=\{s_i^{(1)},\dots,s_i^{(m)}\}
\end{equation}
where $s_i^{(j)}$ is the sampled photon-number in the $j$th mode associated with string $i$, of which there are an exponential number. In the limit of large $m$, $s_i^{(j)}\in\{0,1\}$. On the other hand, with \PASSV we are sampling from the same set of strings, with the same probability distribution, where now $s_i^{(j)}\in\{-1,1\}$.  This proves that \PASSV implements the same logical sampling problem as \BS, independent of the squeezing parameter.

\subsection{Complexity Concerns and Discussion} \label{subsec:passvcomplex}
We previously mentioned, while discussing the evolution of the input state, whether choosing an orthogonal matrix has any implications for the complexity of \PASSV.  Since we have now shown that the \PASSV model samples permanents of submatrices in the same way as \BS, this is the only barrier to completing our proof that the two models are in the same complexity class.

The first consideration is whether or not a Haar-random matrix in $SO(m)$ might have an efficiently computable exact or approximate permanent. The exact permanent case is known to be \#\textbf{P}-complete even for binary entries, \mbox{$U_{i,j}\in\{0,1\}$} \cite{bib:Valiant79}. There is also a known algorithm for efficiently approximating a permanent if the matrix has entries consisting of only non-negative real numbers. In the same work, it is shown that for a matrix with even a single negative entry, an efficient approximation algorithm would allow one to compute an \textit{exact} \mbox{$\{0,1\}$}-permanent efficiently \cite{bib:Jerrum04}. Although having to compute a difficult permanent is a necessary but not sufficient condition for computational hardness, since $SO(m)$ is considered to be universal for linear optics \cite{bib:Bouland}, there is no such complexity gap between unitary and orthogonal matrices.

More concretely, it has been shown that $SU(m)\subset SO(2m)$ \cite{bib:Georgi99}, i.e. for a $2m$-mode interferometer, the set of all orthogonal transformations includes all unitary $m$-mode transformations as a subgroup.  Thus, the complexity of sampling the output from a \BS device implementing an arbitrary matrix from $SO(2m)$ is at least as hard as sampling matrices from $SU(m)$, and for only a linear cost in the number of modes. Since trivially $SO(2m)\subset SO(2m+1)$, the same complexity extends to an odd number of modes as well.  Note that this also carries the implication that \BS itself remains hard under orthogonal transformations.

We can now conclude that \PASSV is in the same complexity class as \BS.  Suppose that \textbf{A} is some complexity class containing \BS (that is closed under polynomial reductions).  Since the output of \PASSV is completely independent of the squeezing parameter $\xi$, we may assume without loss of generality that $\xi=0$.  In this limit, however, $\ket{\xi_i}=\ket{0_i}$ and thus, by construction, any instance of \PASSV reduces to an instance of \BS since $SO(m)\subset SU(m)$.  Thus, the class \textbf{A} also contains \PASSV.  Conversely, suppose \textbf{B} is some complexity class containing \PASSV.  Again choosing $\xi=0$, the inclusion  $SU(m)\subset SO(2m)$ similarly implies \textbf{B} also contains \BS.

Our result can be distilled to a relatively simple idea which is most evident in light of Eq.~(\ref{eq:aaoutput1}), where the ket acts as a `background' signal whose form is invariant under the evolution of $\hat{U}_{SV}$. Since the leading operator in Eq.~(\ref{eq:svoutput1b}) takes exactly the same form as Eq.~(\ref{eq:aaoutput1}), we would like the ket to also be independent of the choice of $\hat{U}_{SV}$ under \emph{some} measurement, while still being distinguishable from a state which has an added or subtracted photon. It may be possible to use the same technique to characterize other states which implement a logically equivalent classically intractable sampling problem. A desirable goal would be to prove an even more experimentally friendly set of states and measurements that implements the same problem.

One criticism of \PASSV is that the use of photon-number resolving detectors to implement the parity measurement is experimentally harder than on/off detection. Whilst this is true, one does not need to distinguish between \emph{arbitrarily} large even and odd photon-number Fock states. For any given $\xi$ and error rate, one can truncate the maximum number of necessarily distinguishable Fock states. Indeed, \PASSV can be regarded as a generalization of \BS, since in the limit of small squeezing (\mbox{$\xi\rightarrow 0$}), the SV reduces to a vacuum state and an on/off detector suffices. For large squeezing, additional experimental hurdles may arise in reducing squeezing parameter error and in the increased sensitivity of squeezed states to noise.  We do not address these issues here.  Rather, despite PASSV states being more difficult to experimentally prepare, our goal is to theoretically demonstrate the non-uniqueness of Fock states for computationally hard sampling problems.

After having spent some effort showing that orthogonal matrices are sufficiently complex for \PASSV, a natural question is whether or not choosing a unitary matrix could change the complexity of the sampling problem.  Because Eq.~(\ref{eq:svoutput1b}) no longer holds, we cannot establish a straightforward relationship between the output probabilities and submatrix permanents.  Conventional wisdom seems to suggest that the problem would not become easier.  In the limit of zero squeezing, we know there is no complexity divide because \PASSV reduces to \BS.  Thus, if a complexity divide did exist, then we would expect a complexity phase transition at $\xi=0$.  It may be possible to construct a more complicated measurement scheme which produces the same sampling probabilities.

We have shown a direct mapping between \BS and \PASSV. An open question in the field is `what characterizes quantum states of light that yield hard sampling problems with linear optics?' This result, in conjunction with previous results on photon-added coherent states and generalized cat states, demonstrates that there exists a large class of non-Fock states, which yield sampling problems of equal computational complexity.

Importantly, unlike \cc{PACSampling}, \PASSV operates in \emph{all} parameter regimes. Thus there are no bounds on the amount of squeezing and no approximations are made.

Whilst \PASSV may be experimentally more challenging than \BS, this result certainly confirms that there is nothing unique about the computational complexity of Fock states. In fact, there is a plethora of other quantum states exhibiting similar sampling complexity, and computational complexity appears to be a ubiquitous property of sampling quantum states of light.

We hope that future research will enable us to fully characterize what it is that makes a quantum optical system computationally hard, and what classes of states are required for computational complexity.

\pagebreak

\singlespacing
\chapter{Super-Sensitive Metrology}
\doublespacing
So far, we have talked a great deal about complexity theory results relating to \BS.  We would now like to turn our attention to the physical intuition that we can gain from these results.  It is clear that Fock states evolved by passive, multimode interferometers have surprisingly powerful (or at least non-classical) properties.  While the development of quantum computing is certainly a major investment by the quantum information community, there may be other benefits to applying the lessons we have learned up to this point.  Our goal in this chapter is to develop a quantum metrology protocol that is based on the same architecture as \BS.  To help the reader understand the important features of this protocol and some of the more subtle points, we first give a brief background on quantum metrology.  We then introduce a promising protocol that shows how single photons with only passive unitary evolution can very nearly approach the best sensitivity possible allowable with quantum mechanics.  A generalization of this protocol follows, showing what single photon metrology can hope to achieve in the future.

\section{Introduction to Quantum Metrology} \label{sec:metro}

Metrology is by definition the science of measurement.  In physics, making measurements of various quantities plays an integral role in discovering the properties of phenomena, and is necessary to confirm the physical laws that theorists may postulate.  Moreover, precision measurement is now hugely important in a wide variety of industrial applications and military technologies.  It is no surprise that the development of quantum mechanics would have serious implications for measurement theory.  Although uncertainty principles give us certain limitations on what can be known about physical systems, the field of quantum metrology has shown us that quantum mechanics also allows enhancements in measurement that classical measurement theory could never produce.  In this section, we will discuss how quantum optics can enhance the precision of interferometric measurements, and the ultimate limitations thereof.

The discovery that light has an intrinsic phase has enabled interferometric methods for making the measurement of many systems practical.  Moreover, a number of famous discoveries---from the Michelson-Morley experiment to the recent discovery of gravity waves by the LIGO collaboration---would not have taken place without the use of interferometry.  While there are other ways to make measurements based on the interference of light, phase estimation has been enormously successful at discovering the properties of materials which interact closely with light (e.g. the index of refraction of glass).

The underlying concept of phase estimation has a simple and elegant description.  We know that (in the classical electromagnetic picture) the intrinsic phase of light propagates proportional to the frequency of the light.  When two electromagnetic waves of differing phase interfere, the difference between the phases can be inferred from the frequency of the resultant wave.  This fact is exploited to make inferences of the properties of e.g. a material by preparing a system with a known phase difference, and then perturbing the system by inserting the material.  If one knows how the material interacts with light, then the phase that is accumulated by the interaction is a witness for the properties of that material.  We can take the same approach with quantum states of light, provided that we understand how the phase in different modes affects the evolution of the state.

\begin{figure}[!htb]
\centering
\includegraphics[scale=0.8]{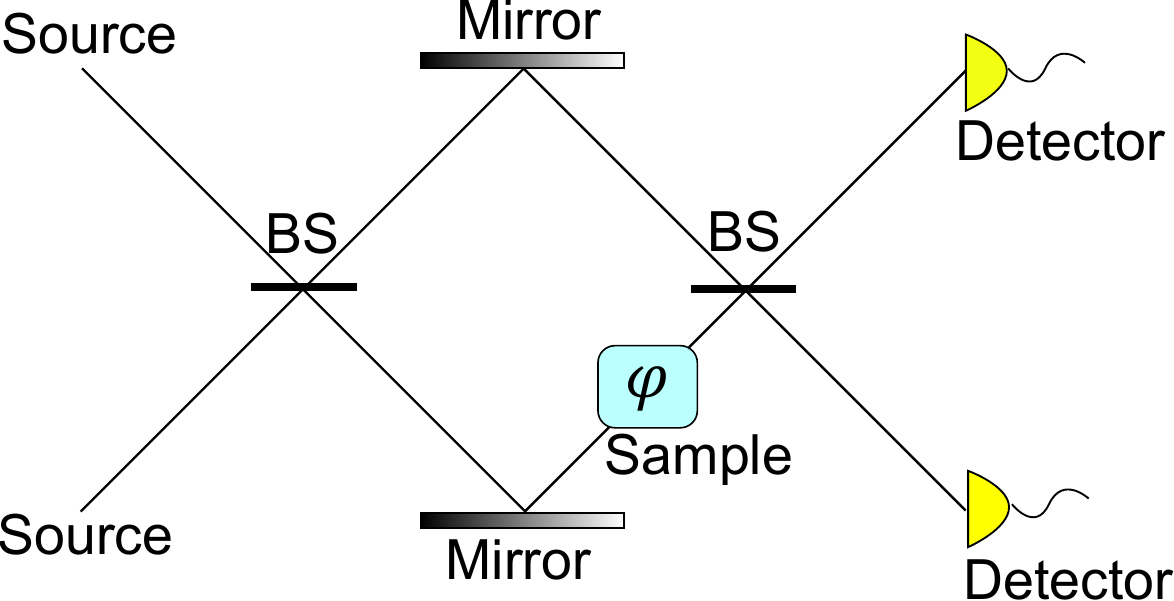}
\caption{Architecture of the Mach-Zehnder Interferometer.  Two input modes are first interfered on a beamsplitter.  One mode experiences an additional unknown phase shift $\varphi$ before the two modes are interfered on a second beamsplitter.  The output state is then measured on detectors in both modes.} \label{fig:mzi}
\end{figure}

To explain this in terms of the optical networks we have discussed so far in this thesis, we begin by giving an example of perhaps the simplest such device, the \textit{Mach-Zehnder interferometer} (MZI), shown in Figure \ref{fig:mzi}.  If both the beamsplitters $\hat{B}$ are 50:50 (see Sec.~\ref{subsec:optics}), the action of the entire network can be described by the matrix, 
\begin{equation}
\hat{U}=\hat{B}\hat{\Phi}\hat{B}=\frac{1}{2}
\left[
\begin{array}{ll}
      1 & i \\
      i & 1 
\end{array} 
\right]
\left[
\begin{array}{ll}
      1 & 0 \\
      0 & e^{i\varphi} 
\end{array} 
\right]
\left[
\begin{array}{ll}
      1 & i \\
      i & 1
\end{array} 
\right]=\frac{1}{2}
\left[
\begin{array}{ll}
      1-e^{i\varphi} & i(1+e^{i\varphi}) \\
      i(1+e^{i\varphi}) &-(1-e^{i\varphi})
\end{array} 
\right], \label{fig:mzimatrix}
\end{equation}
where $\hat{\Phi}$ is understood to be the action of the unknown phase shift $\varphi$ on the second mode.

Consider the case where only a single photon is input into the first mode, $\ket{\psi_{in}}=\ket{1,0}$.  If we measure in the Fock basis, the output statistics can be easily computed by taking the squared norm of the $\{1,1\}$ and $\{1,2\}$ entries (i.e. the squared norm of a trivial $1\times 1$ matrix permanent) so that we find,
\begin{equation}
P[\ket{1,0}]=\frac{1}{2}(1-\cos(\varphi)) \qquad P[\ket{0,1}]]=\frac{1}{2}(1+\cos(\varphi)).
\end{equation}
If we made infinitely many measurements on the output (assuming the system had no noise due to error), we would be able to ascertain the value of $\varphi$ by this relationship, and thereby make some prediction about the properties of the material that generated the phase $\varphi$.  Realistically, of course, we cannot make infinitely many measurements to establish the output statistics with perfect precision.  The key question then becomes: given $m$ measurements of the output, what is our best guess for $\varphi$, and how accurate is this guess?  

It is natural to assume that the sample distribution is the best estimate of the true distribution, and therefore the underlying estimate of the variable is simply a known function of the sample statistics.  The more difficult question is how we determine the precision, or uncertainty, of our estimate.  Our uncertainty about the probability distribution must propagate somehow into uncertainty about the unknown variable.  To do this, we employ the \textit{standard error propagation formula} defined by,
\begin{equation}
\Delta\varphi=\frac{\sqrt{\ip{\hat{O}^2}-\ip{\hat{O}}^2}}{\sqrt{n}\cdot|\frac{d\ip{\hat{O}}}{d\varphi}|}, \label{eq:errorprop}
\end{equation}
where $\hat{O}$ is some observable that allows us to estimate $\varphi$, and $n$ is the number of repeated independent measurements.  If we choose $\hat{O}=\ket{10}\bra{10}$, which is the projection operator corresponding to observing the output $\ket{10}$, then since $\ip{\hat{O}}=\frac{1}{2}(1-\cos(\varphi))$ and $\hat{O^2}=\hat{O}$, we can substitute this into Eq.~(\ref{eq:errorprop}) to give,
\begin{eqnarray}
\Delta\varphi&=&\frac{\sqrt{\ip{\hat{O}}-\ip{\hat{O}}^2}}{\sqrt{n}\cdot|\frac{d\ip{\hat{O}}}{d\varphi}|} \\
&=&\frac{\sqrt{\frac{1}{2}(1-\cos(\varphi))-\frac{1}{4}(1-\cos(\varphi))^2}}{\sqrt{n}\cdot\frac{1}{2}\sin(\varphi)} \\
&=&\frac{\sqrt{1-\cos^2(\varphi)}}{\sqrt{n}\cdot\sin(\varphi)} \\
&=&\frac{1}{\sqrt{n}}\;.
\label{eq:errorprop2}
\end{eqnarray}
It is shown in Ref.~\cite{bib:GerryKnight05} that, for a coherent state input into one mode $\ket{\psi_{in}}=\ket{\alpha,0}$, $\ip{\hat{O}}=\hat{a}_1^\dagger\hat{a}_1-\hat{a}_2^\dagger\hat{a}_2$, and 50:50 beamsplitters, the uncertainty $\Delta\varphi^\alpha$ is,
\begin{equation}
\Delta\varphi^{\alpha}=\frac{1}{\sqrt{\bar{n}}}\;. \label{eq:errorprop3}
\end{equation}

We do not prove, but it can be shown that in both cases, this is the lowest achievable uncertainty.  The theory describing these lower bounds can be described with the \textit{quantum Fisher information} and is given by the \textit{quantum Cram\'{e}r-Rao bound}.  For more on this topic, a reader should consult Ref.~\cite{bib:Caves1994}.  It is with some regret that I cannot include this, but I do not believe I could give the subject sufficient justice in a short review, and a longer analysis would be ill-suited for a thesis whose emphasis is primarily complexity theory.

The coincidental form of Eq.~(\ref{eq:errorprop2}) and Eq.~(\ref{eq:errorprop3}) is perhaps suggestive that $1/\sqrt{n}$ may be the lower limit in uncertainty for any state with an average of $n$ photons.  In fact, this is not the case, as we will shortly prove.  Instead, the property that these two states share is that they are rather classical systems.  In the case of a single photon, the statistics could have simply been described by repeated measurements of a probabilistic classical particle.  Meanwhile, coherent states are arguably engineered to give a classical description.  This bound is often referred to as the \textit{shotnoise limit}, as it represents the noise due to uncertainty when only single photon ``shots" are employed.  The term is often used to refer to the sensitivity of ``the best classical scheme" for measuring a system, though what is exactly meant by this is often debatable depending on the system in question; we urge the reader to excercise caution when it is used in the literature.

Consider the case where $\ket{\psi_{in}}=\ket{1,1}$ and $\hat{O}=\ket{1,1}\bra{1,1}$.  We can easily find the probability of the outcome by applying the permanent method (described in Sec.~\ref{sec:permanent}) to Eq.~(\ref{fig:mzimatrix}).  Namely, the probability $P[\ket{1,1}]=|\gamma|^2$ where,
\begin{eqnarray}
\gamma&=&\perm(\hat{U})=\;\perm\Bigg[\frac{1}{2}\left[
\begin{array}{ll}
      1-e^{i\varphi} & i(1+e^{i\varphi}) \\
      i(1+e^{i\varphi}) & -(1-e^{i\varphi})
\end{array} 
\right]\Bigg] \\
&=&-\frac{1}{2}(1+e^{2i\varphi}).
\end{eqnarray}
One can use permanents to obtain the probability of the outputs $\ket{2,0}$ and $\ket{0,2}$ as well, though a symmetry argument can be employed instead. Since the probability of all outcomes should sum to 1, and since the probability of outcomes $\ket{2,0}$ and $\ket{0,2}$ should be the same (by inspection of the matrix), we see that,
\begin{equation}
P[\ket{1,1}]=\frac{1}{2}(1+\cos(2\varphi)) \quad P[\ket{2,0}]=\frac{1}{4}(1-\cos(2\varphi)) \quad P[\ket{0,2}]=\frac{1}{4}(1-\cos(2\varphi)).
\end{equation}

We can now begin to see non-classical effects in the output of the interferometer.  Note that for $\varphi=\frac{\pi}{4}$, only the $\ket{2,0}$ and $\ket{0,2}$ outputs can be observed, since $P[\ket{1,1}]=0$.  If photons were classical particles, this outcome could never be observered since the paths of the two particles would have to be independent of each other, since they are non-interacting.  Similarly, this could not be simulated by two runs of an experiment where only a single photon entered the interferometer at a particular time.  This ``bunching" of photons at the output of the interferometer was first observed by three physicists in 1987, and is now referred to as the \textit{Hong-Ou-Mandel effect} \cite{bib:HOM87}.   

If we compute the uncertainty of $\Delta\varphi$ for $\hat{O}=\ket{1,1}\bra{1,1}$ using Eq.~(\ref{eq:errorprop}), we arrive at, $\Delta\varphi=\frac{1}{2\sqrt{n}}$.  We must be careful to interpret this result, however, if we compare it to the result in Eq.~(\ref{eq:errorprop2}).  In order to achieve this uncertainty, we have used two photons instead of one.  To be fair, we should instead consider the case where we are restricted to using at most $n$ photons, rather than considering $n$ independent runs of the experiment.  If we make this correction, we can make only $n/2$ runs of the experiment, so that the uncertainty becomes,
\begin{equation}
\Delta\varphi=\frac{1}{2\sqrt{n/2}}=\frac{1}{\sqrt{2n}}.
\end{equation}
This is still, however, an improvement in the sensitivity by a factor of $\sqrt{2}$.  It can be shown that for 2 photon experiments, this is optimal.  In fact, for an MZI, it can be shown that \cite{bib:dowling2008quantum},
\begin{equation}
\Delta\varphi\leq\frac{1}{N\sqrt{\mu}}
\end{equation}
for an $N$ photon input and $\mu$ independent runs of the experiment.  This ultimate quantum limit, called the \textit{Heisenberg limit}, has been computed for a number of different systems to show the lowest uncertainty one can achieve with quantum mechanics.  The lowerbound for the MZI is achieved by using the input state $\ket{\psi}=\frac{1}{\sqrt{2}}(\ket{N,0}+\ket{0,N}$), referred to as the $N$-photon \textit{NOON state}.  This state, however, cannot be prepared efficiently using only passive linear optics.  In fact, the only known ways to construct this state require technology that is similar to the requirements for building universal quantum computers.  We believe a very relevant question, then, is whether sensitivity beating the shotnoise limit can be achieved with only passive linear optics.  This will be the topic of the remaining sections of this chapter. 

\section{MORDOR Interferometer} \label{sec:mordor}
\footnote{This section previously appeared as: K. R. Motes, J. P. Olson, E. Rabeaux, J. P. Dowling, S. J. Olson, and P. P. Rohde. Linear optical quantum metrology with single photons -- Exploiting spontaneously generated entanglement to beat the shot-noise limit.  \textit{Phys. Rev. Lett.}, 114:170802, 2015. It is reprinted by permission of APS.}In this section, we discuss a \BS-like (similar in terms of architecture, not computational complexity) optical network that can be used for metrology.  This scheme was originally presented in Ref.~\cite{bib:Mordor2015}, and is repeated here with additional discussion and clarifications.  We have seen earlier in this thesis that such passive linear optical devices can generate a superexponentially large amount of number-path entanglement.  We show that a simple, passive, linear-optical interferometer---fed with only uncorrelated, single-photon inputs, coupled with simple, single-mode, disjoint photodetection---is capable of significantly beating the shotnoise limit. This result implies a potential pathway forward to practical quantum metrology with readily available technology.

Ever since the early work of Yurke \& Yuen it has been understood that quantum number-path entanglement is a resource for super-sensitive quantum metrology, allowing for sensors that beat the shotnoise limit \cite{bib:yurke1986input, bib:yuen1986generation}. Such devices would then have applications to super-sensitive gyroscopy \cite{bib:dowling1998correlated}, gravimetry \cite{bib:yurtsever2003interferometry}, optical coherence tomography \cite{bib:nasr2003demonstration}, ellipsometry \cite{bib:toussaint2004}, magnetometry \cite{bib:jones2009magnetic}, protein concentration measurements \cite{bib:crespi2012measuring}, and microscopy \cite{bib:rozema2014scalable, bib:israel2014supersensitive}. This line of work culminated in the analysis of the bosonic NOON state (\mbox{$(\ket{N,0}+\ket{0,N})/\sqrt{2}$}, where $N$ is the total number of photons, which was shown to be optimal for local phase estimation with a fixed, finite number of photons, and in fact allows one to hit the Heisenberg limit and the Quantum Cram{\'e}r-Rao Bound \cite{bib:holland1993interferometric, bib:lee2002quantum, bib:durkin2007local, bib:dowling2008quantum}.

Let us consider the NOON state, where for this state in a two-mode interferometer we have the condition of all $N$ particles in the first mode (and none in the second mode) superimposed with all $N$ particles in the second mode (and none in the first mode). While such a state is known to be optimal for sensing, its generation is also known to be highly problematic and resource intensive. There are two routes to preparing high-NOON states: the first is to deploy very strong optical nonlinearities \cite{bib:gerry2001generation, bib:kapale2007bootstrapping}, and the second is to prepare them using measurement and feed-forward \cite{bib:lee2002linear, bib:vanmeter2007general, bib:cable2007efficient}. In many ways then NOON-state generators have had much in common with all-optical quantum computers and therefore are just as difficult to build \cite{bib:kok2007linear}. In addition to the complicated state preparation, typically a complicated measurement scheme, such as parity measurement at each output port, also had to be deployed \cite{bib:seshadreesan2013phase}. 

\begin{figure}[!htb]
\centering
\includegraphics[scale=1.8]{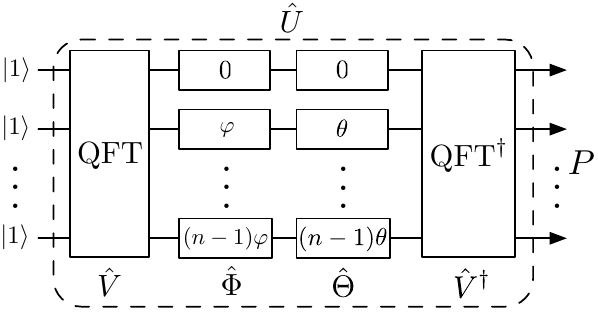}
\caption{Architecture of the MORDOR interferometer for metrology using single-photon states. The input state comprises $n$ single photons, \mbox{$\ket{1}^{\otimes n}$}. The state evolves via the passive linear optics unitary \mbox{$\hat{U} = \hat{V} \cdot \hat{\Phi} \cdot \hat{\Theta} \cdot \hat{V}^\dag$}, where $\hat{V}$ is the quantum Fourier transform, $\hat\Phi$ is an unknown, linear phase gradient, and $\hat\Theta$ is a reference phase gradient used for calibration. At the output we perform a coincidence photodetection projecting on exactly one photon per output mode, measuring the observable $\hat{O}=(\ket{1}\bra{1})^{\otimes n}$, which, over many measurements, yields the probability distribution $P(\varphi)$ that acts as a witness for the unknown phase $\varphi$.} \label{fig:arch}
\end{figure}

Recently two independent lines of research, the study of quantum random walks with multi-photon walkers in passive linear-optical interferometers \cite{bib:mayer2011counting, bib:gard2013quantum, bib:gard2014inefficiency}, as well as the quantum complexity analysis of \BS devices \cite{bib:AA10, bib:Chapter}, has led to a somewhat startling yet inescapable conclusion---passive, multi-mode, linear-optical interferometers, fed with only uncorrelated single photon inputs in each mode (Figure \ref{fig:arch}), produce quantum mechanical states of the photon field with path-number entanglement that grows exponentially fast in the two resources of mode and photon-number. What is remarkable is that this large degree of number-path entanglement is not generated by strong optical nonlinearities, nor by complicated measurement and feed-forward schemes, but by the natural evolution of the single photons in the passive linear optical device. Whilst such devices are often described to have `non-interacting' photons in them, there is a type of photon-photon interaction generated by the demand of bosonic state symmetrization, which gives rise to the superexponentially large number-path entanglement via multiple applications of the Hong-Ou-Mandel effect \cite{bib:gard2014inefficiency}. It is known that linear optical evolution of single photons, followed by projective measurements, can give rise to `effective' strong optical nonlinearities, and we conjecture that there is indeed a hidden Kerr-like nonlinearity at work also in these interferometers \cite{bib:lapaire2003conditional}. Like \BS \cite{bib:AA10}, and unlike universal quantum computing schemes such as that by Knill, Laflamme, and Milburn \cite{bib:KLM01}, this protocol is deterministic and does not require any ancillary photons.

The advantage of such a setup for quantum metrology is that resources for generating and detecting single photons have become quite standardized and relatively straightforward to implement in the lab \cite{bib:matthews2011heralding, bib:spring2013boson, bib:Broome2012, bib:crespi2013integrated, bib:ralph2013quantum, bib:motes2013spontaneous, bib:spagnolo2014experimental}. The community then is moving towards single photons, linear interferometers, and single-photon detectors all on a single, integrated, photonic chip, which then facilitates a roadmap for scaling up devices to large numbers of modes and photons. If all of this work could be put to use for quantum metrology, then a road to scalable metrology with number states would be at hand. 

It then becomes a natural question to ask---since number-path entanglement is known to be a resource for quantum metrology---can a passive, multi-mode interferometer, fed only with easy-to-generate uncorrelated single photons in each mode, followed by uncorrelated single-photon measurements at each output, be constructed to exploit this number-path entanglement for super-sensitive (sub-shotnoise) operation? The answer is indeed yes, as we shall now show.

Recall from the previous section that the phase-sensitivity, $\Delta\varphi$, of a metrology device can be defined in terms of the standard error propagation formula as, 
\begin{eqnarray} \label{eq:phaseSensitivity}
\Delta\varphi = \frac{\sqrt{\ip{\hat{O}^2}-\ip{\hat{O}}^2}}{\left|\frac{\partial\ip{\hat{O}}}{\partial\varphi}\right|}, 
\end{eqnarray}
where $\ip{\hat{O}}$ is the expectation of the observable being measured and $\varphi$ is the unknown phase we seek to estimate.  We have dropped the dependence on the number of identical measurements found in Sec.~\ref{eq:errorprop} for simplicity, and will do so for the remainder of this section.

The photons evolve through a unitary network according to $U a_i^{\dag} U^{\dag} = \sum_j U_{ij} a_j^{\dag}$. In our protocol, we construct the $n$-mode interferometer $\hat{U}$ to be,
\begin{equation} \label{eq:U}
\hat{U} = \hat{V} \cdot \hat{\Phi} \cdot \hat{\Theta} \cdot \hat{V}^{\dag},
\end{equation}
which we will call the ``MORDOR" architecture in reference to the authors of Ref.~\cite{bib:Mordor2015}.  $\hat{V}$ is the $n$-mode quantum Fourier transform matrix, with matrix elements given by,
\begin{equation}
\mathrm{V}_{j,k}^{(n)} = \frac{1}{\sqrt{n}}\mathrm{exp}\left[\frac{- 2 i j k \pi}{n}\right].
\end{equation} 
$\hat\Phi$ and $\hat\Theta$ are both diagonal matrices with linearly increasing phases along the diagonal represented by,
\begin{eqnarray} \label{eq:PhiTheta}
\Phi_{j,k} = \delta_{j,k} \exp{\Big[i(j-1)\varphi\Big]} \nonumber \\
\Theta_{j,k} = \delta_{j,k} \exp{\Big[i(j-1)\theta\Big]},
\end{eqnarray}
where $\varphi$ is the unknown phase one would like to measure and $\theta$ is the control phase. $\hat\Theta$ is introduced as a reference, which can calibrate the device by tuning $\theta$ appropriately. To see this tuning we combine $\hat\Phi$ and $\hat\Theta$ into a single diagonal matrix with a gradient given by,
\begin{equation}
\Phi_{j,k}\cdot\Theta_{j,k} = \delta_{j,k} \exp{\bigg[i(j-1)(\varphi+\theta)\bigg]}.
\end{equation}
The control phase $\theta$ can shift this gradient to the optimal measurement regime, which can be found by minimizing $\Delta\varphi$ with respect to $n$ and $\varphi$. Since this is a shift according to a known phase, we can for simplicity assume (and without loss of generality) that $\varphi$ is in the optimal regime for measurements and $\theta=0$. Thus, $\hat\Theta=\hat{I}$ and is left out of our analysis for simplicity.

In order to understand how such a linearly increasing array of unknown phase shifts may be arranged in a practical device, it is useful to consider a specific example. Let us suppose that we are to use MORDOR as an optical magnetometer. We consider an interferometric magnetometer of the type discussed in Ref.~\cite{bib:scully1992high} where each of the sensing modes of MORDOR contains a gas cell of Rubidium prepared in a state of electromagnetically induced transparency whereby a photon passing through the cell at the point of zero absorption in the electromagnetically induced transparency spectrum acquires a phase shift   that is proportional to the product of an applied uniform (but unknown) magnetic field and the length of the cell. We assume that the field is uniform across MORDOR, as would be the case if the entire interferometer was constructed on an all optical chip and the field gradient across the chip were negligible. Since we are carrying out local phase measurements (not global) we are not interested in the magnitude of the magnetic field but wish to know if the field changes and if so by how much. (Often we are interested in if the field is oscillating and with what frequency.) Neglecting other sources of noise then in an ordinary Mach-Zehnder interferometer this limit would be set by the photon shotnoise limit. To construct MORDOR with the linear cascade of phase shifters, as shown in Figure \ref{fig:arch}, we simply increase the length of the cell by integer amounts in each mode. The first cell has length $L$, the second length $2L$, and so forth. This will then give us the linearly increasing configuration of unknown phase shifts required for MORDOR to beat the shotnoise limit. 

One might question why one would employ a phase gradient rather than just a single phase.  In fact, the case of a single phase is treated in the next section.  The original motivation to use a linear phase shift was to maximize the exploitation of multi-mode entanglement across the entire network.  We will see that, when resources are reasonably limited, the case of a single phase shift is actually more powerful.  We conjecture that this is because, in the limit of small $\varphi$, splitting the signal among many modes weakens the maximum relative phase between two modes.  For more discussion on this topic, see Sec.~\ref{sec:qufti}.

The interferometer may always be constructed efficiently following the protocol of Reck \emph{et al.} \cite{bib:Reck94}, who showed that an \mbox{$n\times n$} linear optics interferometer may be constructed from $O(n^2)$ linear optical elements (beamsplitters and phase-shifters), and the algorithm for determining the circuit has runtime polynomial in $n$. Thus, an experimental implementation of our protocol may always be efficiently realized.

The input state to the device is \mbox{$\ket{1}^{\otimes n}$}, i.e. single photons inputed in each mode. If \mbox{$\varphi=0$} then $\hat\Phi=\hat{I}$ and thus $\hat{U}=\hat{V}\cdot\hat{I}\cdot\hat{V}^{\dag}=\hat{I}$. In this instance, the output state is exactly equal to the input state, \mbox{$\ket{1}^{\otimes n}$}. Thus, if we define $P$ as the coincidence probability of measuring one photon in each mode at the output, then \mbox{$P=1$} when \mbox{$\varphi=0$}. When \mbox{$\varphi\neq 0$}, in general \mbox{$P<1$}. Thus, intuitively, we anticipate that \mbox{$P(\varphi)$} will act as a witness for $\varphi$. 

In the protocol, assuming a lossless device, no measurement events are discarded. Upon repeating the protocol many times, let $x$ be the number of measurement outcomes with exactly one photon per mode, and $y$ be the number of measurement outcomes without exactly one photon per mode. Then $P$ is calculated as $P=x/(x+y)$. Thus, all measurement outcomes contribute to the signal and none are discarded. Note that, due to preservation of photon-number and the fact that we are considering the anti-bunched outcome, $P(\varphi)$ may be experimentally determined using non-number-resolving detectors if the device is lossless.  If the device is assumed to be lossy, then number-resolving detectors would be necessary to distinguish between an error outcome and one in which more than one photon exits the same mode.  The circuit for the architecture is shown in Figure \ref{fig:arch}.

The state at the output of the device is a highly path-entangled superposition of $\binom{2n-1}{n}$ terms, which grows exponentially with $n$.  This corresponds to the number of ways to add $n$ non-negative integers whose sum is $n$, or equivalently, the number of ways to put $n$ indistinguishable balls into $n$ distinguishable boxes. We conjecture that this exponential path-entanglement yields improved phase-sensitivity as the paths query the phases a exponential number of times.

The observable being measured is the projection onto the state with exactly one photon per output mode, \mbox{$\hat{O} = (\ket{1}\bra{1})^{\otimes n}$}. Thus, \mbox{$\langle\hat{O}\rangle = \langle\hat{O}^2\rangle = P$}. And, the phase-sensitivity estimator reduces to,
\begin{equation} \label{eq:phaseSenP}
\Delta\varphi = \frac{\sqrt{P - P^2}}{\left|\frac{\partial P}{\partial \varphi}\right|}.
\end{equation}

Following the result of Ref.~\cite{bib:Scheel04perm} (see Sec.~\ref{sec:permanent}), $P$ is related to the permanent of $\hat U$ as,
\begin{equation}
P = \big|\perm(U)\big|^2.
\end{equation}
Here the permanent of the full \mbox{$n\times n$} matrix is computed, since exactly one photon is going into and out of every mode.

We will now  examine the structure of this permanent. The matrix form for the $n$-mode unitary $\hat U^{(n)}$ is given by,
\begin{equation} \label{eq:Ujk}
U_{j,k}^{(n)} =\frac{1-e^{i n\varphi}}{n\left(e^{\frac{2 i \pi(j-k)}{n}}-e^{i \varphi}\right)},
\end{equation}
as derived in Appendix \ref{app:Ujk}. Taking the permanent of this matrix is challenging as calculating permanents are in general \mbox{\textbf{\#P}-hard}. However, based on calculating $\perm(\hat U^{(n)})$ for small $n$, we observe the empirical pattern,
\begin{equation} \label{eq:permU}
\perm(\hat{U}^{(n)})= \frac{1}{n^{n-1}}\prod_{j=1}^{n-1}\Big[je^{i n \varphi}+n-j\Big],
\end{equation}
as conjectured in Appendix \ref{app:series}. This analytic pattern we observe is not a proof of the permanent, but an empirical pattern---a conjecture---that has been verified by brute force to be correct up to $n=25$. Although we don't have a proof beyond that point, $n=25$ is well beyond what will be experimentally viable in the near future, and thus the pattern we observe is sufficient for experimentally enabling super-sensitive metrology with technology available in the foreseeable future.

\begin{figure}[!htb]
\centering
\includegraphics[scale=1]{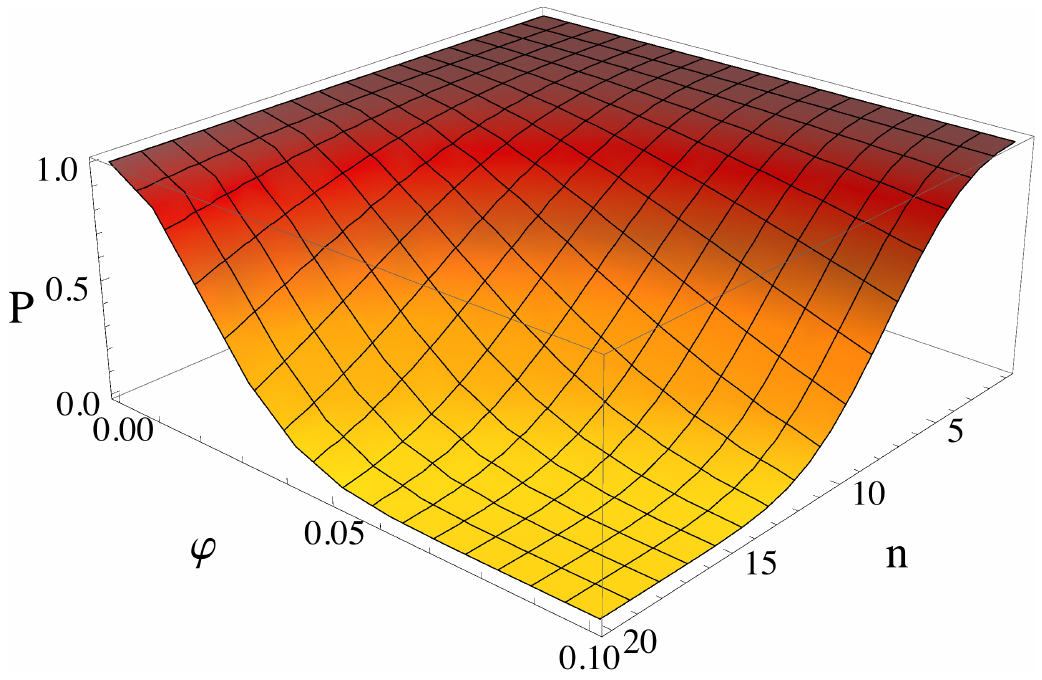}
\caption{Coincidence photodetection probability $P$ against the unknown phase $\varphi$ and the number of photons and modes $n$. As $n$ increases, the dependence of $P$ on $\varphi$ increases, resulting in improved phase-sensitivity.} \label{fig:P}
\end{figure}

Following as a corollary to the previous conjecture, the coincidence probability of measuring one photon in each mode is,
\begin{eqnarray} \label{eq:P_Result}
P &=& \Big|\perm(\hat{U}^{(n)})\Big|^2 \nonumber \\
&=& \frac{1}{n^{2n-2}}\prod_{j=1}^{n-1} \Big[a_n(j)\mathrm{cos}(n\varphi)+b_n(j) \Big],
\end{eqnarray} 
as shown in Appendix \ref{app:P}, where
\begin{eqnarray}
a_n(j) &=& 2j(n-j), \nonumber \\
b_n(j) &=& n^2-2jn+2j^2.
\end{eqnarray}
The dependence of $P$ on $n$ and $\varphi$ is shown in Figure \ref{fig:P}.

It then follows that,
\begin{equation} \label{eq:dP}
\left|\frac{\partial P}{\partial \varphi}\right| = nP\big|\mathrm{sin}(n\varphi)\big|\sum_{j=1}^{n-1} \left|\frac{a_n(j)}{a_n(j)\mathrm{cos}(n\varphi)+b_n(j)}\right|,
\end{equation}
as shown in Appendix \ref{app:dP}.

\begin{figure}[!htb]
\centering
\includegraphics[scale=0.5]{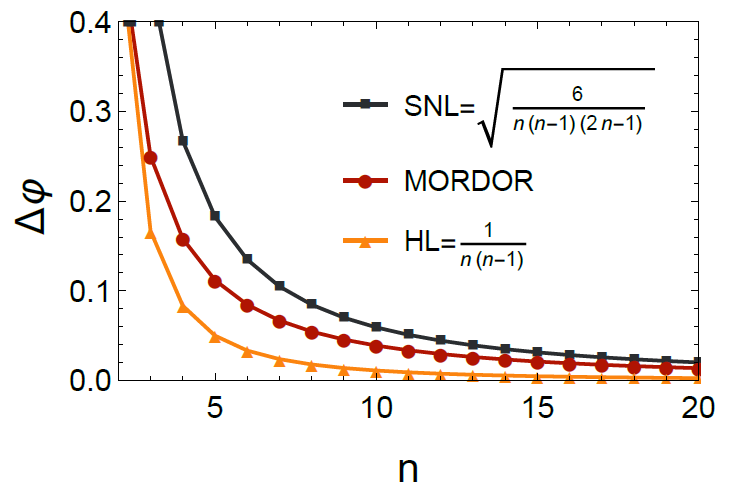}
\caption{Phase-sensitivity $\Delta\varphi$ against the number of photons $n$ (red circles). The shotnoise limit (black squares) and Heisenberg limit (orange triangles) are shown for comparison.  MORDOR exhibits phase-sensitivity significantly better than the shotnoise limit, and only slightly worse than the Heisenberg limit. \label{fig:Delta}}  
\end{figure}

Finally, we wish to establish the scaling of $\Delta\varphi$. With a small $\varphi$ approximation (\mbox{$\mathrm{sin}(\varphi)\approx\varphi$}, \mbox{$\mathrm{cos}(\varphi)\approx 1-\frac{1}{2}\varphi^2$}) we find,
\begin{eqnarray} \label{eq:DeltaVarPhi}
\Delta\varphi &=& \sqrt{\frac{3}{2n(n+1)(n-1)}} \\ \nonumber
&=& \frac{1}{2\sqrt{{{n+1}\choose{3}}}},
\end{eqnarray}
as shown in Appendix \ref{app:dphi}. Thus, the phase sensitivity scales as \mbox{$\Delta\varphi = O(1/n^{3/2})$} as shown in Figure \ref{fig:Delta}.

We would like to compare the performance of MORDOR to an equivalent multimode interferometer baseline for which we will construct the shotnoise limit (SNL) and Heisenberg limit (HL). This is a subtle comparison, due to the linearly increasing unknown phase-shifts, \{\mbox{$0,\varphi,\dots,(n-1)\varphi$}\}, that MORDOR requires to operate. The mathematical relation is shown in Figure \ref{fig:Delta}, where we have written the sensitivity in terms of the number of photons, $n$. There is disagreement on how such resources should be counted.  The method originally referred to in Ref.~\cite{bib:Mordor2015}, called Ordinal Resource Counting (ORC), is one such way to count resources; this method is not the one used to generate Figure \ref{fig:Delta}. A more detailed discussion on this point can be found in Appendix \ref{app:counting}.

While computing the sensitivity (using the standard error propagation formula of Eq.~(\ref{eq:phaseSensitivity})) provides clear evidence that our scheme does indeed beat the SNL, it would be instructive to carry out a calculation of the quantum Fisher information and thereby provide the quantum Cram{\'e}r-Rao bound, which would be a true measure of the best performance of this scheme possible, according to the laws of quantum theory. However, due to the need to compute the permanent of large matrices with complex entries, this calculation currently remains intractable. It is my hope that such an investigation is done for a future work. In general, analytic solutions to matrix permanents are not possible. In this instance, the analytic result is facilitated by the specific structure of the MORDOR unitary. Other inhomogeneous phase gradients may yield analytic results, as is the case with the single phase shifter of the next section.

In Appendix \ref{app:efficiency} we discuss the efficiency of MORDOR and in Appendix \ref{app:dephasing} we analyze dephasing, which is a source of decoherence, and find that MORDOR is far more robust against dephasing than the NOON state is.

We have now shown that a passive linear optics network fed with single-photon Fock states may implement quantum metrology with phase-sensitivity that beats the shotnoise limit. Unlike other schemes that employ exotic states such as NOON states, which are notoriously difficult to prepare, single-photon states may be readily prepared in the laboratory using present-day technology. This new approach to metrology via easy-to-prepare single-photon states and disjoint photodetection provides a road towards improved quantum metrology with frugal physical resources.  In the next section, we will consider an optimization over interferometers sharing the same properties as MORDOR by introducing the Quantum Fourier Transform Interferometer (QuFTI).


\section{General QuFTI} \label{sec:qufti}
\begin{figure}
\centering
\includegraphics[scale=1.7]{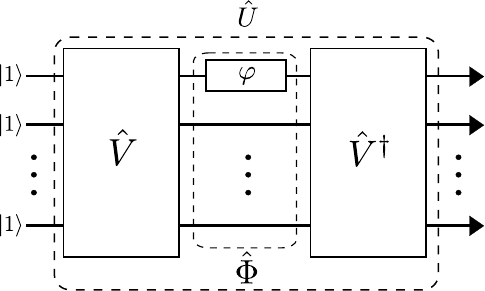}
\begin{caption}{Ideal QuFTI}
The ideal phase strategy for a QuFTI when $\hat{V}$ is an $n$-mode QFT.  All of the unknown phase $\varphi$ is put into a single mode (here, depicted as the first mode, though any mode is sufficient).
\label{fig:single}
\end{caption}
\end{figure}

In very general terms, one can consider the architecture of MORDOR in the previous section as a particular choice of four components of an interferometer---input, unitary evolution, phase evolution, and measurement.  The most compelling aspect of the architecture of MORDOR is the fact that this choice comprises a device which has potential scalability in the near future.  Specifically, single photon sources, bucket photodetectors, and passive optical elements may soon all be implementable on an integrated photonic chip.  The natural question arises, however, whether the MORDOR architecture \textit{optimizes} the phase sensitivity for a device with these properties.  In this section, we first discuss what degrees of freedom we have to make changes to the interferometer without sacrificing any of the desirable properties.  We then provide compelling evidence that the architecture of Figure \ref{fig:single} achieves the best phase sensitivity under these constraints.

\begin{figure}[h]
\centering
\includegraphics[scale=1.7]{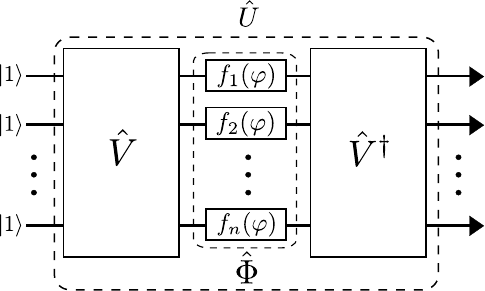}
\begin{caption}{A generalized architecture for the QuFTI.  We consider optimizations over $\hat{V}\in$ SU($n$) and phase strategies $\hat{\Phi}$, together with single photon inputs and photodetection in each mode.  The MORDOR architecture can be restored when $\hat{V}$ is the $n$-mode QFT and $f_i(\varphi)=(i-1)\varphi$.} \label{fig:general}
\end{caption}
\end{figure}

Although NOON states and other exotic quantum states (such as squeezed vacuum) are known to perform well for quantum metrology, one pays a very high price to prepare these states.  Furthermore, many of these states are known to be very sensitive to common sources of noise.  Thus, in keeping with the spirit of MORDOR, we wish to consider interferometers which can take advantage of the emergence of commercially available, high fidelity, high efficiency single photon sources and non-number-resolving detectors.  Hence, for this manuscript, we fix the condition that the input state consists of $n$ single photon sources, together with bucket photodetection at the output.  This leaves us two components to optimize over---unitary evolution, and phase evolution (see Figure \ref{fig:general}).  Somewhat surprisingly, we show that the optimal architecture is not only easier to implement than MORDOR, but also easier to interpret analytically.

We begin our investigation of different phase strategies by fixing the unitary evolution to be the $n$-mode optical quantum Fourier transform (QFT), i.e., the normalized, unitary discrete Fourier transform.  We will return to the topic of unitary evolution later.  The $n$-mode QFT again takes the form,
\begin{equation}
V^{(n)}_{jk}=\frac{1}{\sqrt{n}}\omega_n^{(j-1)(k-1)}
\end{equation}
where $\omega_n=e^{2\pi i/n}$ is a primitive $n^{th}$ root of unity.  Because the interferometer is always of some fixed size of $n$ modes, we may drop superscript or subscript labels when there is no ambiguity.  We will refer to a device fixed with the QFT as a generalized QuFTI.  Consider a general phase strategy $\hat{\Phi}$ that applies a phase $f_j\cdot\varphi$ to mode $j$.  Then $\hat{\Phi}$ can be represented by a diagonal matrix $\Phi$ with entries,
\begin{equation}
\Phi_{jk}=\delta_{jk}e^{i\cdot f_j\cdot\varphi},
\end{equation}
We further assume that,
\begin{equation}
\sum_{j=1}^n f_j=1 \quad \mathrm{where}\; 0\leq f_j < 1.   \label{eq:normalization}
\end{equation}
This assumption is made to ensure that differing phase strategies are fair when compared to one another.  Furthermore, it is not restrictive since any general phase strategy can be normalized or reparameterized to fit this assumption.  We discuss this in more detail at the end of this section.

Recall that our goal is to optimize the phase sensitivity of a QuFTI with respect to all possible phase strategies.  Using Eq.~(\ref{eq:normalization}), we can apply the results of Giovannetti, Lloyd, and Maccone in Ref.~\cite{bib:GLM06} in a relatively straightforward way to determine the shotnoise and Heisenberg limited phase sensitivities of more general schemes with fixed phase strategies.  In this setting, $\hat{V}$ and $\hat{V}^\dagger$ are each replaced by some unitary map.  It is not difficult to see with this analysis (though we give a full proof in Appendix \ref{sec:optphase}) that the optimal phase strategy in this more general setting is when,
\begin{equation}
f_j=\delta_{j1},
\end{equation}
However, because the setting is very general, we cannot guarantee that this is also the optimal phase strategy when considering more specific implementations of optical networks.  We wish to show that, in this case, the same is true for the QuFTI architecture, i.e. when $\hat{V}$ is the $n$-mode optical QFT.  

\begin{figure}[h]
\centering
\includegraphics[scale=.45]{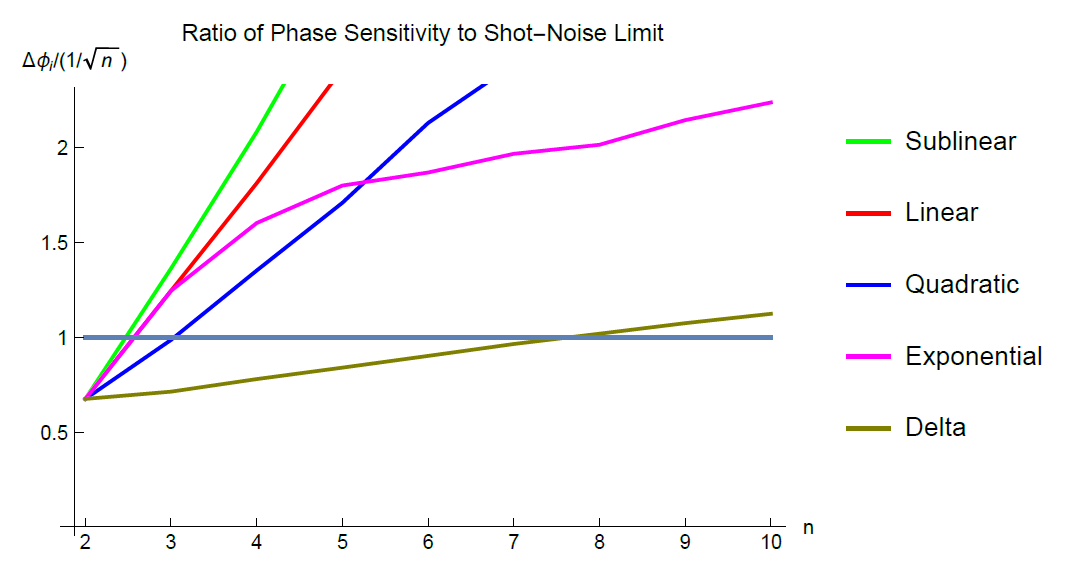}
\begin{caption}{The scaling of different phase strategies for the QuFTI suggests that widening the ``phase gap'' between modes improves the phase sensitivity.  The shot-noise limit used for comparison here is defined to be $1/\sqrt{n}$, which is the best possible classical scheme for $n$ photons and any number of modes $\geq 2$.  Any point below 1 indicates a sub-shot-noise phase sensitiivity.}
\label{fig:stratplot}
\end{caption}
\end{figure}

In order to help understand the dynamics of differing phase strategies, we consider a range of functions representing trial strategies (Table \ref{tab:strategies}).  For each phase strategy, we numerically compute $P=|\mathrm{perm}(\hat{U})|^2=|\mathrm{perm}(\hat{V}\Phi\hat{V}^{\dagger})|^2$, and plot the resulting phase sensitivity in Figure \ref{fig:stratplot}.  From this figure, it is apparent that there is no improvement in phase sensitivity by distributing the phase throughout the modes, and restricting $\varphi$ to one mode is most effective. \\

\begin{table}[h]
\centering
\caption{Functions representing trial phase strategies.  Note that many of strategies are not normalized to satisfy Eq.~(\ref{eq:normalization}), but can easily be made so by dividing each by $\sum_{j=1}^n f_j$.} \label{tab:strategies}
\begin{tabular}{l l}
Constant & $f^{con}_{j}=\frac{1}{n}$ \\
Sub-linear & $f^{sub}_{j}=\sqrt{j}$ \\
Linear & $f^{lin}_{j}=j$ \\
Quadratic & $f^{quad}_{j}=j^2$ \\
Exponential & $f^{exp}_{j}=2^j$ \\
Delta  & $f^{\delta}_{j}=\delta_{j1}$
\end{tabular}
\end{table}

\begin{figure}[h]
\centering
\includegraphics[scale=.45]{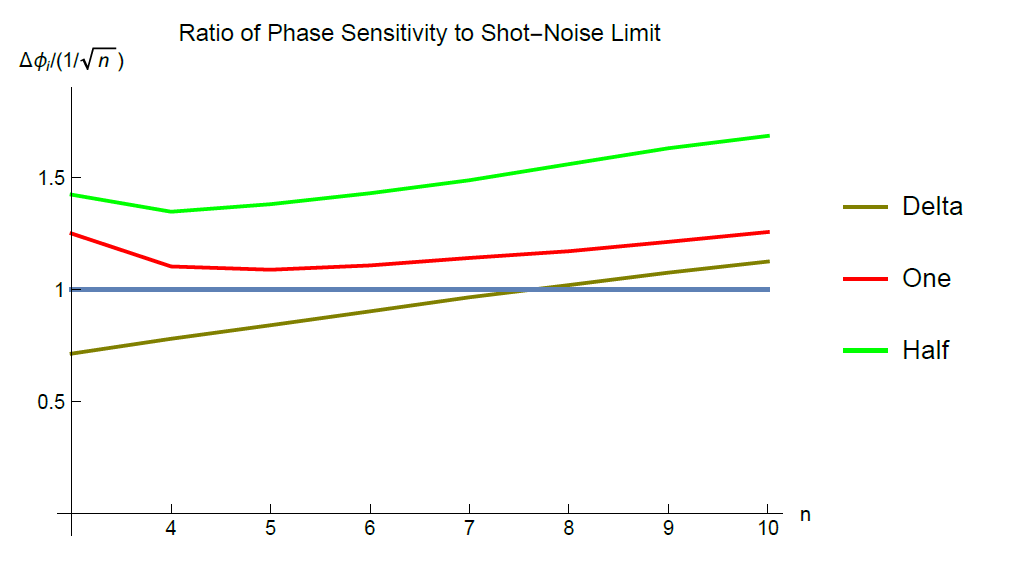}
\begin{caption}{Phase Sensitivity in One vs. Two Modes}
Restricting $\varphi$ to one mode is strictly better than mixing $\varphi$ in two modes.  
\label{fig:onehalf}
\end{caption}
\end{figure}

In order to more firmly establish this, we consider two more cases:
\begin{eqnarray}
f^{one}_{j}&=& \left\{
        \begin{array}{ll}
            (1-\frac{1}{n})\varphi & j=1 \\
            (1/n)\varphi & j=2 \\
             0 & j>2 
        \end{array}
    \right. \\
f^{half}_{j}&=& \left\{
        \begin{array}{ll}
            (1/2)\varphi & j=1 \\
            (1/2)\varphi & j=2 \\
             0 & j>2 
        \end{array}
    \right.
\end{eqnarray}
It stands to reason that if there is any advantage to be gained by distributing $\varphi$ into two modes instead of one, it would be achieved by one of these two strategies---i.e. a strategy that either balances the two modes, or weighs one more heavily.  It is easy to see from Figure \ref{fig:onehalf} that this is clearly not the case, and both are outperformed by having $\varphi$ in a single mode.  Finally, we remark that if the phase sensitivity is strictly lower by distributing $\varphi$ into two modes, then the same is surely true for a phase strategy where $\varphi$ is distrbuted into three or more modes.  We thus conclude that the optimal phase strategy for a QuFTI is as given in Figure \ref{fig:single}.

With this in mind, we would like to compare the architecture described in MORDOR to the optimal QuFTI strategy described above. However, we have already made this comparison, since MORDOR possesses the linear phase strategy $f_j=(j-1)$, whose normalized strategy is plotted against the ideal strategy in Figure \ref{fig:stratplot}.  This may seem at first contradictory to the results in MORDOR, which show that for all $n$, the phase sensitivity of MORDOR beats the shotnoise limit.  This is because the shotnoise limit as defined in MORDOR is the best possible classical sensitivity \textit{given the linear phase strategy}.  Thus, one may summarize the results of MORDOR in the following way: if one were restricted to using a linear phase gradient to approximate an unknown $\varphi$, then there exists a passive optical quantum strategy which is much more efficient than any classical strategy.   This is unfortuantely a rather restrictive condition.

\begin{figure}[!htb]
\centering
\includegraphics[scale=0.45]{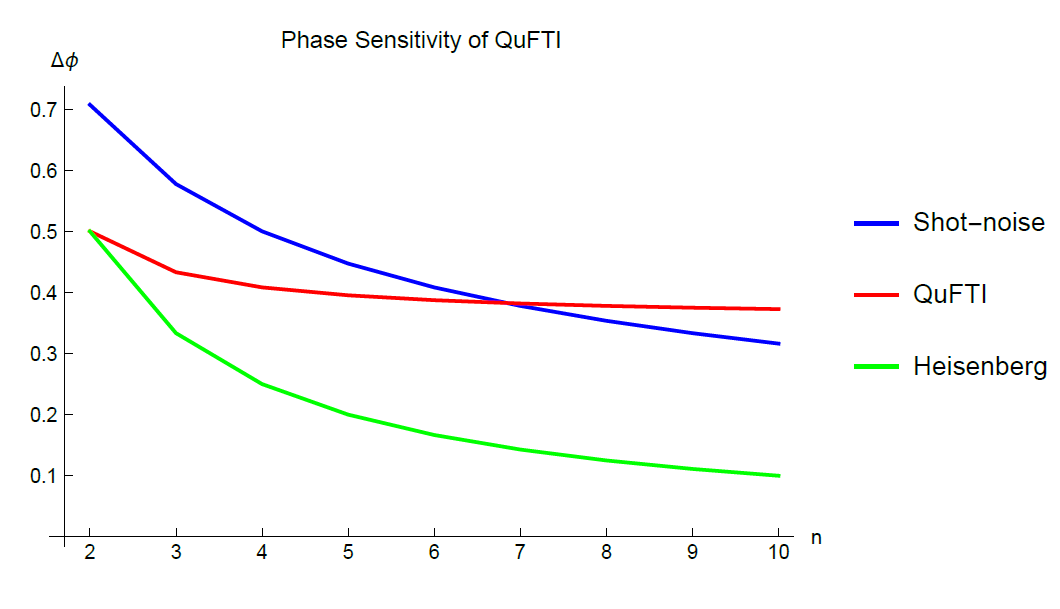}
\begin{caption}{Phase Sensitivity of Ideal QuFTI}
\label{fig:ideal}
\end{caption}
\end{figure}

In many applications, the experimental device, including the distribution of $\varphi$, is controllable.  As our goal is to produce an efficient, scalable device that is useful in a more general setting, we are more interested in comparing the sensitivity of a QuFTI to the best (classical or quantum) strategy available.  As we have ample evidence to conclude that the ideal QuFTI is as shown in Figure \ref{fig:single}, we wish to characterize the phase sensitivity of this device analytically.  The operator $\hat{\Phi}_{\delta}^{(n)}$ describing the single mode phase strategy $f^{\delta}$ is given by the diagonal matrix,
\begin{equation}
\hat{\Phi}_{\delta}^{(n)}\equiv \Phi_{j,k}=\delta_{j,k}e^{i\varphi\delta_{j,1}}.
\end{equation}
This implies that the matrix describing the entire inteferometer is given by,
\begin{equation}
\hat{U}\equiv\hat{V}\hat{\Phi}\hat{V}^\dagger=\frac{1}{n}\Big[e^{i \varphi}+\delta_{j,k}n-1\Big],
\end{equation}
an explicit derivation of which can be found in Appendix \ref{sec:Uentries}.  Because of the simpler form of $\hat{\Phi}$, an analytic derivation of perm($\hat{U}$) is also possible---a result that was only postulated in Ref.~\cite{bib:Mordor2015}.  We show in Appendix \ref{sec:permU} that,
\begin{equation}
\textrm{perm}(\hat{U}) = \frac{1}{n^n}\sum_{k=0}^n D_{n,k} [e^{i\varphi}+n-1]^k[e^{i\varphi}-1]^{n-k},
\end{equation}
where 
\begin{equation}
D_{n,k}=\frac{n!}{k!}\sum_{j=0}^{n-k}\frac{(-1)^j}{j!},
\end{equation}
 is referred to as the \textit{rencontres numbers}, which enumerate all permutations in $S_n$ with $k$ fixed points. We subsequently derive in Appendix \ref{sec:deltaphi} that the phase sensitivity $\Delta\varphi$ when $\varphi\ll n\varphi$ is given by,
\begin{equation}
\Delta\varphi = \frac{1}{2\sqrt{2}\sqrt{\frac{n-1}{n}}}.
\end{equation}
We now compare this in Figure \ref{fig:ideal} directly to the `usual' shotnoise ($1/\sqrt{n}$) and Heisenberg ($1/n$) limit as defined in Ref.~\cite{bib:GLM06}, which characterizes the maximally achievable phase sensitivity for classical and quantum strategies satisfying Eq. \ref{eq:normalization} (see Appendix \ref{sec:optphase}).  One can easily see that for $2\leq n<7$, the QuFTI provides sub-shotnoise sensitivity, but is limited to $1/\sqrt{8}$ in the asymptotic limit.

We now address the role of Eq.~(\ref{eq:normalization}) and the relevance of considering a normalized phase strategy.  Recall that the error propagation formula for some observable $\hat{O}$ as a function of $\varphi$ is given by,
\begin{equation}
\Delta\varphi=\frac{\sqrt{\ip{O^2}-\ip{O}^2}}{\big|\frac{\partial \ip{O}}{\partial \varphi}\big|}.
\end{equation}
Suppose we consider a reparameterization of $\varphi$ defined by $\tau=k\varphi$ for some positive integer $k$.  If one compares the phase sensitivity $\Delta\tau$ to that of $\Delta\varphi$, one can see that
\begin{equation}
\Delta\varphi=\frac{\sqrt{\ip{O^2}-\ip{O}^2}}{k\big|\frac{\partial \ip{O}}{\partial \tau}\big|} \quad \Rightarrow \quad \Delta\varphi=\frac{1}{k}\Delta\tau,
\end{equation}
which may tempt one to think that perhaps the sensitivity of measuring with respect to $\varphi$ is better than that of $\tau$.  This is indicative of the fact that, by replacing $\varphi$ with $k\varphi$ in an experiment (e.g. in the case of measuring the index of refraction of glass, by putting $k$ copies of a glass slab into your interferometer) has a real effect on the output of the device.  In terms of the sensitivity, however, this only compresses the coordinates by a factor of $k$, so that the uncertainty with respect to $\varphi$ is equally compressed.  A fact that is sometimes overlooked is that this also scales the shotnoise and Heisenberg limit by an equal factor, so any comparison between the sensitivity and either limit is maintained.

Thus, in order to remove the illusion of arbitrarily high phase sensitivity, we wish to consider the phase $\varphi$ to be a limited resource possessed by the experimenter.  The constraint of Eq.~(\ref{eq:normalization}) then can be interpreted as allowing the experimenter the freedom to distribute fractions of $\varphi$ among the modes of his choosing.  This allows each strategy to be compared directly to the usual notion of shotnoise and Heisenberg limit without scaling the limit for every strategy.

We remark that Eq.~(\ref{eq:normalization}) is not reflective of every possible experimental setup.  For example, it may be the case that an experimenter is able to have an arbitrary number of modes access $\varphi$ at no cost, in which case the classical and quantum limits may not aptly describe the ideal architecture under this constraint.

Earlier, we discussed optimization over the phase strategies in a generalized QuFTI.  However, we can also consider the choice of unitary evolution as an additional degree of freedom in the device.  That is, we wish to maximize the phase sensitivity with respect to an arbitrary $\hat{V}\in$ SU($n$), which characterizes the set of all passive unitary transformations on $n$ modes, any of which can be efficiently implemented with at most O($n^2$) passive optical elements \cite{bib:Reck94}.

Ideally, we would like to consider optimizing over the phase strategy $\hat{\Phi}$ and the unitary $\hat{V}$ simultaneously, so that every possible case is considered.  However, we believe this is unnecessarily rigorous.  In Ref.~\cite{bib:GLM06}, it is noted that the lowerbound is attained by ``an equally weighted superposition of the eigenvectors relative to the maximum and minimum eigenvalues of the global generator $h$.''  Since we show in Appendix \ref{sec:optphase} that the maximum difference in eigenvalue is only achievable with the strategy $f^{\delta}$, we will fix this strategy and consider optimizing only over $\hat{V}$.  It is perhaps suggestive already that the optimal $\hat{V}$ should be the QFT (or any other unitary matrix satisfying $|V_{ij}|=\frac{1}{n}$), since it is the relative maximum in SU($n$) for producing a superposition of states which have the largest amplitude corresponding to these eigenvalues.

\begin{figure}[h]
\centering
\includegraphics[scale=0.45]{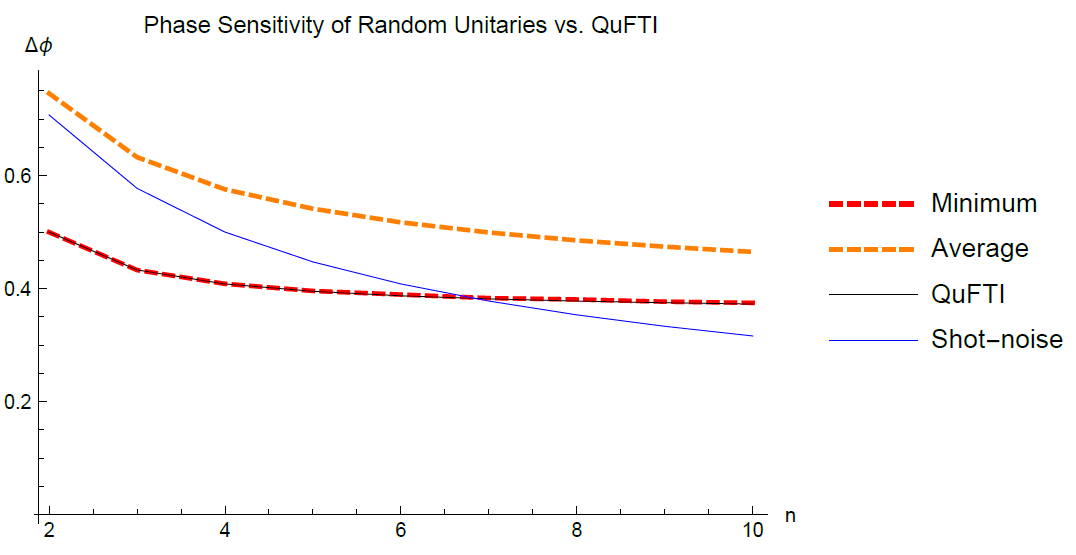}
\begin{caption}{QFT is Optimal for Delta Strategy}
\label{fig:minuplot}
\end{caption}
\end{figure}

If $\hat{V}$=QFT is not optimal, then it is either a relative maximum, or there will exist a $\hat{V}'$ in a neighborhood of $\hat{V}$ such that the phase sensitivity of $\hat{V}'$ supersedes that of $\hat{V}$.  In order to suggest both assertions could not be correct, we computed the phase sensitivity of 10,000 random unitaries in SU($n$) (for each $n$), and plotted the best phase sensitivity (i.e. minimum $\Delta\varphi$) of this set against the phase sensitivity of the QFT (see Figure \ref{fig:minuplot}).  For all $2\leq n\leq 7$, the sensitivity of the QFT exceeds that of every random unitary, providing solid evidence that it is indeed the optimal unitary for the $f^{\delta}$ strategy.  We further remark that it is not \textit{trivially} optimal---Figure \ref{fig:minuplot} also shows the phase sensitivity of the average case, which does not attain sub-shotnoise sensitivity for any $n$.

\pagebreak

\singlespacing
\chapter{Conclusion}
\doublespacing
I would like to leave the reader with an impression of what I believe the results in this thesis suggest about the future of quantum technologies.  We have seen examples of powerful theoretical models which provide clear advantages over their classical counterparts.  But with these advancements often come major engineering challenges that need to be overcome.  Whether it be implementing fault-tolerant LOQC or generating high-NOON states, there seems to be no obvious short-term solution that would enable these mechanisms to be realistically scalable.  It seems that the full potential of quantum mechanics eludes us for the time being. 

On the other hand, new developments suggest that in the mean time we may be able to exploit some of the more accessible properties of quantum theory.  In this thesis, we have seen how optical multi-mode networks are a natural environment for generating entanglement.  In the setting of \BS, we saw that there is a fundamental connection between the evolution of bosonic Fock states and matrix permanents.  This connection spans a colossal gap between the complexity of what we believe classical computers can do and what it seems that Nature does automatically.  We have seen in \cc{DisplacedSampling} and \cc{PASSVSampling} that this complexity persists in more general systems than Fock states alone, widening the potential applicability of these devices.  At the same time, the entanglement generated from a multimode network can be utilized by a similar device---the QuFTI---to make more precise measurements than could be made classically.  Results from other groups seem to tell the same story; for example, a \BS-like network can be used to simulate certain systems in quantum chemistry \cite{bib:HGPMA14}.

Before scalable quantum computing becomes a reality, there would appear to be enormous potential for an \textit{intermediate} regime of quantum processing, where novel schemes based on particular quantum systems exhibit properties that classical devices cannot imitate.  Yet at present, it seems that this potential remains relatively untapped.  My hope is that this thesis provides a stepping stone to devising a truly practical post-classical device, and enables and inspires the reader to become involved in achieving this goal.  I will now outline some future projects that I believe may hold to the key to doing so.

In this thesis, we discussed an application to quantum metrology---a field with far-reaching applications in many other areas.  Because of the inherent experimental overhead in constructing quantum devices, it seems unlikely that the metrological devices discussed in this thesis would find direct application in many industrial settings.  However, the results suggest that there may exist \textit{particular} systems that could greatly benefit from the application of quantum devices---perhaps those where the number of photons probing a system is very limited.  If reliable ways to create more exotic quantum states of light become available, it may be that derivatives of the schemes discussed in this section might show improved scaling.  For example, initial investigation suggests that Fock states of photon number $\ket{n}$ may be used in place of single photons in the QuFTI to improve sensitivity closer to the Heisenberg limit.

Most public-key cryptosystems rely on the intractability of computing particular quantities; most famously, RSA requires an eavesdropper to factor large semiprime integers.  The hardness of \BS and computing matrix permanents may imply that it possible to construct a quantum cryptosystem based on far stronger complexity assumptions, such as the unlikely collapse of the polynomial hierarchy or $\BPP\neq\sharpP$.  If possible, such a system may only need to rely on relatively simple optical elements and photon sources.

Interest has been growing rapidly on the topic of quantum machine learning.  Many machine learning tasks can be mapped to graph-theoretic problems, of which matrix permanents have some natural connections.  Namely, permanents of binary matrices can be thought of as counting the number of perfect matchings in a graph, or counting the number of vertex cycle covers of a graph (where the matrix in question represents the adjacency matrix of the graph).  It may be possible that a network similar to \BS could map to a useful training algorithm for certain learning environments.

Finally, the mathematics of linear optics is certainly not unique to the quantum world.  Bosonic systems can be found in many other subfields of physics, from particle physics to condensed matter.  Quantum optics itself is a very rich field, and certainly not restricted to optical networks.  It may be that considering \BS or the QuFTI in other environments within optics may allow one to take advantage of other effects that expand the scope of the problem, or reduce the experimental overhead for implementing these schemes.

\pagebreak

\pagebreak
\singlespacing
\addtocontents{toc}{\vspace{12pt}}
\addcontentsline{toc}{chapter}{\hspace{-1.6em} REFERENCES} 
\vspace{0.9em}

\makeatletter
\let\oldmake\@makechapterhead
\let\oldmakes\@makeschapterhead

\def\@makechapterhead#1{%
  \vspace*{-20\p@}%
  {\parindent \z@ \raggedright \normalfont
    \ifnum \c@secnumdepth >\m@ne
        \huge\bfseries \@chapapp\space \thechapter
        \par\nobreak
        \vskip 20\p@
    \fi
    \interlinepenalty\@M
    \Huge \bfseries #1\par\nobreak
    \vskip 12\p@
  }}
\def\@makeschapterhead#1{%
  \vspace*{-20\p@}%
  {\parindent \z@ \raggedright
    \normalfont
    \interlinepenalty\@M
    \Huge \bfseries  #1\par\nobreak
    \vskip 12\p@
  }}
\makeatother

\bibliographystyle{plain}
\bibliography{thesisbib}

\begin{thebibliography}{10}

\bibitem{bib:AA13response}
S.~Aaronson and A.~Arkhipov.
\newblock Boson{S}ampling is far from uniform.
\newblock arXiv:1309.7460v2, 2013.

\bibitem{Agarwal_91}
G.~S. Agarwal and K.~Tara.
\newblock Nonclassical properties of states generated by the excitations on a
  coherent state.
\newblock {\em Phys. Rev. A}, 43:492--497, Jan 1991.

\bibitem{bib:AA10}
A.~Arkhipov and S.~Aaronson.
\newblock The computational complexity of linear optics.
\newblock {\em Proc. ACM STOC (New York)}, page 333, 2011.

\bibitem{BW99}
K.~Banaszek and K.~W\'odkiewicz.
\newblock Testing quantum nonlocality in phase space.
\newblock {\em Phys. Rev. Lett.}, 82:2009--2013, Mar 1999.

\bibitem{bib:Bardhan2013}
B.~R. Bardan, K.~Jiang, and J.~P. Dowling.
\newblock Effects of phase fluctuations on phase sensitivity and visibility of
  path-entangled photon {F}ock states.
\newblock {\em Phys. Rev. A}, 88(4):023857, 2013.

\bibitem{bib:BartSand03}
S.~D. Bartlett and B.~C. Sanders.
\newblock Requirement for quantum computation.
\newblock {\em J. Mod. Opt.}, 50:2331–2340, 2003.

\bibitem{bib:Bell1964}
J.~S. Bell.
\newblock On the {E}instein-{P}odolsky-{R}osen paradox.
\newblock {\em Physics}, 1:195--200, 1964.

\bibitem{bib:RepLinearGroups}
R.~Berndt.
\newblock {\em Representations of linear groups}.
\newblock Vieweg, 2007.

\bibitem{bib:Bouland}
A.~Bouland and S.~Aaronson.
\newblock Generation of universal linear optics by any beamsplitter.
\newblock 2013.

\bibitem{bib:Caves1994}
S.~L. Braunstein and C.~M. Caves.
\newblock Statistical distance and the geometry of quantum state.
\newblock {\em Phys. Rev. Lett.}, 72:3439, 1994.

\bibitem{bib:Broome2012}
M.~A. Broome, A.~Fedrizzi, S.~Rahimi-Keshari, J.~Dove, S.~Aaronson, T.~C.
  Ralph, and A.~G. White.
\newblock Photonic boson sampling in a tunable circuit.
\newblock {\em Science}, 339:6121, 2013.

\bibitem{bib:cable2007efficient}
H.~Cable and J.~P. Dowling.
\newblock Efficient generation of large number-path entanglement using only
  linear optics and feed-forward.
\newblock {\em Phys. Rev. Lett.}, 99(16):163604, 2007.

\bibitem{bib:cerf}
N.J. Cerf, C.~Adami, and P.G. Kwiat.
\newblock Optical simulation of quantum logic.
\newblock {\em Phys. Rev. A}, 57:R1477--R1480, 1998.

\bibitem{bib:cerny}
V.~Cerny.
\newblock Quantum computers and intractable ({NP}-complete) computing problems.
\newblock {\em Phys. Rev. A}, 48:116--119, 1993.

\bibitem{bib:Christandl06}
M.~Christandl.
\newblock The structure of bipartite quantum states - insights from group
  theory and cryptography.
\newblock 2006.

\bibitem{bib:clauser}
J.~F. Clauser and J.~P. Dowling.
\newblock Factoring integers with {Y}oung's {N}-slit interferometer.
\newblock {\em Phys. Rev. A}, 53:4587--4590, 1996.

\bibitem{bib:crespi2012measuring}
A.~Crespi, M.~Lobino, J.~C.~F. Matthews, A.~Politi, C.~R. Neal, R.~Ramponi,
  R.~Osellame, and J.~L. O'Brien.
\newblock Measuring protein concentration with entangled photons.
\newblock {\em App. Phys. Lett.}, 100:233704, 2012.

\bibitem{bib:crespi2013integrated}
A.~Crespi, R.~Osellame, R.~Ramponi, D.~J. Brod, E.~F. Galv{\~a}o, N.~Spagnolo,
  C.~Vitelli, E.~Maiorino, P.~Mataloni, and F.~Sciarrino.
\newblock Integrated multimode interferometers with arbitrary designs for
  photonic boson sampling.
\newblock {\em Nature Phot.}, 7:545, 2013.

\bibitem{Dakna_98}
M.~Dakna, L.~Knoll, and D.-G. Welsch.
\newblock Photon-added state preparation via conditional measurement on a beam
  splitter.
\newblock {\em Opt. Comm.}, 145(1–6):309 -- 321, 1998.

\bibitem{Dakna_98_2}
M.~Dakna, L.~Knoll, and D.-G. Welsch.
\newblock Quantum state engineering using conditional measurement on a beam
  splitter.
\newblock {\em Eur. Phys. J. D}, 3(3):295--308, 1998.

\bibitem{bib:Dita01}
P.~Dita.
\newblock Factorization of unitary matrices.
\newblock 2001.

\bibitem{bib:dowling1998correlated}
J.~P. Dowling.
\newblock Correlated input-port, matter-wave interferometer: {Q}uantum-noise
  limits to the atom-laser gyroscope.
\newblock {\em Phys. Rev. A}, 57:4736, 1998.

\bibitem{bib:dowling2008quantum}
J.~P. Dowling.
\newblock Quantum optical metrology \--- the lowdown on high-{NOON} states.
\newblock {\em Contemp. Phys.}, 49:125, 2008.

\bibitem{bib:DummitFoote}
D.~S. Dummit and R.~M. Foote.
\newblock {\em Abstract algebra}.
\newblock Wiley, 2003.

\bibitem{bib:durkin2007local}
G.~A. Durkin and J.~P. Dowling.
\newblock Local and global distinguishability in quantum interferometry.
\newblock {\em Phys. Rev. Lett.}, 99(7):070801, 2007.

\bibitem{bib:fukuda2011titanium}
D.~Fukuda, G.~Fujii, G.~Numata, K.~Amemiya, A.~Yoshizawa, H.~Tsuchida,
  H.~Fujino, H.~Ishii, T.~Itatani, S.~Inoue, et~al.
\newblock Titanium-based transition-edge photon number resolving detector with
  98\% detection efficiency with index-matched small-gap fiber coupling.
\newblock {\em Opt. Express}, 19(2):870--875, 2011.

\bibitem{bib:gard2013quantum}
B.~T. Gard, R.~M. Cross, P.~M. Anisimov, H.~Lee, and J.~P. Dowling.
\newblock Quantum random walks with multiphoton interference and high-order
  correlation functions.
\newblock {\em JOSA B}, 30(6):1538--1545, 2013.

\bibitem{bib:Chapter}
B.~T. Gard, K.~R. Motes, J.~P. Olson, P.~P. Rohde, and J.~P. Dowling.
\newblock {\em Chapter 8: {A}n introduction to boson-sampling}, pages 167--192.
\newblock World Scientific Publishing Co, 2015.

\bibitem{bib:gard2014inefficiency}
B.~T. Gard, J.~P. Olson, R.~M. Cross, M.~B. Kim, H.~Lee, and J.~P. Dowling.
\newblock Inefficiency of classically simulating linear optical quantum
  computing with {F}ock-state inputs.
\newblock {\em Phys. Rev. A}, 89(2):022328, 2014.

\bibitem{bib:GentWalsh94}
I.~Gent and T.~Walsh.
\newblock The {SAT} phase transition.
\newblock {\em Proceedings of ECAI}, 94:105--109, 1994.

\bibitem{bib:Georgi99}
H.~M. Georgi.
\newblock {\em Lie algebras in particle physics}.
\newblock Perseus, 1999.

\bibitem{bib:gerry2001generation}
C.~C. Gerry and R.~A. Campos.
\newblock Generation of maximally entangled photonic states with a
  quantum-optical {F}redkin gate.
\newblock {\em Phys. Rev. A}, 64:063814, 2001.

\bibitem{bib:GerryKnight05}
C.~C. Gerry and P.~L. Knight.
\newblock {\em Introductory quantum optics}.
\newblock Cambridge University Press, 2005.

\bibitem{bib:GLM06}
V.~Giovannetti, S.~Lloyd, and L.~Maccone.
\newblock Quantum metrology.
\newblock {\em Phys. Rev. Lett.}, 96:010401, Jan 2006.

\bibitem{bib:Gogo13}
C.~Gogolin, M.~Kliesch, L.~Aolita, and J.~Eisert.
\newblock Boson-sampling in the light of sample complexity.
\newblock arXiv:1306.3995, 2013.

\bibitem{bib:Gurvits}
L.~Gurvits.
\newblock On the complexity of mixed discriminants and related problems.
\newblock {\em MFCS}, pages 447--458, 2005.

\bibitem{bib:Hardy01}
L.~Hardy.
\newblock Quantum theory from five reasonable axioms.
\newblock 2001.

\bibitem{bib:LFBT}
B.~Hensen, H.~Bernien, A.~E. Dréau, A.~Reiserer, N.~Kalb, M.~S. Blok,
  J.~Ruitenberg, R.~F.~L. Vermeulen, R.~N. Schouten, C.~Abellán, W.~Amaya,
  V.~Pruneri, M.~W. Mitchell, M.~Markham, D.~J. Twitchen, D.~Elkouss,
  S.~Wehner, T.~H. Taminiau, and R.~Hanson.
\newblock Loophole-free {B}ell inequality violation using electron spins
  separated by 1.3 kilometres.
\newblock {\em Nature (London)}, 526:682--686, 2015.

\bibitem{bib:holland1993interferometric}
M.~J. Holland and K.~Burnett.
\newblock Interferometric detection of optical phase shifts at the {H}eisenberg
  limit.
\newblock {\em Phys. Rev. Lett.}, 71:1355, 1993.

\bibitem{bib:HOM87}
C.~K. Hong, Z.~Y. Ou, and L.~Mandel.
\newblock Measurement of sub-picosecond time intervals between two photons by
  interference.
\newblock {\em Phys. Rev. Lett.}, 59:2044, 1987.

\bibitem{bib:Horodecki09}
R.~Horodecki, P.~Horodecki, M.~Horodecki, and K.~Horodecki.
\newblock Quantum entanglement.
\newblock {\em Rev. Mod. Phys.}, 81:865.

\bibitem{bib:HGPMA14}
J.~Huh, G.~G. Guerreschi, B.~Peropadre, J.~R. McClean, and A.~Aspuru-Guzik.
\newblock Boson sampling for molecular vibronic spectra.
\newblock {\em Nature Photonics}, 9:615--620, 2015.

\bibitem{bib:israel2014supersensitive}
Y.~Israel, S.~Rosen, and Y.~Silberberg.
\newblock Supersensitive polarization microscopy using {NOON} states of light.
\newblock {\em Phys. Rev. Lett.}, 112:103604, 2014.

\bibitem{bib:Jerrum04}
M.~Jerrum, A.~Sinclair, and E.~Vigoda.
\newblock A polynomial-time approximation algorithm for the permanent of a
  matrix with nonnegative entries.
\newblock {\em J. of the ACM}, 51:673, 2004.

\bibitem{bib:PhysRevA.88.044301}
Z.~Jiang, M.~D. Lang, and C.~M. Caves.
\newblock Mixing nonclassical pure states in a linear-optical network almost
  always generates modal entanglement.
\newblock {\em Phys. Rev. A}, 88:044301, Oct 2013.

\bibitem{bib:jones2009magnetic}
J.~A. Jones, S.~D. Karlen, J.~Fitzsimons, A.~Ardavan, S.~C. Benjamin, G.~A.~D.
  Briggs, and J.~J.~L. Morton.
\newblock Magnetic field sensing beyond the standard quantum limit using
  10-spin {NOON} states.
\newblock {\em Science}, 324:1166, 2009.

\bibitem{bib:toussaint2004}
K.~Toussaint Jr., G.~D. Giuseppe, K.~J. Bycenski, A.~V. Sergienko, B.~E.~A.
  Saleh, and M.~C. Teich.
\newblock Quantum ellipsometry using correlated-photon beams.
\newblock {\em Phys. Rev. A}, 70:023801, 2004.

\bibitem{bib:kapale2007bootstrapping}
K.~T. Kapale and J.~P. Dowling.
\newblock Bootstrapping approach for generating maximally path-entangled photon
  states.
\newblock {\em Phys. Rev. Lett.}, 99:053602, 2007.

\bibitem{bib:KLM01}
E.~Knill, R.~Laflamme, and G.~Milburn.
\newblock A scheme for efficient quantum computation with linear optics.
\newblock {\em Nature (London)}, 409:46, 2001.

\bibitem{bib:kok2007linear}
P.~Kok, W.~J. Munro, K.~Nemoto, T.~C. Ralph, J.~P. Dowling, and G.~J. Milburn.
\newblock Linear optical quantum computing with photonic qubits.
\newblock {\em Rev. Mod. Phys.}, 79:135, 2007.

\bibitem{bib:lapaire2003conditional}
G.~G. Lapaire, P.~Kok, J.~P. Dowling, and J.~E. Sipe.
\newblock Conditional linear-optical measurement schemes generate effective
  photon nonlinearities.
\newblock {\em Phys. Rev. A}, 68(4):042314, 2003.

\bibitem{bib:lee2002linear}
H.~Lee, P.~Kok, N.~J. Cerf, and J.~P. Dowling.
\newblock Linear optics and projective measurements alone suffice to create
  large-photon-number path entanglement.
\newblock {\em Phys. Rev. A}, 65:030101, 2002.

\bibitem{bib:lee2002quantum}
H.~Lee, P.~Kok, and J.~P. Dowling.
\newblock A quantum {R}osetta {S}tone for interferometry.
\newblock {\em J. Mod. Opt.}, 49:2325, 2002.

\bibitem{bib:seshadreesan2013phase}
Phys.~Rev. Lett.adreesan, S.~Kim, J.~P. Dowling, and H.~Lee.
\newblock Phase estimation at the quantum {C}ram\'{e}r-{R}ao bound via parity
  detection.
\newblock {\em Phys. Rev. A}, 87:043833, 2013.

\bibitem{bib:Lund13}
A.~P. Lund, A.~Laing, S.~Rahimi-Keshari, T.~Rudolph, J.~L. O'Brien, and T.~C.
  Ralph.
\newblock Boson sampling from a {G}aussian state.
\newblock {\em Phys. Rev. Lett.}, 113:100502, Sep 2014.

\bibitem{bib:Maier14}
S.~Maier, P.~Gold, A.~Forchel, N.~Gregersen, J.~M{\o}rk, S.~H\"{o}fling,
  C.~Schneider, and M.~Kamp.
\newblock Bright single photon source based on self-aligned quantum dot--cavity
  systems.
\newblock {\em Opt. Express}, 22(7):8136--8142, Apr 2014.

\bibitem{bib:matthews2011heralding}
J.~C.~F. Matthews, A.~Politi, D.~Bonneau, and J.~L. O'Brien.
\newblock Heralding two-photon and four-photon path entanglement on a chip.
\newblock {\em Phys. Rev. Lett.}, 107:163602, 2011.

\bibitem{bib:mayer2011counting}
K.~Mayer, M.~C. Tichy, F.~Mintert, T.~Konrad, and A.~Buchleitner.
\newblock Counting statistics of many-particle quantum walks.
\newblock {\em Phys. Rev. A}, 83:062307, 2011.

\bibitem{bib:motes2013spontaneous}
K.~R. Motes, J.~P. Dowling, and P.~P. Rohde.
\newblock Spontaneous parametric down-conversion photon sources are scalable in
  the asymptotic limit for boson sampling.
\newblock {\em Phys. Rev. A}, 88(6):063822, 2013.

\bibitem{bib:Mordor2015}
K.~R. Motes, J.~P. Olson, E.~Rabeaux, J.~P. Dowling, S.~J. Olson, and P.~P.
  Rohde.
\newblock Linear optical quantum metrology with single photons -- exploiting
  spontaneously generated entanglement to beat the shot-noise limit.
\newblock {\em Phys. Rev. Lett.}, 114:170802, 2015.

\bibitem{bib:nasr2003demonstration}
M.~B. Nasr, B.~E.~A. Saleh, A.~V. Sergienko, and M.~C. Teich.
\newblock Demonstration of dispersion-canceled quantum-optical coherence
  tomography.
\newblock {\em Phys. Rev. Lett.}, 91:083601, 2003.

\bibitem{bib:LPOR201400404}
L.~A. Ngahi, O.~Alibart, L.~Labonté, V.~D'Auria, and S.~Tanzilli.
\newblock Ultra-fast heralded single photon source based on telecom technology.
\newblock {\em Laser \& Photonics Reviews}, 9:L1--L5, 2015.

\bibitem{bib:NielsenChuang00}
M.~A. Nielsen and I.~L. Chuang.
\newblock {\em Quantum computation and quantum information}.
\newblock Cambridge University Press, Cambridge, 2000.

\bibitem{RLR_14}
S.~Rahimi-Keshari, A.~P. Lund, and T.~C. Ralph.
\newblock What can quantum optics say about complexity theory?
\newblock {\em arXiv:1408.3712v1}, 2014.

\bibitem{bib:ralph2013quantum}
T.~C. Ralph.
\newblock Quantum computation: Boson sampling on a chip.
\newblock {\em Nature Phot.}, 7(7):514, 2013.

\bibitem{bib:Reck94}
M.~Reck, A.~Zeilinger, H.~J. Bernstein, and P.~Bertani.
\newblock Experimental realization of any discrete unitary operator.
\newblock {\em Phys. Rev. Lett.}, 73:58, 1994.

\bibitem{bib:RSA}
R.~Rivest, A.~Shamir, and L.~Adleman.
\newblock A method for obtaining digital signatures and public-key
  cryptosystems.
\newblock {\em Comm. of the ACM}, 21(2):120–126, 1978.

\bibitem{bib:Rho14}
P.~P. Rohde.
\newblock Boson-sampling with photons of arbitrary spectral structure.
\newblock {\em Phys. Rev. A}, 91:012307, 2015.

\bibitem{bib:RohdeCat}
P.~P. Rohde, K.~R. Motes, P.~A. Knott, J.~Fitzsimons, W.~J. Munro, and J.~P.
  Dowling.
\newblock Evidence for the conjecture that sampling generalized cat states with
  linear optics is hard.
\newblock {\em Phys. Rev. A}, 91:012342, Jan 2015.

\bibitem{bib:rozema2014scalable}
L.~A. Rozema, J.~D. Bateman, D.~H. Mahler, R.~Okamoto, A.~Feizpour, A.~Hayat,
  and A.~M. Steinberg.
\newblock Scalable spatial superresolution using entangled photons.
\newblock {\em Phys. Rev. Lett.}, 112:223602, 2014.

\bibitem{bib:Scheel04perm}
S.~Scheel.
\newblock Permanents in linear optical networks.
\newblock 2004.
\newblock quant-ph/0508189.

\bibitem{bib:scully1993quantum}
M.~O. Scully and J.~P. Dowling.
\newblock Quantum-noise limits to matter-wave interferometry.
\newblock {\em Phys. Rev. A}, 48(4):3186, 1993.

\bibitem{bib:scully1992high}
M.~O. Scully and M.~Fleischhauer.
\newblock High-sensitivity magnetometer based on index-enhanced media.
\newblock {\em Phys. Rev. Lett.}, 69(9):1360, 1992.

\bibitem{bib:Sesh15}
K.~P. Seshadreesan, J.~P. Olson, K.~R. Motes, P.~P. Rohde, and J.~P. Dowling.
\newblock Boson sampling with displaced single-photon {F}ock states versus
  single-photon-added coherent states: The quantum-classical divide and
  computational-complexity transitions in linear optics.
\newblock {\em Phys. Rev. A}, 91:022334, 2015.

\bibitem{bib:Shches16}
V.~S. Shchesnovich.
\newblock Universality of generalized bunching and efficient assessment of
  boson sampling.
\newblock {\em Phys. Rev. Lett.}, 2016.

\bibitem{bib:ShorAlg}
P.~Shor.
\newblock Polynomial-time algorithms for prime factorization and discrete
  logarithms on a quantum computer.
\newblock {\em SIAM J. Comput.}, 26(5):1484–1509, 1997.

\bibitem{bib:spagnolo2014experimental}
N.~Spagnolo, C.~Vitelli, M.~Bentivegna, D.~J. Brod, A.~Crespi, F.~Flamini,
  S.~Giacomini, G.~Milani, R.~Ramponi, and P.~Mataloni.
\newblock Experimental validation of photonic boson sampling.
\newblock {\em Nature Phot.}, 8:615, 2014.

\bibitem{bib:spring2013boson}
J.~B. Spring, B.~J. Metcalf, P.~C. Humphreys, W.~S. Kolthammer, X.~Jin,
  M.~Barbieri, A.~Datta, N.~Thomas-Peter, N.~K. Langford, D.~Kundys, J.~C.
  Gates, B.~J. Smith, P.~G.~R. Smith, and I.~A. Walmsley.
\newblock Boson sampling on a photonic chip.
\newblock {\em Science}, 339(6121):798--801, 2013.

\bibitem{bib:Stockmeyer76}
L.~J. Stockmeyer.
\newblock The polynomial-time hierarchy.
\newblock {\em Theor. Comp. Sci.}, 3:1--22, 1976.

\bibitem{bib:Toda91}
S.~Toda.
\newblock {PP} is as hard as the polynomial-time hierarchy.
\newblock {\em SIAM J. Comput.}, 20(5):865--877, 1991.

\bibitem{bib:Valiant79}
L.~G. Valiant.
\newblock The complexity of computing the permanent.
\newblock {\em Theor. Comp. Sci.}, 8:189, 1979.

\bibitem{bib:vanmeter2007general}
N.~M. VanMeter, P.~Lougovski, D.~B. Uskov, K.~Kieling, J.~Eisert, and J.~P.
  Dowling.
\newblock General linear-optical quantum state generation scheme: Applications
  to maximally path-entangled states.
\newblock {\em Phys. Rev. A}, 76:063808, 2007.

\bibitem{bib:Wig1932}
E.~P. Wigner.
\newblock On the quantum correction for thermodynamic equilibrium.
\newblock {\em Phys. Rev.}, 40:749, 1932.

\bibitem{bib:Wilde}
M.~M. Wilde.
\newblock {\em Quantum information theory}.
\newblock Cambridge University Press, 2013.

\bibitem{WLD07}
C.~F. Wildfeuer, A.~P. Lund, and J.~P. Dowling.
\newblock Strong violations of {B}ell-type inequalities for path-entangled
  number states.
\newblock {\em Phys. Rev. A}, 76:052101, Nov 2007.

\bibitem{bib:yuen1986generation}
H.~P. Yuen.
\newblock Generation, detection, and application of high-intensity
  photon-number-eigenstate fields.
\newblock {\em Phys. Rev. Lett.}, 56:2176, 1986.

\bibitem{bib:yurke1986input}
B.~Yurke.
\newblock Input states for enhancement of fermion interferometer sensitivity.
\newblock {\em Phys. Rev. Lett}, 56:1515, 1986.

\bibitem{bib:yurtsever2003interferometry}
U.~Yurtsever, D.~Strekalov, and J.P. Dowling.
\newblock Interferometry with entangled atoms.
\newblock {\em Euro. Phys. J. D}, 22:365, 2003.

\bibitem{Zavatta_04}
A.~Zavatta, S.~Viciani, and M.~Bellini.
\newblock Quantum-to-classical transition with single-photon-added coherent
  states of light.
\newblock {\em Science}, 306(5696):660--662, 2004.

\bibitem{Zavatta_05}
A.~Zavatta, S.~Viciani, and M.~Bellini.
\newblock Single-photon excitation of a coherent state: {C}atching the
  elementary step of stimulated light emission.
\newblock {\em Phys. Rev. A}, 72:023820, Aug 2005.

\end{thebibliography}
\pagebreak

\makeatletter
\def\@makechapterhead{\oldmake}
\def\@makeschapterhead{\oldmakes}
\makeatother

\singlespacing
\addtocontents{toc}{\vspace{12pt} \hspace{-1.8em} APPENDIX \vspace{-1em}}
\appendix
\chapter{Reuse and Permissions}
\vspace{0.5em}
\doublespacing
\url{http://journals.aps.org/copyrightFAQ.html} 

As the author of an APS-published article, may I
include my article or a portion of my article in my
thesis or dissertation?

Yes, the author has the right to use the article or a portion of the article
in a thesis or dissertation without requesting permission from APS,
provided the bibliographic citation and the APS copyright credit line are
given on the appropriate pages.

\pagebreak

\singlespacing
\chapter{Derivations}
\vspace{0.5em}
\doublespacing
\section*{Proof of $U_{j,k}^{(n)}$} \label{app:Ujk}

Beginning from Eq.~(\ref{eq:U}) and setting $\hat{\Theta}=\hat{I}$,
\begin{eqnarray}
U_{j,k}^{(n)} &=& (\hat{V}\hat{\Phi}\hat{V^{\dag}})_{j,k} \nonumber \\ 
&=& \sum_{l,m=1}^{n}V_{j,l}\Phi_{l,m}V_{m,k}^{\dag} \nonumber \\
&=& \sum_{l,m=1}^{n} \underbrace{\frac{e^{- 2 i j l \pi/n}}{\sqrt{n}}}_{V_{j,l}}\underbrace{\delta_{l,m}e^{i(l-1)\varphi}}_{\Phi_{l,m}}\underbrace{\frac{e^{2 i m k \pi/n}}{\sqrt{n}}}_{V_{m,k}^{\dag}} \nonumber \\
&=& \frac{1}{n} \sum_{l=1}^{n} e^{\frac{- 2 i j l \pi}{n}} e^{i (l-1) \varphi} e^{\frac{2 i l k \pi}{n}}\nonumber \\
&=& \frac{1}{n} \sum_{l=1}^{n} e^{\frac{2 i l(k-j) \pi}{n} + i(l-1)\varphi} \nonumber \\
&=& e^{\frac{2 i (k-j) \pi}{n}}\frac{1}{n} \sum_{l=0}^{n-1} (e^{\frac{2 i (k-j) \pi}{n} + i\varphi})^l. \nonumber 
\end{eqnarray}
From the geometric series, it follows,
\begin{eqnarray}
U_{j,k}^{(n)}&=& \frac{1}{n(e^{\frac{2 i (j-k) \pi}{n}})}\frac{1-e^{i n\varphi}}{\left(1-e^{\frac{2 i (k-j) \pi}{n} +i \varphi}\right)}, \label{eq:U} \nonumber \\
&=& \frac{1-e^{i n\varphi}}{n\left(e^{\frac{2 i \pi(j-k)}{n}}-e^{i \varphi}\right)}
\end{eqnarray}
which is what we set out to prove.
which is Eq.~(\ref{eq:Ujk}) that we set out to prove, where the last line follows from the geometric series.

\section*{Conjecture for the Analytic Form of Per($\hat U^{(n)}$)} \label{app:series}

Our goal is to find the analytic form for Per($\hat U^{(n)}$) where $U_{j,k}^{(n)}$ is as in Eq.~(\ref{eq:U}).  We can perform a brute force calculation to obtain the analytic form for small $n$.  Doing so up to $n=6$ yields:
\begin{table}[h]
\centering
\begin{tabular}{|c|c|}

\hline
 $n$ & Per$(\hat U^{(n)})$ \nonumber \\
 \hline
 1 & 1 \nonumber \\
 2 & $e^{i \phi } \cos (\phi )$ \\
 3 & $\ \frac{1}{9} \left(2+e^{3 i \phi   }\right) \left(1+2 e^{3 i \phi   }\right)$ \\
 4 & $\frac{1}{32} \left(1+e^{4 i \phi   }\right) \left(3+e^{4 i \phi }\right)    \left(1+3 e^{4 i \phi }\right)$ \\
 5 & $\frac{1}{625} \left(4+e^{5 i \phi
   }\right) \left(3+2 e^{5 i \phi   }\right) \left(2+3 e^{5 i \phi   }\right) \left(1+4 e^{5 i \phi   }\right)$ \\
 6 & $\frac{1}{648} \left(1+e^{6 i \phi    }\right) \left(2+e^{6 i \phi }\right)    \left(5+e^{6 i \phi }\right)    \left(1+2 e^{6 i \phi }\right)    \left(1+5 e^{6 i \phi }\right)$ 
 \\
 \hline
\end{tabular}
\end{table} 
One can see the pattern that emerges is of the form:
\begin{equation} \label{eq:permU2} 
\mathrm{Per}(\hat{U}^{(n)})= \frac{1}{n^{n-1}}\prod_{j=1}^{n-1}\Big[je^{i n \varphi}+n-j\Big],
\end{equation}
which is Eq.~(\ref{eq:permU}) that we set out to show. This equation has been verified analytically up to $n=16$ and up to $n=25$ numerically..

\section*{Calculation of $P$} \label{app:P}
Assuming our conjecture in Eq.~(\ref{eq:permU}) holds, we can compute the coincidence probability of measuring one photon in each mode at the output,
\begin{eqnarray}
P &=& \big|\mathrm{Perm}(U^{(n)})\big|^2 \nonumber \\
&=& \left|\frac{1}{n^{n-1}} \prod_{j=1}^{n-1}\left(je^{i n \varphi}+n-j\right)\right|^2 \nonumber \\
&=& \frac{1}{n^{2n-2}} \prod_{j=1}^{n-1}\Big|\left(je^{i n \varphi}+n-j\right)\Big|^2 \nonumber \\
&=& \frac{1}{n^{2n-2}} \prod_{j=1}^{n-1}\Big|j\mathrm{cos}(n\varphi)+ij\mathrm{sin}(n\varphi)+n-j\Big|^2 \nonumber \\
&=& \frac{1}{n^{2n-2}} \prod_{j=1}^{n-1}\Big|\underbrace{j\mathrm{cos}(n\varphi)+(n-j)}_\mathrm{Re}+i\underbrace{j\mathrm{sin}(n\varphi)}_\mathrm{Im}\Big|^2. \nonumber \\
\end{eqnarray} 
Invoking the property that \mbox{$|z|^2= \mathrm{Re}(z)^2+\mathrm{Im}(z)^2$}, where \mbox{$z\in \mathbb{C}$}, 
\begin{eqnarray} \label{eq:Pproof}
P &=& \frac{1}{n^{2n-2}} \prod_{j=1}^{n-1} \Big[\big(j\mathrm{cos}(n\varphi)+(n-j)\big)^2+j^2\mathrm{sin}^2(n\varphi)\Big] \nonumber \\
&=& \frac{1}{n^{2n-2}} \prod_{j=1}^{n-1} \Big[\underbrace{j^2\mathrm{cos}^2(n\varphi)+j^2\mathrm{sin}^2(n\varphi)}_{=j^2} \nonumber \\
&+& 2j(n-j)\mathrm{cos}(n\varphi)+(n-j)^2 \Big] \nonumber \\
&=& \frac{1}{n^{2n-2}} \prod_{j=1}^{n-1} \Big[j^2 + 2j(n-j)\mathrm{cos}(n\varphi)+(n-j)^2 \Big]\nonumber \\
&=& \frac{1}{n^{2n-2}} \prod_{j=1}^{n-1} \Big[\underbrace{2j(n-j)}_{a_n(j)}\mathrm{cos}(n\varphi)+\underbrace{n^2-2jn+2j^2}_{b_n(j)} \Big] \nonumber \\
&=& \frac{1}{n^{2n-2}} \prod_{j=1}^{n-1} \Big[a_n(j)\mathrm{cos}(n\varphi)+b_n(j) \Big], \nonumber \\
\end{eqnarray} 
which is Eq.~(\ref{eq:P_Result}) that we set out to show.

\section*{Calculation of $\left|\frac{\partial P}{\partial \varphi}\right|$} \label{app:dP}

From Eq.~(\ref{eq:Pproof}), exploiting the logarithm product rule,
\begin{eqnarray}
\mathrm{ln}(P) &=& \underbrace{\mathrm{ln}\left(\frac{1}{n^{2n-2}}\right)}_{C} + \mathrm{ln}\left(\prod_{j=1}^{n-1} \Big[a_n(j)\mathrm{cos}(n\varphi) + b_n(j) \Big] \right) \nonumber \\
&=& C + \sum_{j=1}^{n-1} \mathrm{ln}\Big[a_n(j)\mathrm{cos}(n\varphi)+b_n(j) \Big],
\end{eqnarray} 
where $C$ is a constant. Now the derivative becomes, 
\begin{eqnarray}
\frac{1}{P}\frac{\partial P}{\partial \varphi} &=& -\sum_{j=1}^{n-1} \frac{na_n(j)\mathrm{sin}(n\varphi)}{a_n(j)\mathrm{cos}(n\varphi)+b_n(j)} \nonumber \\
\frac{\partial P}{\partial \varphi} &=& -nP\mathrm{sin}(n\varphi)\sum_{j=1}^{n-1} \frac{a_n(j)}{a_n(j)\mathrm{cos}(n\varphi)+b_n(j)}.\nonumber \\
\end{eqnarray} 
Thus,
\begin{equation}
\left|\frac{\partial P}{\partial\varphi}\right|=nP\big|\mathrm{sin}(n\varphi)\big|\sum_{j=1}^{n-1} \left|\frac{a_n(j)}{a_n(j)\mathrm{cos}(n\varphi)+b_n(j)}\right|,
\end{equation}
which is Eq.~(\ref{eq:dP}) that we set out to show.

\section*{Calculation of $\Delta\varphi$ in the small angle approx.} \label{app:dphi}
We wish to compute $\Delta\varphi$ in the limit that $n\varphi\ll1$. Then $P$ in the small angle regime of Eq.~(\ref{eq:P_Result}) becomes,
\begin{eqnarray}
\label{eq:papprox}
P&\approx& \frac{1}{n^{2n-2}}\prod_{j=1}^{n-1} \bigg[a_n(j)\Big(1-\frac{1}{2}(n\varphi)^2\Big)+b_n(j)\bigg] \nonumber \\ 
&=& \frac{1}{n^{2n-2}}\prod_{j=1}^{n-1}\bigg[ n^2-(nj-j^2)n^2\varphi^2 \bigg]\nonumber \\ 
&=& \prod_{j=1}^{n-1}\Big[1-(nj-j^2)\varphi^2\Big],
\end{eqnarray}
where $\cos(n\varphi)$ is expanded to the first nonconstant term in its Taylor series. This product has the form of a binomial expansion. Dropping terms above order $\varphi^2$, $P$ reduces to,
\begin{eqnarray}
\label{eq:pfinal}
P&\approx& 1-\varphi^2\sum_{j=1}^{n-1}\Big[nj-j^2\Big] \nonumber \\
&=& 1-\varphi^2\Big[\frac{1}{6}(n-1)n(n+1)\Big] \nonumber \\
&=&1-k(n)\varphi^2,
\end{eqnarray}
where $k(n)=\frac{1}{6}n(n-1)(n+1)\geq0$ $\forall$ $n\geq1$.  From Eq.~(\ref{eq:pfinal}) we can easily compute $P^2$ and $\big|\frac{\partial P}{\partial \varphi}\big|$ to be,
\begin{eqnarray}
P^2&\approx& 1-2k(n)\varphi^2 \\
\left|\frac{\partial P}{\partial \varphi}\right|&=&2k(n)|\varphi|,
\end{eqnarray}
where we have again dropped terms above order $\varphi^2$. Using Eq.~(\ref{eq:phaseSenP}) the phase sensitivity $\Delta\varphi$ in the small angle regime is,
\begin{eqnarray}
\Delta\varphi &=& \frac{\sqrt{P-P^2}}{\left|\frac{\partial P}{\partial\varphi}\right|} \nonumber \\
&=&\frac{\sqrt{\Big(1-k(n)\varphi^2\Big)-\Big(1-2k(n)\varphi^2\Big)}}{2k(n)|\varphi|} \nonumber \\
&=&\frac{\sqrt{k(n)\varphi^2}}{2k(n)|\varphi|} \nonumber \\
&=& \frac{1}{2\sqrt{k(n)}} \nonumber \\
&=&\sqrt{\frac{3}{2{(n-1)n(n+1)}}},
\end{eqnarray}
which is Eq.~(\ref{eq:DeltaVarPhi}) that we set out to show.

\section*{Discussion of Ordinal Resource Counting (ORC)} \label{app:counting}

We would like to compare the performance of MORDOR to an equivalent multimode interferometer baseline for which we will construct the shotnoise limit (SNL) and Heisenberg limit (HL). This is a subtle comparison, due to the linearly increasing unknown phase-shifts, \{\mbox{$0,\varphi,\dots,(n-1)\varphi$}\}, that MORDOR requires to operate. There is a long and muddled history of increasing the interrogation time (or here length) of the probe particles with the unknown phase-shift followed by an incorrect reckoning of the true resources. Here we shall discuss a protocol described in Ref.~\cite{bib:Mordor2015} called Ordinal Resource Counting (ORC) whereby all resources, such as number of `calls' to the phase-shifter $\varphi$, are converted to the `currency' of the resource that is most precious to us, namely photon-number.

\begin{figure}[!htb]
\centering
\includegraphics[scale=0.9]{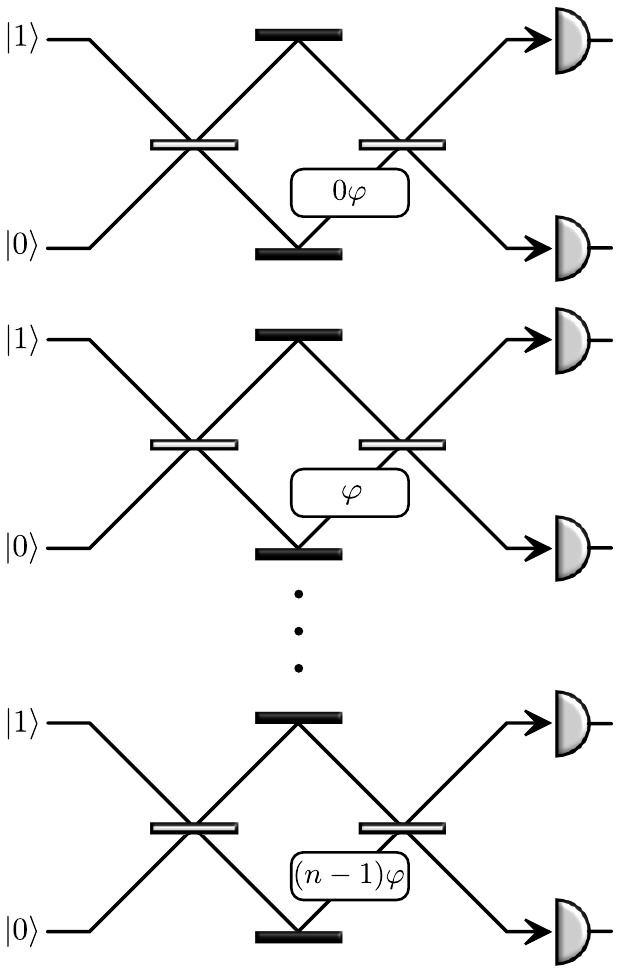}
\caption{$n$ instances of two-mode Mach-Zehnder interferometers, with a linearly increasing phase gradient. This system has the same configuration of phases as MORDOR, but the photons are not allowed to interfere, and thus has minimal number-path entanglement.} \label{fig:resources1}
\end{figure}

First we must construct a multimode interferometer with $n$ photon inputs that provides the baseline if the photons remain uncorrelated and the number-path entanglement remains minimal. Such a comparator is shown in Figure \ref{fig:resources1}, and consists of $n$, two-mode Mach-Zehnder Interferometers (MZI) in a vertical cascade, fed with single-photon inputs, with the same linearly increasing unknown phase-shift sequence as MORDOR. Since the MZIs are disconnected, the number-path entanglement remains constant and minimal, and of the form \mbox{$(\ket{1,0}+\ket{0,1})/\sqrt{2}$} inside each MZI. 

\begin{figure}[!htb]
\centering
\includegraphics[scale=0.9]{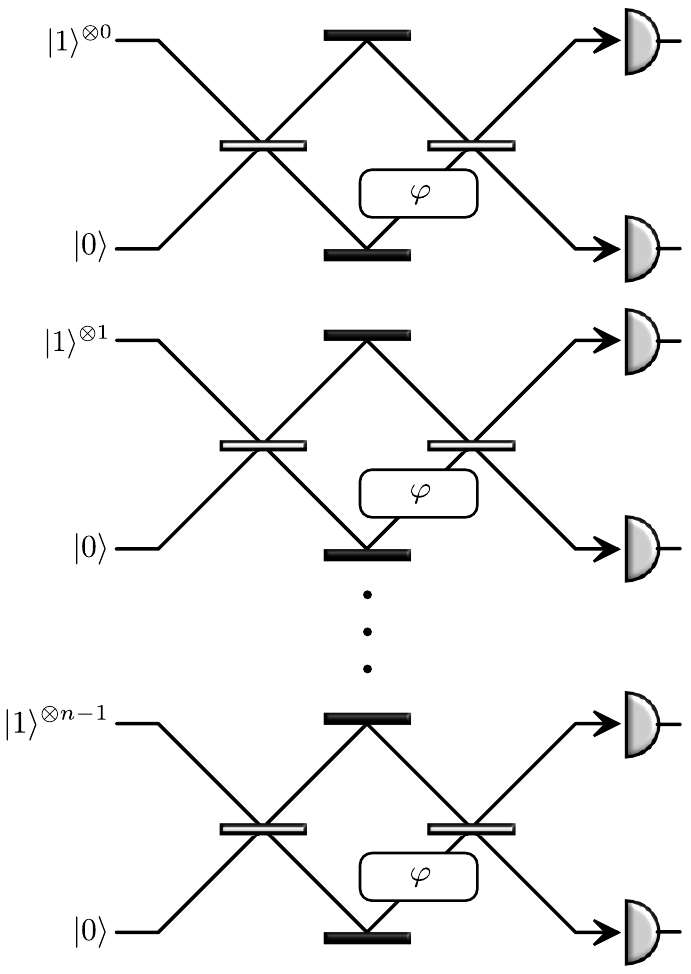}
\caption{Noting that a single photon interrogating a phase-shift of $n\varphi$ is equivalent to $n$ independent interrogations of $\varphi$, Figure \ref{fig:resources1} can be represented in terms of the resource of photons as shown here. Here $\ket{1}^{\otimes j}$ means that $j$ independent (i.e distinguishable) photons have been prepared.} \label{fig:resources2}
\end{figure}

Now to convert the linearly increasing interrogation lengths of the unknown phase-shifts, we note that a single photon interrogating a phase-shift of say $2\varphi$ is equivalent to a single photon interrogating a single phase-shift $\varphi$ twice, which is in turn equivalent to two uncorrelated photons entering the same port of the MZI containing a single phase-shift of $\varphi$. In this way we may convert `number of interrogations of the phase-shifter' into the currency of `number of photons' to carry out a fair reckoning of the resources. Following this logic we are led to Figure \ref{fig:resources2} showing a cascade of MZIs where the linearly increasing phase-shifters are replaced with a single phase-shifter of $\varphi$ and the single photons at the MZI inputs are replaced with a linearly increasing number of photons. Then the `number of interrogations of the phase-shifter' becomes $n(n-1)/2$, but there is an additional photon that is part of the MORDOR resources so our total number of resources becomes,
\begin{equation} \label{eq:N}
N\equiv1+\frac{n(n-1)}{2}.
\end{equation}

\begin{figure}[!htb]
\centering
\includegraphics[scale=1.1]{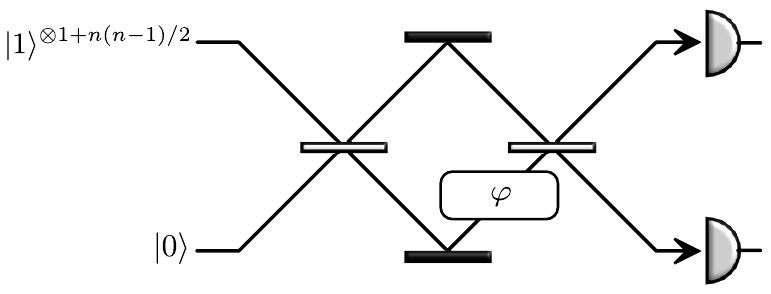}
\caption{Grouping all the independent interferometers in Figure \ref{fig:resources2} together and including the extra photon from MORDOR, we obtain a single MZI with \mbox{$1+n(n-1)/2$} independent photons as input. This configuration achieves the shotnoise limit, and thus provides a benchmark for comparing MORDOR against the shotnoise and Heisenberg limits, with photons as the resource being counted.} \label{fig:resources3}
\end{figure}

Next we note that this cascade of $n$ MZIs in Figure \ref{fig:resources2} may be replaced with a single MZI, shown in Figure \ref{fig:resources3}, where the input is now an ordinal grouped ranking of the uncorrelated photons following the same pattern as in Figure \ref{fig:resources2}. Hence in the configuration in Figure \ref{fig:resources3} we have a single MZI with vacuum entering the lower port, a stream of $N$ uncorrelated photons entering the upper port, and a single phase-shifter $\varphi$ between the beamsplitters. It is well-known that for this configuration the sensitivity of this system scales as the SNL \cite{bib:scully1993quantum, bib:dowling1998correlated}, namely,
\begin{equation}
\Delta\varphi_\mathrm{SNL} = \frac{1}{\sqrt{N}} = \frac{1}{\sqrt{1+\frac{n(n-1)}{2}}}.
\end{equation} 
This then provides the scaling used in Ref.~\cite{bib:Mordor2015} to construct the SNL for comparison to MORDOR.

Finally, if instead we were to maximally path-number entangle these resources into a NOON state of the form \mbox{$(\ket{N,0}+\ket{0,N})/\sqrt{2}$} (just to the right of the first beam splitter but before the phase-shifter) the sensitivity then becomes Heisenberg limited,
\begin{equation}
\Delta\varphi_\mathrm{HL} = \frac{1}{N} = \frac{1}{1+\frac{n(n-1)}{2}},
\end{equation}
which is a sensitivity known to saturate the Quantum Cram{\'e}r-Rao Bound (CRB) for sensitivity in local phase estimation with $N$ photons \cite{bib:lee2002quantum, bib:durkin2007local}. As the CRB is the best one may do, according to the laws of quantum mechanics, then in this case the HL is optimal. As discussed, the performance of MORDOR falls between the SNL and the HL, but with the feature of not having to do anything resource intensive such as preparing a high-NOON state. 

In Ref.~\cite{bib:Mordor2015}, it was stated that this provided the fairest comparison of sensitivity performance of MORDOR with such ambiguities such as how to handle `number of calls to the phase-shifter' removed by replacing such a notion with `number of photons' inputted into the interferometer.  While it may be the case that ORC correctly computes the HL, it appears upon further analysis that this may not be the case for the SNL.  For example, suppose for $n=2$ (i.e. an MZI) that $\varphi$ is replaced by $2\varphi$ in the interferometer, and we wish to compute the phase sensitivity $\Delta\varphi$.  ORC predicts that an equivalent number of resources for MORDOR should be $N=3$, and so the SNL corresponds to a phase sensitivity of $1/\sqrt{3}$.

On the other hand, for a single experimental run of the MZI with only a single photon, one can see that the phase sensitivity corresponds to $1/\sqrt{2}$.  Once we take into account an additional experimental run (since we still have a second photon), we see that $\Delta\varphi=1/(\sqrt{2}\cdot\sqrt{2})=1/\sqrt{4}$, which suggests that $N=4$ is the correct equivalence.  The SNL plotted within this thesis (relative to the MORDOR architecture)  is derived from the classical limit calculation in Ref.~\cite{bib:GLM06}, i.e. the SNL scales as $1/\sqrt{N}$ where $N=\sum_{i=0}^{n-1}i^2=\frac{1}{6}n(n-1)(2n-1)$.

\section*{Efficiency} \label{app:efficiency}

In the presence of inefficient photon sources and photo-detectors the success probability of the protocol will drop exponentially with the number of photons. Specifically, if $\eta_s$ and $\eta_d$ are the source and detection efficiencies respectively, the success probability of the protocol is $\eta = (\eta_s \eta_d)^n$. Current cutting edge transition edge detectors operate at 98\% efficiency, with negligible dark count \cite{bib:fukuda2011titanium}. SPDC sources are the standard photon-source technology but they are non-deterministic. However, there are techniques that can greatly improve the heralding efficiency up to 42\% at 2.1 MHz \cite{bib:LPOR201400404}. Also, other source technologies, such as quantum dot sources are becoming viable with efficiencies also up to 42\% \cite{bib:Maier14}. For $n=10$, which is already well beyond current experiments, this yields $\eta= (0.98*0.42)^{10} \approx 0.00014$, which is about 300 successful experimental runs per second when operating with 2.1 MHz sources. 

\section*{Dephasing} \label{app:dephasing}

A form of decoherence to consider is dephasing. Dephasing in our work may be modelled with the result of Bardhan \emph{et al.} \cite{bib:Bardhan2013}, whereby dephasing occurs on each mode separately. When considering our example of a magnetometer, dephasing would occur in the magnetic field cells where atomic fluctuations may occur that differ between cells.  In the rest of the interferometer, dephasing can be made very close to zero, particularly on an all optical chip. 

\begin{figure}[t]
\centering
\includegraphics[scale=0.5]{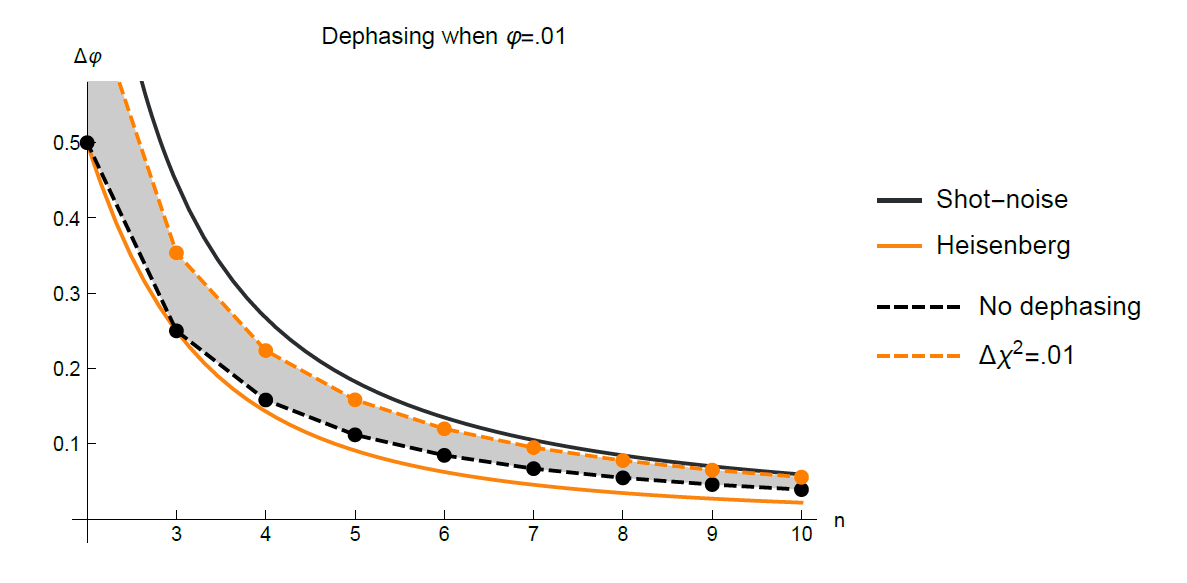}
\caption{Dephasing for $\varphi=0.01$.  The shaded region represents the phase sensitivity for MORDOR where $0\leq\chi\leq 0.01$.} \label{fig:dephasing}
\end{figure}

\begin{figure}[h]
\centering
\includegraphics[scale=0.55]{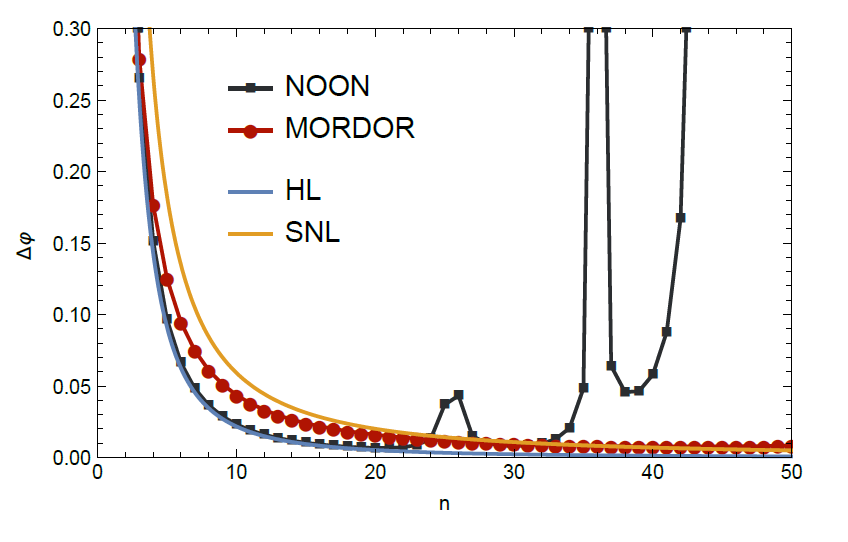}
\caption{The effect of dephasing on the NOON state and MORDOR where $\varphi=0.01, \chi=0.005$.  The NOON state is plotted with respect to $N$ for fair resource counting.} \label{fig:dephasingNOON}
\end{figure}

To model dephasing we investigate a random phase shift $\Delta\chi$ added to each mode separately. $\Delta\chi$ is a Gaussian random variable of zero mean but nonzero second order moment. The phase shift in the $j$th mode then becomes,
\begin{eqnarray}
e^{\pm i j \varphi}&\rightarrow& e^{\pm i j (\varphi + \Delta\chi)} \nonumber \\
&=& e^{\pm i j \varphi}e^{\pm i j \Delta\chi} \nonumber \\
&=& e^{\pm i j \varphi}\left(1 \pm i j\Delta\chi - \frac{1}{2}j\Delta\chi^2\pm\dots\right).
\end{eqnarray}
Using $\ip{\Delta\chi}=0$, $\ip{\Delta\chi^2}\neq0$, and that $\Delta\chi \ll \phi$ we simplify this to be, 
\begin{eqnarray}
e^{\pm i j \varphi}&\rightarrow& e^{\pm i j \varphi}\left(1- \frac{1}{2}j\Delta\chi^2 \pm\dots\right) \nonumber \\
&\approx& e^{\pm i j \varphi}e^{-\frac{1}{2} j^2 \Delta\chi^2}.
\end{eqnarray}

The signal $P$ in Eq. 10 from our work then changes in the presence of dephasing. The dependence that $P$ has on the unknown phase $\varphi$ does not depend on the mode number $j$. Then the term that depends on $\varphi$ becomes,
\begin{eqnarray}
\mathrm{cos}(n\phi) &=& \frac{1}{2}\left(e^{i n \varphi}+e^{-i n \varphi}\right) \nonumber \\
&\rightarrow& \frac{1}{2}\left(e^{i n \phi}+e^{- i n \phi}\right)e^{-\frac{�1}{2}n^2\Delta\chi^2} \nonumber \\
&=& \mathrm{cos}(n\phi)e^{-\frac{�1}{2}n^2\Delta\chi^2}
\end{eqnarray}
Using this substitution $P$ becomes,
\begin{eqnarray} \label{}
P &=& \Big|\mathrm{Per}(\hat{U}^{(n)})\Big|^2 \nonumber \\
&=& \frac{1}{n^{2n-2}}\prod_{j=1}^{n-1} \Big[a_n(j)\mathrm{cos}(n\phi)e^{-\frac{�1}{2}n^2\Delta\chi^2}+b_n(j) \Big].
\end{eqnarray} 
The factor $e^{-\frac{1}{2}n^2\Delta\chi^2}$ can be absorbed into $a_n(j)$ so that the derivation of $|\frac{\partial P}{\partial\phi}|$ in Eq.~(\ref{eq:dP}) is identical.
Using this result we numerically plot the phase sensitivity with dephasing in Figure \ref{fig:dephasing}.

In order to meaningfully analyze the dephased sensitivity, we would like to compare with other well known metrological schemes.  In Figure \ref{fig:dephasingNOON}, we compare MORDOR to the NOON state (with $N$ input photons for a fair resource comparison) and see that MORDOR is far more robust against dephasing.

\section*{Entries of $U_{ij}$} \label{sec:Uentries}
Consider a linear optical network similar to the MORDOR protocol, except where the original phase gradient $\hat{\Phi}$ has been replaced by a single unknown phase shift $\varphi$ in the uppermost arm, together with no phase shift in the other arms.  We denote this operator by $\hat{\mathbf{X}}^{(n)}$, whose matrix form is given by,
\begin{equation}
[\hat{\mathbf{X}}^{(n)}]_{j,k}\equiv X_{j,k}=\delta_{j,k}(e^{i\varphi})^{\delta_{j,1}},
\end{equation}
i.e., $\hat{\mathbf{X}}^{(n)}$ is the identity operator $\hat{I}_{n}$ with only the $(1,1)$ entry replaced by $e^{i\varphi}$.  Analogous to the MORDOR protocol, we also choose the control phase $\hat{\Theta}^{(n)}$ to be of the same form, 
\begin{equation}
[\hat{\Theta}^{(n)}]_{j,k}\equiv X_{j,k}=\delta_{j,k}(e^{-i\theta})^{\delta_{j,1}},
\end{equation}
though for simplicity of the proof we may assume $\theta=0$ and thus $\hat{\Theta}^{(n)}=\hat{I}_n$.  We drop the superscript $n$ on most operators when it is clear from context that all operators have the same index.

We now compute the matrix entries of the entire network, $\hat{U}^{(n)}=\hat{V}\hat{\mathbf{X}}\hat{V}^\dag$.  
\begin{eqnarray}
U_{j,k}^{(n)} &=& (\hat{V}\hat{X}\hat{V^{\dag}})_{j,k} \nonumber \\ 
&=& \sum_{l,m=1}^{n}V_{j,l}X_{l,m}V_{m,k}^{\dag} \nonumber \\
&=& \sum_{l,m=1}^{n} \underbrace{\frac{1}{\sqrt{n}}\omega_n^{(j-1)(l-1)}}_{V_{j,l}}\underbrace{\delta_{l,m}e^{i\varphi\delta_{l,1}}}_{X_{l,m}}\underbrace{\frac{1}{\sqrt{n}}\omega_n^{(m-1)(1-k)}}_{V_{m,k}^{\dag}} \nonumber 
\end{eqnarray}
\begin{eqnarray}
&=& \frac{1}{n}\Big[e^{i\varphi}+\sum_{l=2}^{n}\omega_{n}^{(j-1)(l-1)}\omega_n^{(l-1)(1-k)}\Big] \nonumber \\
&=& \frac{1}{n}\Big[e^{i\varphi}+\sum_{l=2}^{n}(\omega_{n}^{(j-k)})^{(l-1)}\Big] \nonumber \\
&=& \frac{1}{n}\Big[ e^{i\varphi}+\sum_{l=1}^{n-1}(\omega_{n}^{(j-k)})^{l}\Big]. \label{eq:sum}
\end{eqnarray}
\begin{eqnarray}
&=& 
\begin{cases}
\displaystyle\frac{1}{n}\Big[ e^{i\varphi}+n-1] & j=k \\ \\
\displaystyle\frac{1}{n}\Big[ e^{i\varphi}-1 \Big] & j\neq k 
\end{cases} \nonumber \\
&=& \frac{1}{n} \Big[e^{i\varphi}+(\delta_{j,k})n-1 \Big]. \label{eq:entries} 
\end{eqnarray}
For $j=k$, it is easy to see the sum in Eq.~(\ref{eq:sum}) should be $n-1$ since each term is simply $1^l=1$.  For $j\neq k$, the result follows from the fact that the sum of all $n^{th}$ roots of unity is zero, i.e.,
\begin{equation}
0=\sum_{l=1}^{n}\omega_n^l.\label{eq:sumroots}
\end{equation}
The proof for the above follows directly from the geometric series, and it easy to see that it extends to a sum over $\omega_n^{(j-k)}$ as well.

\section*{Permanent of $U$} \label{sec:permU}
The permanent of $\hat{U}^{(n)}$ is, by definition,
\begin{eqnarray}
\textrm{perm}(\hat{U}) &=& \sum_{\sigma\in S_n}\prod_{j=1}^n \frac{1}{n} \Big[e^{i\varphi}+(\delta_{j,\sigma(j)})n-1 \Big] \nonumber \\
&=&  \frac{1}{n^n}\sum_{\sigma\in S_n}\prod_{j=1}^n \Big[e^{i\varphi}+(\delta_{j,\sigma(j)})n-1 \Big]. \label{eq:permsp}
\end{eqnarray}
Suppose $\sigma_k$ is some permutation with $k$ fixed points, recalling that a \textit{fixed point} of a permutation is a value $j\in\{1,..,n\}$ such that $\sigma(j)=j$ (also referred to as a \textit{partial derangement}).  Then the product $\prod_{j=1}^n$ in Eq.~(\ref{eq:permsp}) corresponding to $\sigma_k$ is,
\begin{equation}
\prod_{j=1}^n \Big[e^{i\varphi}+(\delta_{j,\sigma_k(j)})n-1 \Big]=[e^{i\varphi}+n-1]^k[e^{i\varphi}-1]^{n-k}
\end{equation}
The sum in Eq.~(\ref{eq:permsp}) can thus be rewritten in terms of a sum over the number of fixed points in a permutation, whose coefficient $D_{n,k}$ enumerates all permutations in $S_n$ with $k$ fixed points.  The quantity $D_{n,k}$  is referred to as the \textit{rencontres numbers}, where,
\begin{equation}
D_{n,k}=\frac{n!}{k!}\sum_{j=0}^{n-k}\frac{(-1)^j}{j!}.
\end{equation}
The permanent is thus,
\begin{equation}
\textrm{perm}(\hat{U}) = \frac{1}{n^n}\sum_{k=0}^n D_{n,k} [e^{i\varphi}+n-1]^k[e^{i\varphi}-1]^{n-k}. \label{eq:permfinal}
\end{equation}
\section*{Calculation of $\Delta\varphi$} \label{sec:deltaphi}
We are mostly interested in the behavior of $\textrm{perm}(\hat{U})$ for small $\varphi$, where the phase sensitivity is optimal.  To simplify the remaining calculations, we focus our attention on the Taylor expansion of $F_n[\varphi]=\textrm{perm}(\hat{U}^{(n)})$ up to second order,
\begin{equation}
F_n[\varphi]\approx F_n[0]+F_n'[0]\varphi+\frac{1}{2}F_n''[0]\varphi^2.\label{eq:taylor}
\end{equation}
We can find $F_n[0]$ easily by noting that, because of the product with $[e^{i\varphi}-1]^{n-k}$ the only non-zero term in Eq.~(\ref{eq:permfinal}) corresponds to $k=n$,
\begin{equation}
F_n[0]=\frac{1}{n^n}D_{n,n}[1+n-1]^n=\frac{1}{n^n}\cdot1\cdot[n]^n=1. 
\end{equation}
Similarly, the only non-zero terms in $F_n'[0]$ must be derivatives of either $k=n$ or $k=n-1$.  Since $D_{n,n-1}=0$, we need only concern ourselves with the derivative of the $k=n$ term.  Applying the chain rule,
\begin{eqnarray}
F_n'[0] &=& \Bigg[\frac{1}{n^n}D_{n,n}[e^{i\varphi}+n-1]^n\Bigg]'_{\varphi=0} \nonumber \\
&=& \Bigg[\frac{1}{n^n}D_{n,n}n[e^{i\varphi}+n-1]^{n-1}ie^{i\varphi}\Bigg]_{\varphi=0} \label{eq:der} \\
&=& \Bigg[\frac{1}{n^n}\cdot1\cdot n[1+n-1]^{n-1}\cdot i\Bigg]=\frac{n^n}{n^n}\cdot i \nonumber \\
&=& i.
\end{eqnarray}
Evaluating $F_n''[0]$ is only marginally more difficult.  The $k=n$ term can be evaluated by straightforward application of the product rule to Eq.~(\ref{eq:der}).  Also, although the second derivative of the $k=n-2$ term may be non-zero and contains a product, it is only so for the second derivative of $[e^{i\varphi}-1]^{2}$---the other terms originating from the product rule are zero.  Hence, $F_n''[0]$ has only three non-zero terms,
\begin{eqnarray}
& & F_n''[0]=\Bigg[\frac{1}{n^n}D_{n,n}n[e^{i\varphi}+n-1]^{n-1}ie^{i\varphi}\Bigg]'_{\varphi=0}+ \nonumber \\
& &\Bigg[\frac{1}{n^n}D_{n,n-2}[e^{i\varphi}+n-1]^{n-2}[e^{i\varphi}-1]^2\Bigg]''_{\varphi=0} \nonumber \\
& &=\Bigg[\frac{1}{n^n}D_{n,n}n(n-1)[e^{i\varphi}+n-1]^{n-2}(ie^{i\varphi})^2\Bigg]_{\varphi=0}+ \nonumber \\
& &\Bigg[\frac{1}{n^n}D_{n,n}n[e^{i\varphi}+n-1]^{n-1}(ie^{i\varphi})^2\Bigg]_{\varphi=0} + \nonumber \\
& &\Bigg[\frac{1}{n^n}D_{n,n-2}2[e^{i\varphi}+n-1]^{n-2}(ie^{i\varphi})^2\Bigg]_{\varphi=0} \nonumber \\
& &=-\Bigg[\frac{1}{n^n}(n-1)n^{n-1}\Bigg]-\Bigg[\frac{1}{n^n}n^n\Bigg]- \nonumber \\
& &\Bigg[\frac{1}{n^n}2D_{n,n-2}n^{n-2}\Bigg] \nonumber 
\end{eqnarray}
\begin{eqnarray}
& & = -\Big[\frac{n-1}{n}+1+\frac{2D_{n,n-2}}{n^2}\Big] \nonumber \\
& & = -\Big[\frac{n-1}{n}+1+\frac{n(n-1)}{n^2}\Big] \nonumber \\
& & = -\Big[\frac{2n-2}{n}+1\Big] \nonumber \\
& & = -\frac{3n-2}{n}
\end{eqnarray}
Thus, Eq.~(\ref{eq:taylor}) becomes the simple expression,
\begin{equation}
\textrm{perm}(\hat{U}^{(n)})\approx 1+i\varphi-(\frac{3n-2}{2n})\varphi^2
\end{equation}
Recall that the probability $P$ of observing $n$ photons each exit individual ports is $P=|\textrm{perm}(\hat{U}^{(n)})|^2$.  For small $\varphi$, then,
\begin{eqnarray}
P &=& \Big|1+i\varphi-(\frac{3n-2}{2n})\varphi^2\Big|^2 \nonumber \\
&=& \Big(1+i\varphi-(\frac{3n-2}{2n})\varphi^2\Big)\Big(1-i\varphi-(\frac{3n-2}{2n})\varphi^2\Big) \nonumber \\
&=& 1+i\varphi-i\varphi-2(\frac{3n-2}{2n})\varphi^2-i^2\varphi^2+O(\varphi^4) \nonumber \\
&=& 1-\frac{2n-2}{n}\varphi^2+O(\varphi^4).
\end{eqnarray}
Finally, $\Delta\varphi$ becomes,
\begin{eqnarray}
\Delta\varphi &=& \frac{\sqrt{P-P^2}}{|\frac{\partial P}{\partial \varphi}|} \nonumber \\
&=& \frac{\sqrt{1-\frac{2n-2}{n}\varphi^2-1+\frac{4n-4}{n}\varphi^2}}{\frac{4n-4}{n}\varphi} \nonumber \\
&=& \frac{\sqrt{\frac{2n-2}{n}\varphi^2}}{2\cdot\frac{2n-2}{n}\varphi} \nonumber \\
\Delta\varphi &=& \frac{1}{2\sqrt{2}\cdot\sqrt{\frac{n-1}{n}}}.
\end{eqnarray}
Tha ratio between $\Delta\varphi$ and the shotnoise-limited phase sensitivity for $n$ photons is then,
\begin{equation}
\frac{\Delta\varphi}{\sqrt{n}}=\frac{\sqrt{8(n-1)}}{n}
\end{equation}
which is greater than one (i.e. gives an advantage over shotnoise) for $2\leq n \leq 6$.

\section*{Optimum Phase Strategy} \label{sec:optphase}
Here, we wish to show that the phase strategy $f^{\delta}$ represents the best possible strategy in the setting discussed in Ref.~\cite{bib:GLM06}, which we now briefly summarize.  In this setting, $N$ parallel probes are prepared in a state $\ket{\Psi}$, where each probe is acted on by a unitary transformation $U_\varphi\equiv \exp(-i\varphi H)$.  The parallel strategy is thus described by $U_{\varphi}^{\otimes N}$ generated by $h=\sum_{j=1}^N H_j$, where $H_j$ is a Hermitian operator acting on the $j$th probe.

In our scenario, we note that a single mode optical phase shift on the $j$th mode has the form $\exp(-i f_j \varphi \hat{a}_j^\dagger\hat{a}_j)$ (where $\hat{a}^\dagger, \hat{a}$ are the creation and annihilation operators), so that $H_j=f_j \hat{a}_j^\dagger\hat{a}_j$.  It is easy to see that the maximum eigenvalue for any $H_j$ is the case that all $n$ photons probe the $j$th mode, which produces the eigenvalue $f_j\cdot n$.  Trivially, the minimum eigenvalue is $0$ when no photons probe the mode.  Thus, for the $f^{\delta}$ strategy, it is straightforward that $h$ has maximum eigenvalue $n$ and minimum eigenvalue $0$.  However, recall that every strategy must satisfy the constraint,
\begin{equation}
\sum_{j=1}^n f_j=1 \quad \mathrm{where}\; 0\leq f_j< 1,  \label{eq:normalizationapp}
\end{equation}
so that for every strategy other than $f^{\delta}$, $n$ photons cannot fully and simultaneously probe more than one mode, meaning that there is no input state that achieves the maximum eigenvalue for every mode in $h$.

\pagebreak


\chapter*{Vita}
\doublespacing
\setlength{\parindent}{1.75em}
\vspace{0.2em}
\addtocontents{toc}{\vspace{12pt}}
\addcontentsline{toc}{chapter}{\hspace{-1.5em} VITA}
Jonathan ``Jonny" Olson was born in Nampa, Idaho, USA. He received his bachelor's degrees in Physics and Mathematics in 2010, and his Master's degree in Mathematics in 2012 all from the University of Idaho. His primary area of interest as a mathematics student was the study of abstract algebra and analysis.  Jonny joined the Quantum Sciences and Technologies (QST) research group at LSU as a Ph.D. student under Jonathan Dowling in 2012.  After graduation, Jonny plans to move to Boston and work as a postdoc at Harvard University under Al\'{a}n Aspuru-Guzik.

\end{document}